\documentclass[aps,prx,10pt,amsmath,amssymb,floatfix,superscriptaddress,notitlepage,twocolumn]{revtex4-2}
\usepackage{lmodern}
\usepackage[T1]{fontenc}
\usepackage[utf8]{inputenc}
\usepackage[english]{babel}

\usepackage{graphicx}
\usepackage[caption=false]{subfig}
\usepackage{amsmath}
\usepackage{mathrsfs}
\usepackage{amsfonts}
\usepackage{amsthm}
\usepackage{amssymb}
\usepackage{tikz-cd}
\usepackage{enumerate}
\usepackage{wrapfig}
\usepackage{mathtools}
\usepackage[colorlinks=true,linkcolor=blue]{hyperref}
\usepackage{xcolor}
\usepackage{physics}
\usepackage{listings}

\definecolor{codegreen}{rgb}{0,0.6,0}
\definecolor{codegray}{rgb}{0.5,0.5,0.5}
\definecolor{codepurple}{rgb}{0.58,0,0.82}
\definecolor{backcolour}{rgb}{0.95,0.95,0.92}
\definecolor{orchid}{RGB}{218,112,214}

\lstdefinestyle{mystyle}{
    backgroundcolor=\color{backcolour},   
    commentstyle=\color{codegreen},
    keywordstyle=\color{magenta},
    numberstyle=\tiny\color{codegray},
    stringstyle=\color{codepurple},
    basicstyle=\ttfamily\footnotesize,
    breakatwhitespace=false,         
    breaklines=true,                 
    captionpos=b,                    
    keepspaces=true,                   
    showspaces=false,                
    showstringspaces=false,
    showtabs=false,                  
    tabsize=2
}

\newtheorem{teo*}{Teorema}
	
\newtheorem{lem*}{Lema}					

\newtheorem{deff*}{Definición}

\newtheorem{prop*}{Proposición}

\theoremstyle{definition}

\usepackage{stackengine}
\stackMath

\usepackage[normalem]{ulem}

\begin{document}

\title{Theory of superconducting proximity effect in hole-based hybrid semiconductor-superconductor devices}
 \author{D. Michel Pino}
 \affiliation{Instituto de Ciencia de Materiales de Madrid (ICMM), Consejo Superior de Investigaciones Científicas (CSIC), Sor Juana Inés de la Cruz 3, 28049 Madrid, Spain}
  \author{Rub\'en Seoane Souto}
 \affiliation{Instituto de Ciencia de Materiales de Madrid (ICMM), Consejo Superior de Investigaciones Científicas (CSIC), Sor Juana Inés de la Cruz 3, 28049 Madrid, Spain}
  \author{ Maria Jos\'e Calder\'on}
 \affiliation{Instituto de Ciencia de Materiales de Madrid (ICMM), Consejo Superior de Investigaciones Científicas (CSIC), Sor Juana Inés de la Cruz 3, 28049 Madrid, Spain}
  \author{Ram\'on Aguado}
  \email{ramon.aguado@csic.es}
 \affiliation{Instituto de Ciencia de Materiales de Madrid (ICMM), Consejo Superior de Investigaciones Científicas (CSIC), Sor Juana Inés de la Cruz 3, 28049 Madrid, Spain}
  \author{Jos\'e Carlos Abadillo-Uriel}
\email{jc.abadillo.uriel@csic.es}
 \affiliation{Instituto de Ciencia de Materiales de Madrid (ICMM), Consejo Superior de Investigaciones Científicas (CSIC), Sor Juana Inés de la Cruz 3, 28049 Madrid, Spain}
 \begin{abstract}
 Hybrid superconductor-semiconductor systems have received a great deal of attention in the last few years because of their potential for quantum engineering, including novel qubits and topological devices. The proximity effect, the process by which the semiconductor inherits superconducting correlations, is an essential physical mechanism of such hybrids. Recent experiments have demonstrated the proximity effect in hole-based semiconductors, but, in contrast to electrons, the precise mechanism by which the hole bands acquire superconducting correlations remains an open question. In addition, hole spins exhibit a complex strong spin-orbit interaction, with largely anisotropic responses to electric and magnetic fields, further motivating the importance of understanding the interplay between such effects and the proximity effect. In this work, we analyze this physics with focus on germanium-based two-dimensional gases. Specifically, we develop an effective theory supported by full numerics, allowing us to extract various analytical expressions and predict different types of superconducting correlations including non-standard forms of singlet and triplet pairing mechanisms with non-trivial momentum dependence; as well as different Zeeman and Rashba spin–orbit contributions. This, together with their precise dependence on electric and magnetic fields, allows us to make specific experimental predictions, including the emergence of f-type superconductivity, Bogoliubov Fermi surfaces, and gapless regimes caused by large in-plane magnetic fields.
 \end{abstract}

\maketitle
\section{Introduction}
In recent years, superconductor-semiconductor hybrid devices are becoming an exciting playground for exploring new physical phenomena, such as Majorana-based topological protected phases \cite{Alicea_RPP2012,LeijnseReview,AguadoReview,LutchynReview,Prada_review,10.1063/PT.3.4499,flensberg2021engineered} and novel qubit designs \cite{APL-Aguado,Seoane-Aguado2024}. 
The latter, in particular, are becoming an interesting alternative to more traditional qubits due to the benefits arising from the combination of semiconductor and superconductor properties. On the one hand, semiconducting spin qubits \cite{burkard2023semiconductor} are often highly localized, hence potentially more scalable. The natural mechanism for mediating two-qubit gates is the exchange interaction \cite{huang2019fidelity, noiri2022fast, mills2022two, xue2019benchmarking}, which is fast yet quite short-ranged, making it difficult to entangle distant qubits. Superconducting qubits, on the other hand, are easy to couple and read-out by using circuit quantum electrodynamics (cQED) techniques \cite{blais2021circuit}, but physically they are much larger, hence potentially less scalable and more susceptible to crosstalk and noise.

To date, the majority of experiments with hybrids have been performed in semiconductors from group III-V materials, such as InAs and InSb, proximitized by a superconductor. In such semiconductors, nuclear spins are unavoidable, resulting in very short coherence times, as recently demonstrated superconducting in spin qubits \cite{Hays_Science2021,Pita-Vidal_NatPhys2023} based on Andreev levels \cite{Lee-NatureNano14,Janvier_Science2015,HaysNaturePhysics2020,PRXQuantum.3.030311,PhysRevLett.131.097001}. This has recently led to a shift towards group IV semiconductors, particularly germanium, which naturally comprises over $92\%$ nuclear spinless isotopes and can be isotopically purified to significantly reduce hyperfine noise. Besides, many metals have a Fermi-level pinning to the valence band of Ge \cite{dimoulas2006fermi}, making Ge an interesting candidate for the proximity effect on holes. Importantly, holes in the valence band of Ge exhibit a large spin-orbit interaction (SOI), which is a key ingredient in topological superconductivity. Hole bands have a p-orbital character, which together with the spin, leads to total angular momentum $J=3/2$. The SOI in the valence band manifests in different forms, the predominant term being a cubic in momentum Rashba interaction \cite{Winkler:684956}, with small interface-mediated linear corrections \cite{philippopoulos2020pseudospin, liu2022emergent, rodriguez2023linear}, as well as anisotropic and inhomogeneous g-tensors \cite{crippa2018electrical, michal2021longitudinal, terrazos2021theory, zhang2021anisotropic,froning2021ultrafast, liles2021electrical, martinez2022hole, jirovec2022dynamics, AbadilloPRL2023}. These mechanisms enable all-electrical manipulation of the spin \cite{maurand2016cmos, hendrickx2021four} and strong coupling to photons in superconducting cQED setups \cite{yu2023strong, de2023strong, janik2024strong, michal2023tunable, bosco2022fully}.  

Even though the use of Ge for hybrid devices is quite recent, several important achievements have already been demonstrated, including hard-gap superconductivity \cite{Tosato2023}; gate-tunable transmon (gatemon) qubits \cite{Sagi2024,Kiyooka_2024};  parity-conserving Cooper-pair transport and ideal superconducting diode effect in planar germanium \cite{Valentini2024,leblanc2024}; as well as readout of Andreev levels, using both cQED in planar Ge Josephson junctions \cite{PRXQuantum.5.030357} and transport spectroscopy in quantum dots \cite{lakic2024}.

Despite these early successes from the experimental front, the theoretical understanding of the proximity effect in Ge-based hybrids is quite limited. Similarly to the usual procedure with electron states, the simplest route is to add a constant superconductor pairing term to the effective lowest-hole-band theory \cite{ laubscher2024germanium}. The same approach can be extended to assume constant pairings in both heavy and light-hole bands, yielding new effects~\cite{adelsberger2023microscopic}. Such models capture part of the hole physics but miss the interplay of the different bands in the superconducting pairing and, hence, a relevant piece of the physics of the device. A more elaborated theory including both superconductor and semiconductor states coupled through constant tunneling terms has been found to lead to g-factor renormalization effects, as demonstrated for heavy-holes \cite{luethi2023planar}. In fact, based on symmetry, it has been argued that superconducting correlations may only be induced directly to the conduction band \cite{moghaddam2014exporting}, perturbatively exporting superconductivity to the valence band. Real devices, however, have interfaces and disorder that break symmetries in all directions, hence, such symmetry arguments are not expected to hold to their full extent.

In this work, we follow a more agnostic route and, starting from a full $8\times 8$ $\bf k \cdot p$ (8KP) Kane model, consider the effective theory of a proximitized semiconductor involving all the possible superconducting pairing terms arising
from direct coupling between the superconductor and the distinct conduction and hole bands. From there, we extract the effective theory of hole states in strained Ge devices with vertical confinement that applies to most experimental situations with two-dimensional hole gases (2DHG). Importantly, by considering all the different bands, we find that the pairing terms are rather involved, going beyond s-type superconductivity by including non-trivial k-dependencies coupling the different bands and terms originated from the cubic Rashba contributions. 
In this context, one of the main
results of our work is the derivation of the general structure of a pairing matrix in Eq.~\eqref{eq:pairing-to-eigen}, where we present in a compact form the effective superconducting pairings of a proximitized 2DHG modelled with a 4-band Kohn-Luttinger Hamiltonian without further approximations. Such form allows to obtain full analytical expressions for pairings in the hole sector containing both singlet and triplet components in the Rashba basis, Eq.~\eqref{Pauli-Rashba}, by considering that either the conduction, Eq.~\eqref{eq:2DHGgaps}, or the valence band, Eq.~\eqref{eq:2DHGgaps-direct}, are proximitized by a superconductor.
Furthermore, our theory includes Zeeman terms and consider orbital corrections, going well-beyond prior theoretical studies on the proximity effect in Ge hybrid devices.

Armed with these results, we discuss various experimental consequences of the proximity effect in a 2DHG in the presence of an external magnetic field. Specifically, we focus on the magnetic field dependence of the density of states (DOS), a quantity directly relevant for tunneling spectroscopy experiments. Our main result here is the emergence of several magnetic-field-dependent logarithmic van Hove singularities, replacing the standard BCS-like square-root singularities (Fig. \ref{fig:4KP_2DHG_DOS}). This behavior is directly linked to the nontrivial properties of the Bogoliubov de Gennes bands in the hole sector (specifically, the distance between the band extrema and the Fermi level has an oscillatory behavior in terms of the in-plane angle of the
momentum, (Fig \ref{fig:phis_Bx}). At large in-plane magnetic fields, the proximitized 2DHG becomes gapless and Bogoliubov
Fermi surfaces emerge (Fig.~\ref{fig:4KP_2DHG_Bogobands}), a behavior that has recently been observed in a two-dimensional Al-InAs hybrid heterostructure \cite{PhysRevLett.128.107701} and in a proximitized topological insulator film (bismuth telluride placed on top of superconducting niobium diselenide) \cite{doi:10.1126/science.abf1077}, see also \cite{PhysRevB.97.115139,10.21468/SciPostPhys.16.5.115}. Importantly, the intricate form of the proximity effect in the 2DHG results in qualitative differences with respect to a 2DEG, notably various features originating from the gap anisotropy.

The remainder of this paper is organized as follows. In Sec. \ref{sec:8KP}, we show the 8KP Ge bulk model in the Bogoliubov-deGennes formalism with pairing terms coming from the conduction band as the starting point of our theory. In Sec. \ref{sec:4KP}, we derive the resulting 4KP theory for proximitized hole bands after integrating out the conduction band. These results are used in Sec. \ref{sec:2DHG-4KP} to extract an effective model for proximitized 2DHG, from which we obtain analytical expressions for the different superconducting pairing terms (Eq.~\eqref{eq:pairing-to-eigen}) with contributions from the conduction  (Eq.~\eqref{eq:2DHGgaps}) and valence bands (Eq.~\eqref{eq:2DHGgaps-direct}). In Sec. \ref{sec:Signatures}, we discuss different experimental signatures, such as the role of magnetic fields in the gaps  including anisotropic g-factors in Subsec.~\ref{sec:Magnetic}. In Subsec. \ref{BdGFermi-DOS}, we apply our theory to calculate the magnetic field dependence of the DOS and discuss its various features in terms of the anisotropy of the BdG bands, including the emergence of Bogoliubov Fermi surfaces at large in-plane magnetic fields. Finally, we finish the paper in Sec.~\ref{sec:Conclusions} with a summary of our main
results.
We have tried to leave in the main text only the most important derivations, trying at all times to lighten the technical burden and reduce the number of equations for the most part to the minimum necessary, and moved all the more dense mathematical developments to Appendices.

\section{Full BdG 8KP model}
\label{sec:8KP}
\begin{figure*}[ht]
    \centering    \includegraphics[width=\linewidth]{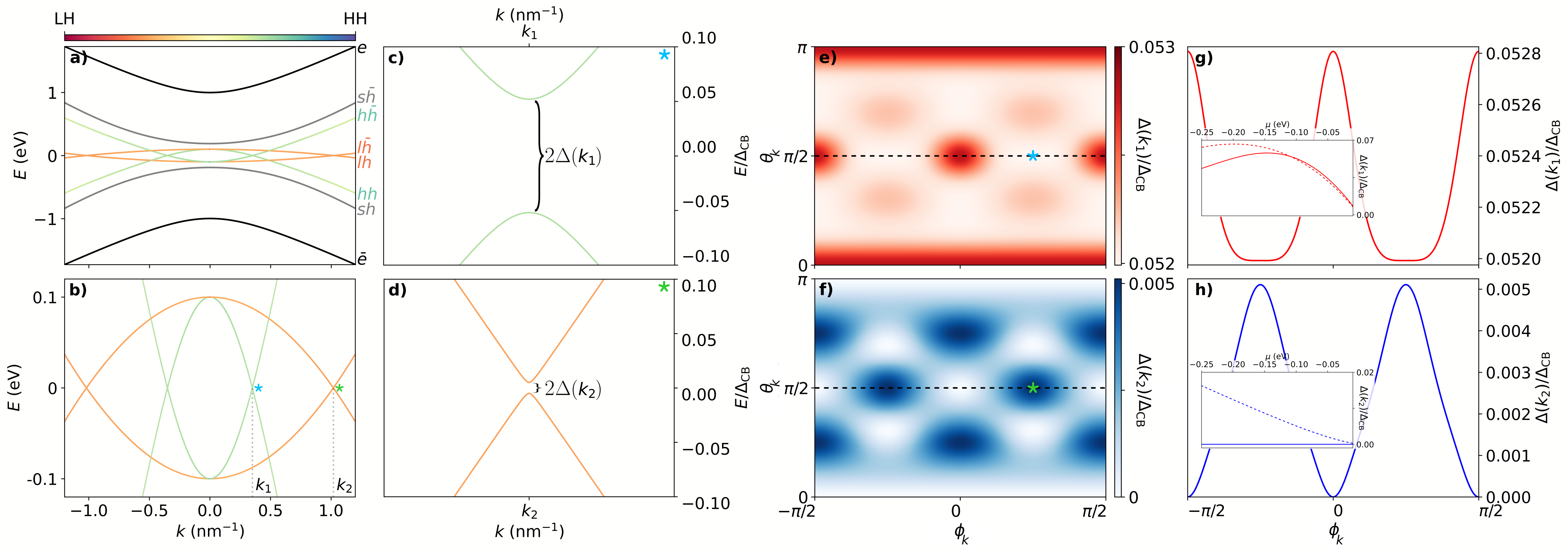}
    \caption{\textbf{8KP model.} (a-b) Energy dispersion of the BdG 8KP model along the $k\,(k_x=k_y)$ in-plane direction (perpendicular to the growth direction). Bands are labeled as $\{e/\bar{e},\; hh/h\bar{h},\; lh/l\bar{h},\; sh/s\bar{h}\}$ respectively for electrons, heavy holes, light holes, split-off holes, and their respective particle and time-reversed counterparts.
    \textbf{CB pairing.} (c-d) Gaps opening along the $k_x=k_y$ direction; at (c) $k_1(\pi/4,\pi/2,\mu)$ and (d) $k_2(\pi/4,\pi/2,\mu)$. The color scale determines the LH/HH nature of the band with the quantity $|\bra{\mathrm{HH}}\ket{\psi}|^2 - |\bra{\mathrm{LH}}\ket{\psi}|^2 \in [-1,1]$. 
    (e-f) Gaps as a function of the spherical coordinates $\phi_k$ and $\theta_k$, such as $(k_x,k_y,k_z)=k(\sin\theta_k\cos\phi_k,\sin\theta_k\sin\phi_k,\cos\theta_k)$; at (e) $k_1(\phi_k,\theta_k,\mu)$ and (f) $k_2(\phi_k,\theta_k,\mu)$. (g-h) Gaps along the black dashed line in (e-f) at (g) $k_1(\phi_k,\pi/2,\mu)$ and (h) $k_2(\phi_k,\pi/2,\mu)$, respectively. (g-h Inset) Gaps as a function of $\mu$; at (g) $k_1$ and (h) $k_2$, respectively; for $\phi_k=0$, $\theta_k=\pi/2$ (solid lines) and $\phi_k=\pi/4$, $\theta_k=\pi/2$ (dashed lines).  Parameters used: $\Delta_\mathrm{CB} = 200$ $\mu$eV; $\mu=-0.1$ eV; Table~\ref{table:Ge-parameters}.}
    \label{fig:panel1_8KP}
\end{figure*}
The Hamiltonian of the hybrid system is given by $H_\text{hybrid}=H_\text{super}+H_\text{semi}+H_\text{tunnel}$. In the Bogoliubov de Gennes formalism, the superconducting Hamiltonian is $H_\text{super}=\frac{1}{2}\sum_\mathbf{k}\Psi_{\sigma,\mathbf{k}}^{s\dagger} H_\text{super}^\text{BdG}\Psi^{s}_{\sigma,\mathbf{k}'}$, where $\Psi^s_{\sigma,\mathbf{k}}$ is the Nambu spinor for superconductor states with momentum $\mathbf{k}$, and spin $\sigma$, and  $H_\text{super}^\text{BdG}$ is given by
\begin{equation}
\begin{aligned}
    H_\text{super}^\text{BdG}&= \\ \begin{pmatrix}
        \frac{p^2}{2m_s}-\mu_s+\frac{1}{2}g_s\mu_B \mathbf{B}\cdot\boldsymbol{\sigma} & i\sigma_y\Delta_s  \\
        -i\sigma_y\Delta_s & -\frac{p^2}{2m_s}+\mu_s-\frac{1}{2}g_s\mu_B \mathbf{B}\cdot\boldsymbol{\sigma}^*
    \end{pmatrix},
\end{aligned}
\label{eq:superconductingH}
\end{equation}
where $m_s$, $\mu_s$, $g_s$, and $\Delta_s$ are the effective mass, chemical potential, g-factor, and parent pairing potential of the superconductor, respectively. For the semiconductor, we consider an $8\times 8$ Kane model:
\begin{equation}
\label{eq:8kp}
    H_\text{semi}=\left( \begin{matrix}
        H_{cb} - \mu & H_{cb-v}  \\
        H_{cb-v} & H_{v} - \mu 
    \end{matrix} \right),
\end{equation}
which includes the first conduction band (CB, with symmetry $\Gamma_{6c}$), described by $H_{cb}$; 
the valence band $H_{v}$, which includes light and heavy holes (LH and HH, $\Gamma_{8v}$), states with total angular momentum $J=3/2$ described by a $H_\mathrm{4KP}$ Kohn-Luttinger Hamiltonian, as well as the split-off (SH, $\Gamma_{7v}$) bands with total angular momentum $J=1/2$. The conduction and valence bands are coupled by the terms $H_{cb-v}$ (see Appendices~\ref{sec:Appendix_8KP-Hamiltonian} and~\ref{sec:Appendix_Ge-parameters}  for the full expressions and semiconductor constants, respectively).

Given that our main goal is to obtain the proximity-induced pairing terms within the semiconductor device, the first step is to integrate-out the superconducting degrees of freedom to obtain an effective Hamiltonian within the semiconductor subspace. The result of this step depends on the assumed tunneling between the superconductor and the semiconductor $H_\text{tunnel}$. We assume the interface may break the symmetry along any spatial direction yet it preserves spin, such that there are spin-conserving constant tunneling terms between the superconductor band and the different bands of the semiconductor, including conduction and valence bands. The  general form of this tunneling term is given in Appendix~\ref{Appendix:approximation}. We eliminate this coupling to second-order by applying a Schrieffer-Wolff transformation, resulting in an effective semiconductor Hamiltonian with constant pairing terms. In the Bogoliubov de Gennes (BdG) formalism this proximitized semiconductor Hamiltonian $H$ can be written as a $16\times 16$ Hamiltonian 
\begin{equation}
    H=\frac{1}{2}\Psi^\dagger H^\mathrm{BdG}_\mathrm{8KP}\Psi,
\end{equation}
expressed in terms of a Nambu spinor basis $\Psi=$($c_{1/2,\mathbf{k}}$, $c_{-1/2,\mathbf{k}}$, $b_{3/2,3/2,\mathbf{k}}$, $b_{3/2,1/2,\mathbf{k}}$, 
$b_{3/2,-1/2,\mathbf{k}}$, $b_{3/2,-3/2,\mathbf{k}}$, $b_{1/2,1/2,\mathbf{k}}$, $b_{1/2,-1/2,\mathbf{k}}$, $c_{1/2,-\mathbf{k}}^\dagger$, $c_{-1/2,-\mathbf{k}}^\dagger$, $b_{3/2,3/2,-\mathbf{k}}^\dagger$, $b_{3/2,1/2,-\mathbf{k}}^\dagger$, $b_{3/2,-1/2,-\mathbf{k}}^\dagger$, $b_{3/2,-3/2,-\mathbf{k}}^\dagger$, $b_{1/2,1/2,-\mathbf{k}}^\dagger$, $b_{1/2,-1/2,-\mathbf{k}}^\dagger$)$^T$, where $c_{s_z,\mathbf{k}}$ ($b_{j,j_z\mathbf{k}}$) destroys an electron (hole) with angular momentum $j$, spin projection $s_z$ (angular momentum projection $j_z$), and momentum $\mathbf{k}$. For a given chemical potential $\mu$, the BdG Hamiltonian is 
\begin{equation}
\label{8kpBdG}
    H^\mathrm{BdG}_\mathrm{8KP}=\left( \begin{matrix}
        H_{cb} - \mu & H_{cb-v} & H^\Delta_{cb} & H^\Delta_{cb-v} \\
        H_{cb-v}^\dagger & H_{v} - \mu & -(H^{\Delta}_{cb-v})^T & H^\Delta_v \\
        {H^\Delta_{cb}}^\dagger & -(H^{\Delta}_{cb-v})^* & \mu - H_{cb}^* & -H_{cb-v}^* \\
        (H^{\Delta}_{cb-v})^\dagger & {H^\Delta_v}^\dagger & -H_{cb-v}^T & \mu - H_{v}^*
    \end{matrix} \right).
\end{equation}
In Eq.~\eqref{8kpBdG} we find that the superconducting proximity effect can induce finite values in the different pairing terms of the conduction $H^\Delta_{cb}$, valence bands $H^\Delta_v$, and mixed conduction-valence terms $H^\Delta_{cb-v}$, where the direct band pairing contributions are parametrized by three real parameters $\Delta_\mathrm{CB}\neq\Delta_\mathrm{LH}\neq \Delta_\mathrm{HH}$. We neglect mixed conduction-valence pairing contributions in the following due to their reduced impact compared to $H_v^\Delta$ and $H_c^\Delta$ (see Appendix~\ref{Appendix:approximation} for details). Considering the proximity-induced gap for the different bands in this way is a good approximation under the assumption of spin-preserving tunneling between the superconductor and semiconductor regions (see Appendix \ref{Appendix:approximation}). We note that integrating out the superconductor can induce renormalizations of the different constants in the Kohn-Luttinger Hamiltonian, such as effective masses and g-factors. However, these corrections are of higher order than the ones produced by strain and confinement~\cite{Winkler:684956, ares2013nature, AbadilloPRL2023}—factors sensitively dependent on geometry but independent of superconductivity—and are therefore beyond the scope of this work.

Fig.~\ref{fig:panel1_8KP}(a,b) shows the band dispersion along $k_x=k_y$. Along the in-plane direction, the most (less) dispersive valence band is HH-type (LH-type). This behavior is reversed along the $z$ axis, parallel to growth direction, where the LH band is more dispersive than the HH one~\footnote{The terms 'heavy hole' and `light hole' refer to their respectively large and small effective masses for the motion in the growth direction. However, these effective masses are reversed in the in-plane motion~\cite{Winkler:684956}. The names HH (LH) usually refer to their $z$ component of angular momentum $m=\pm 3/2$ ($\pm 1/2$).}. Without proximity effect, the pairing terms on the conduction valence bands, $H^\Delta_{cb}$ and $H^\Delta_v$ respectively, are zero. 

Our next step towards an effective 2DHG theory of proximitized Ge holes is to derive and understand the effect of the conduction band within the hole subspace.
We first focus on the superconducting pairing contribution that comes from the $\Gamma_{6c}$ conduction band, i.e., $H_{cb}^\Delta=i\sigma_y\Delta_\mathrm{CB}$ and $H_v^\Delta=0$. As noted in Ref.~\cite{moghaddam2014exporting}, due to the coupling between CB electrons and holes, mediated by $H_{cb-v}$ in Eq.~(\ref{8kpBdG}), $\Delta_\mathrm{CB}$ induces an effective pairing in the valence band. Importantly, the induced pairing is momentum-dependent, anisotropic and leads to different gaps in the most and least dispersive bands, 
$\Delta(k_1)$ and $\Delta(k_2)$, respectively, see Fig.~\ref{fig:panel1_8KP}(c, d). Here, $k_1$ ($k_2$) are defined as the momentum position of the gap in the most (less) dispersive band and depend on the direction
as $k_{1,2}(\phi_k,\theta_k)=k_{1,2}(\sin\theta_k\cos\phi_k,\sin\theta_k\sin\phi_k,\cos\theta_k).$ 
The size of the gaps changes along different directions: 
in particular, $\Delta(k_1)$ is maximal along $(k_x,0,0)$, $(0,k_y,0)$ and $(0,0,k_z)$, while the least dispersive band remains gapless along these directions, $\Delta(k_2)=0$.
This anisotropy is best illustrated by plotting the gaps for less symmetric momentum directions, 
Fig. \ref{fig:panel1_8KP}(e-h). The largest gap $\Delta(k_1)$ has a slight angular dependence, while $\Delta(k_2)$ has a stronger dependence, between $0$ and $\sim 10\%$ of $\Delta(k_1)$. In addition, the size and positions of these induced gaps depend on $\mu$, see the insets of Fig.~\ref{fig:panel1_8KP}(g,h),
but their values are at best $\Delta_\mathrm{eff}\sim 5-10\%$ of the CB parent gap $\Delta_\mathrm{CB}$.

\section{Effective 4KP model}
\label{sec:4KP}
To get the effect of $\Delta_\mathrm{CB}$ within the hole sector, we use a Schrieffer-Wolff transformation \cite{Winkler:684956} to integrate out the CB contributions and get
a 6KP theory involving heavy holes, light holes, and the split-off band. We will further neglect the split-off band, far away in energy ($290$ meV for Ge), to focus on the HH-LH subspace,  which is the most relevant for strained Ge \cite{Scappucci2021}~\footnote{Other materials, such as Si, have a much smaller split-off energy splitting and a 6KP theory is more adequate for their description.}. The resulting Hamiltonian is
\begin{equation}
    H_\mathrm{4KP}^\mathrm{BdG} = \left(\begin{matrix}
    H_\mathrm{4KP} - \mu & H^\Delta_v + H^\Delta_c \\
    (H^\Delta_v+H^\Delta_c)^\dagger & \mu - H_\mathrm{4KP}^*
    \end{matrix}\right) \;,
    \label{eq:bdg4}
\end{equation}
where $H_\mathrm{4KP}$ is the effective Kohn-Luttinger hamiltonian with renormalized parameters, see Appendix~\ref{sec:Appendix_Ge-parameters}.
Temporarily neglecting the mixed HH-LH contributions that we treat in Sec.~\ref{subsec:disorder}, the pairing term may include direct contributions to LH and HH valence bands of the form 
\begin{equation} \label{eq:VB_pairing-term-main}
\begin{aligned}
    H^\Delta_v =& \left(\begin{matrix}
        0 & 0 & 0 & \Delta_\mathrm{HH} \\
        0 & 0 & \Delta_\mathrm{LH} & 0 \\
        0 & -\Delta_\mathrm{LH} & 0 & 0 \\
        -\Delta_\mathrm{HH} & 0 & 0 & 0
    \end{matrix}\right)\;,
\end{aligned}
\end{equation} 
and an effective contribution mediated by their coupling to the CB,
\begin{equation} \label{eq:4KP_pairing}
\begin{aligned}
    & H_c^\Delta = \frac{\Delta_{CB}}{E_g-2\mu} \times
    \\
    & \left(\begin{matrix}
    0 & -R_\Delta & -S_\Delta & -P_\Delta-Q_\Delta \\
    R_\Delta & 0 & P_\Delta-Q_\Delta & S_\Delta^\dagger \\
    S_\Delta & -P_\Delta+Q_\Delta & 0 & -R_\Delta^\dagger \\
    P_\Delta+Q_\Delta & -S^\dagger_\Delta & R_\Delta^\dagger & 0
    \end{matrix}\right),
    \\
    \\
    & \begin{aligned}
    P_\Delta =& -\frac{\hbar^2}{2m_0}\gamma_\Delta(k_x^2+k_y^2+k_z^2)\,,
    \\
    Q_\Delta =& -\frac{\hbar^2}{4m_0}\gamma_\Delta(k_x^2+k_y^2-2k_z^2)\,,
    \\
    R_\Delta =& \frac{\hbar^2}{4m_0}\sqrt{3}\gamma_\Delta\left[(k_x^2-k_y^2)-2ik_xk_y\right]\,,
    \\
    S_\Delta =& -\frac{\hbar^2}{2m_0}\sqrt{3}\gamma_\Delta k_-k_z\,,
    \end{aligned}
\end{aligned}
\end{equation}
where $E_g$ is the band gap, $m_0$ the bare electron mass, and $\gamma_\Delta=7.38$ is calculated from the band parameters of Ge, see Appendix~\ref{sec:Appendix_Ge-parameters}.

The symmetry of the induced effective pairing in the hole band is inherited from the lowest $\Gamma_{6c}$ conduction band which has spherical symmetry, leading to a single effective Luttinger parameter $\gamma_\Delta$~\cite{moghaddam2014exporting}. On the other hand, the 4KP Hamiltonian has a lower symmetry and is characterized by three Luttinger parameters $\gamma_1$, $\gamma_2$, and $\gamma_3$. This mismatch in symmetries induces non-trivial anisotropies of the effective pairings in the eigenbasis of the 4KP Hamiltonian as shown in Fig.~\ref{fig:panel1_8KP}(e,f). The expressions for the in-plane components ($\theta_k=\pi/2$), in the eigenbasis of $H_{4KP}$,  are
\begin{equation} \label{eq:4KP_analytical-gaps}
\begin{aligned}
    \Delta(\mathbf{k}_1)&\approx \frac{\Delta_\mathrm{CB}\gamma_\Delta\hbar^2|\mathbf{k}_1|^2}{(E_g-2\mu)m_0}\left[1-\frac{3\gamma_-^2\sin^2 2\phi_k}{32\gamma_+^2}\left(5+3\cos 4\phi_k\right)\right] \\
    \Delta(\mathbf{k}_2)&\approx \frac{\Delta_\mathrm{CB}\gamma_\Delta\hbar^2|\mathbf{k}_2|^2}{(E_g-2\mu)m_0}\left[\frac{3\gamma_-^2\sin^2 2\phi_k}{32\gamma_+^2}\left(5+3\cos 4\phi_k\right)\right] ,
\end{aligned}
\end{equation}
where we have expanded in $\gamma_-/\gamma_+$ up to second order, with $\gamma_\pm=\gamma_3 \pm \gamma_2$ the Luttinger parameters of the hole bands. The anisotropic behavior is proportional to $\gamma_-$, which quantifies the lack of spherical symmetry of the system.
Eq.~\eqref{eq:4KP_analytical-gaps} displays the complementary behavior of $\Delta(k_1)$ and $\Delta(k_2)$ shown in Fig.~\ref{fig:panel1_8KP}. Besides, Eq.~\eqref{eq:4KP_pairing} displays pairing terms that couple the different bands. Hence, at points where there is a crossing between holes and their time-reversed states with a different spin, an anticrossing may occur. For a detailed benchmarking between the effective 4KP and the original 8KP model, see Appendix~\ref{Appendix:benchmark}.

\section{Effective 2DHG 4KP model}
\label{sec:2DHG-4KP}
Now we apply this effective theory to the experimentally relevant scenario of a 2DHG in Ge. 
We focus on strained Ge/Ge$_{1-x}$Si$_x$ quantum wells with a depth of $L_W$. 
The HH and LH bands are split by $E_{hl}$ due to their different effective masses in the confinement direction.
The different lattice constants of the GeSi barriers and the Ge well introduce an extra energy splitting $E_{hl}^\text{strain}$~\cite{bir1974symmetry}. Within the hard-wall approximation for confinement, we use the Bastard wavefunctions~\cite{npj2021} to estimate $E_{hl}$ and the spin splitting, see Appendix~\ref{sec:Appendix_confinement}. 

Assuming $L_W=16$ nm and a Si content of 20$\%$, $E_{hl}\approx 70$ meV for experimentally accesible vertical electric fields $|F|<5$ MV/m. 
The vertical electric field breaks the inversion symmetry, and the effective mass difference between HH and LH induces a Rashba SOI within the hole subspace. In particular, for the vertical ground state, we get 
$\bra{0_H(z)}p_z\ket{0_L(z)}=i\alpha_0$, where $\alpha_0$ is proportional to $F$ (see Appendix \ref{sec:Appendix_confinement}). The finite value of $\alpha_0$ induces an effective cubic in momentum SOI within the HH subspace~\cite{Winkler:684956} and leads to a spin-splitting in k-space for both HH and LH.
\begin{figure}[h]
    \centering    \includegraphics[width=\linewidth]{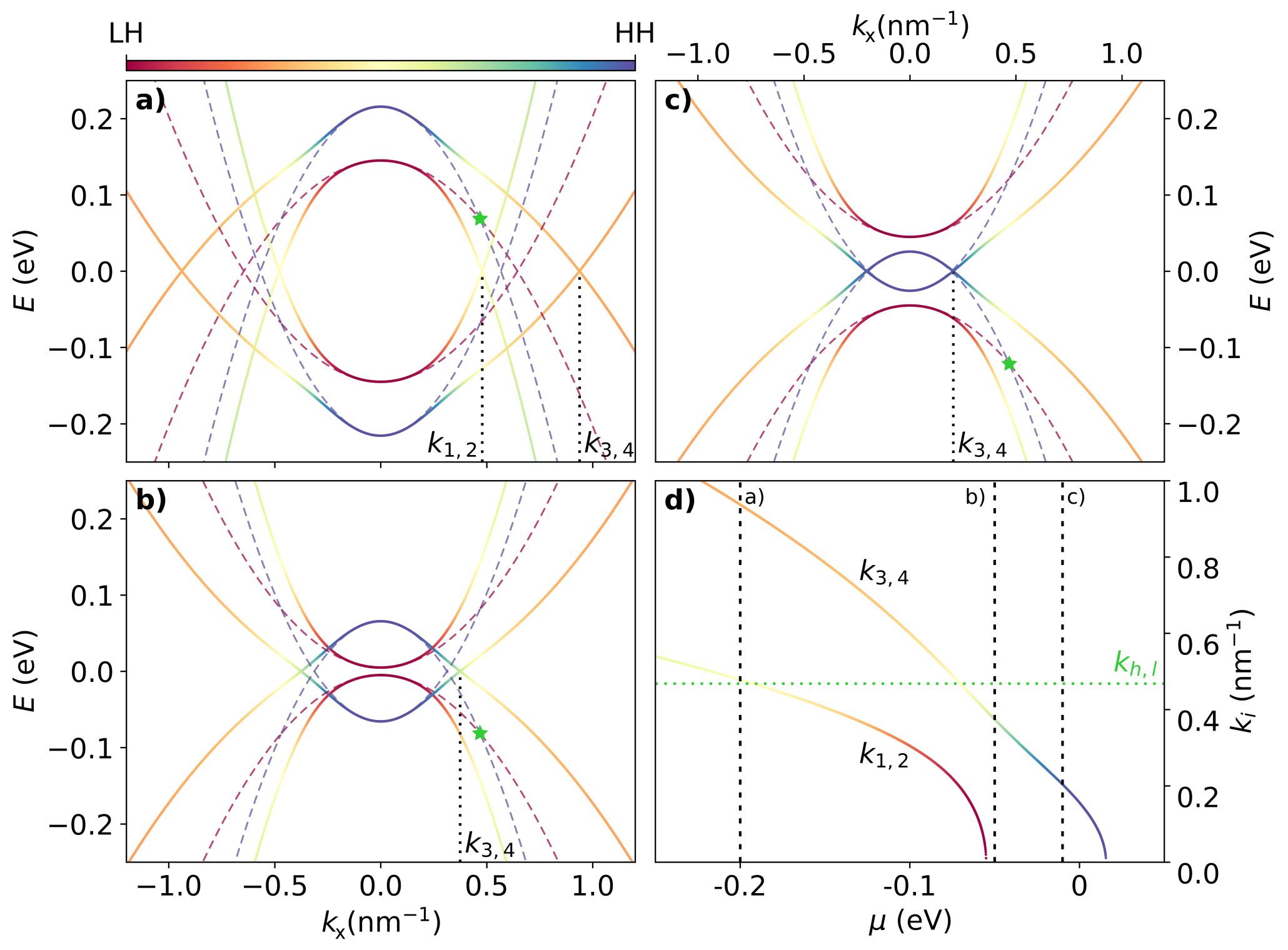}
    \caption{\textbf{2DHG 4KP regimes for different {\boldmath$\mu$}.} (a) Bands along the $k_x$ direction for $\mu=-0.2$ eV. At $\mathbf{k}=0$, the two spin-degenerate hole bands, split in energy by $E_{hl}$, cross the Fermi level at momenta $k_{1,2}$ and $k_{3,4}$, respectively. While these two bands retain their well-defined HH and LH character at low momenta, they gradually mix and hybridize as momentum grows. (b) Bands  along the $k_x$ direction for $\mu=-0.05$ eV, where only one spin-degenerate band crosses the Fermi level, yet this band exhibits strong HH-LH hybridization. (c) Bands along the $k_x$ direction for $\mu=-0.01$ eV, where the crossing band has a well defined HH character. (d) Position of the different gaps in quasi-momentum $k_i$ for $F=0$ along the $k_x$ direction. Black dashed vertical lines indicate the different values of $\mu$ used in panels (a-c). The green dotted line in panel (d) and the green star marker in panels (a-c) indicate when the HH  and the LH band dispersions cross in momentum $\hbar k_{h,l}=\sqrt{E_{hl}m_0/\gamma_2}$, calculated under the assumption of decoupled bands --dashed curves in panels (a-c). Parameters used: $\Delta_\mathrm{CB}=200$ $\mu$eV; $F=0$; Table \ref{table:Ge-parameters}; Appendix \ref{sec:Appendix_confinement}.}
    \label{fig:4KP_2DHG_F-0}
\end{figure}
\begin{figure*}[ht]
    \includegraphics[width=\linewidth]{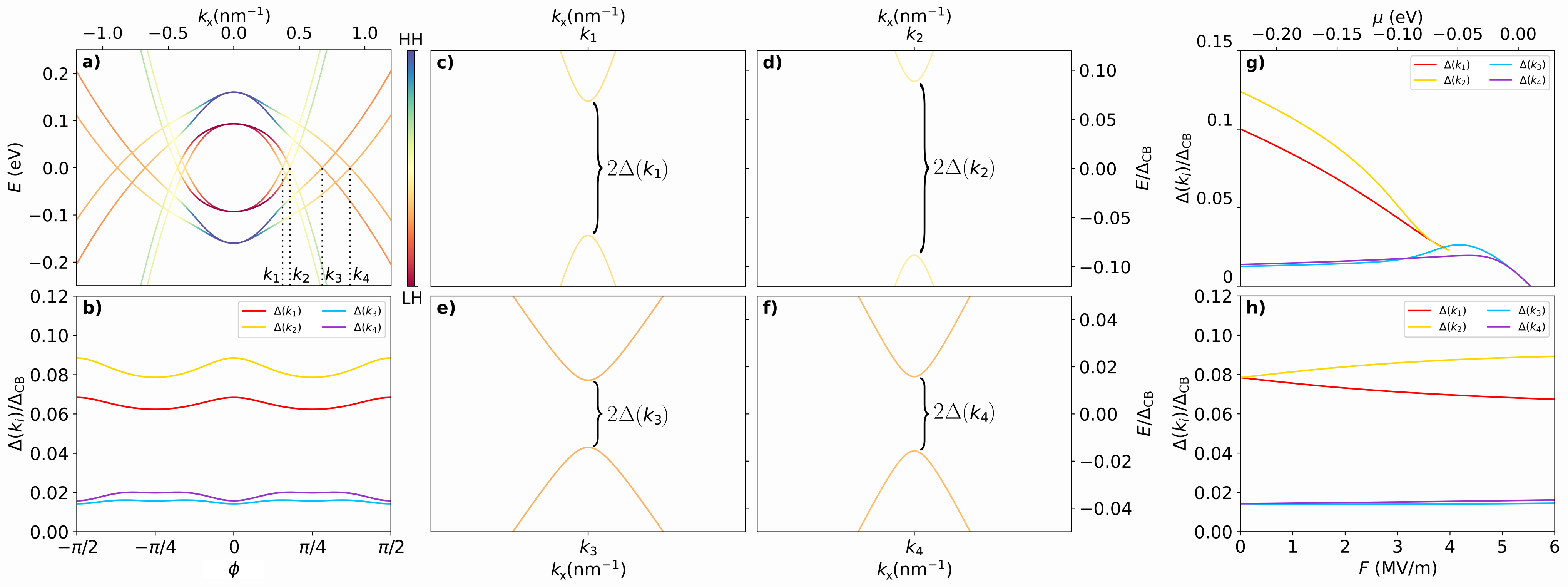}
    \caption{\textbf{Induced pairings in the 2DHG 4KP model with SC proximity coming from the CB only.} (a) Energy dispersion of the BdG 2DHG 4KP model along the $k_x$ direction. \textbf{CB pairing.} (b) SC pairing as a function of the polar coordinate $\phi_k$, such as $(k_x,k_y)=k(\cos\phi_k,\sin\phi_k)$; at $k_{1,2,3,4}(\phi_k,\mu)$. (c-f) Gaps opening along the $k_x$ direction; at $k_{1,2,3,4}(0,\mu)$, respectively. (g-h) Gaps opening at $\phi_k=0$ as a function of (g) $\mu$ with $F=5$ MV/m, and (h) $F$ with $\mu=-0.15$ eV. Parameters used: $\Delta_\mathrm{CB}=200$ $\mu$eV; $\mu=-0.15$ eV and $F=5$ MV/m unless otherwise stated; Table \ref{table:Ge-parameters}; Appendix \ref{sec:Appendix_confinement}.}
    \label{fig:panel_4KP-2DHG}
\end{figure*}

We first identify the different regimes in terms of the chemical potential. In Fig.~\ref{fig:4KP_2DHG_F-0} we show the bands and Fermi-level crossing points $k_i$ as a function of $\mu$ for $F=0$. Fig.~\ref{fig:4KP_2DHG_F-0}(a) shows that, for large negative values of $\mu$, both HH and LH bands cross the Fermi level and there are up to four different crossings. In this regime, the bands exhibit strong HH-LH hybridization, as can be seen from the color code. In Fig.~\ref{fig:4KP_2DHG_F-0}(b) we show an intermediate regime where the value of $\mu$ leads to only two Fermi-level crossings, yet they exhibit a strong HH-LH hybridization. Finally, only when $|\mu|\ll E_{hl}$ the two bands crossing have approximately well-marked HH behavior, see Fig.~\ref{fig:4KP_2DHG_F-0}(c). Due to their different effective masses, HH and LH bands anticross in $k$ space at the value $\hbar k_{h,l}=\sqrt{E_{hl}m_0/\gamma_2}$ (marked as a green star in Fig.~\ref{fig:4KP_2DHG_F-0}); this value of $k$ can be understood as an approximate regime boundary as shown in Fig.~\ref{fig:4KP_2DHG_F-0}(d): as HH-LH bands become strongly hybridized, perturbation theory fails and a full 4-level model must be used.  Conversely, perturbation theory only makes sense if, for a given $\mu$, the bands cross the Fermi level well before this hybridization point $k_{h,l}$. In Fig.~\ref{fig:4KP_2DHG_F-0}(d), case a) would represent a situation well outside the validity range of perturbation theory,  while case c) would correspond to a situation in which perturbation theory could be applied, since the bands crossing the Fermi level have a well defined HH  character. However, and since, in general, the precise value of $\mu$ is largely unknown, we would like to emphasize that even in situations where only HH bands are  expected to participate in the proximitized 2DHG, a full 4-level model, like the one used here, seems to be the correct approach.

\subsection{Analytical pairing expressions}
\label{sec:analytics-2DHG-4KP}
Given that perturbation theory is only valid in a small window of $\mu$, we analyze our problem with the full 4-bands Hamiltonian and the vertical Bastard wavefunctions. Besides, an exact treatment circumvents the issue with diverging Fermi surfaces in effective HH theories with superconductivity~\cite{adelsberger2023microscopic}. Within this approximation, the 4-band Kohn-Luttinger Hamiltonian $H_{4KP}$ in Eq.~\eqref{eq:bdg4} can be diagonalized analytically at zero magnetic field. The procedure of this exact diagonalization is explained in Appendix \ref{Appendix:exactdiag} and yields
\begin{equation}
\label{eq:H-to-eigen}
    H_{\text{4KP}}^{(d)}=UH_{\text{4KP}}U^\dagger=\text{diag}\left(E_2, E_1, E_4, E_3\right)\,,
\end{equation}
where $U$ is the unitary diagonalizing the hole Hamiltonian. Conversely, the time-reversed Hamiltonian is diagonalized by $U'=U^*(-\mathbf{k})$. The energies $E_i$ are associated with the states that cross the Fermi level at $k_i$ in Figs.~\ref{fig:4KP_2DHG_F-0} and \ref{fig:panel_4KP-2DHG}.  Hence, at $\mathbf{k}=0$, $E_{1,2}$ are eigenvalues of the LH states and $E_{3,4}$ are eigenvalues of the HH band. Note that these energies are not written in an explicit manner including their full dependence on parameters $E_i(k_x,k_y,F,\mu)$ for simplicity. %
The eigenbasis can be expressed in terms of a rotated Nambu basis as $\tilde{\Psi}=$$(\tilde{b}_{2,\tilde{\uparrow}}$$,\tilde{b}_{1,\tilde{\downarrow}}$$,\tilde{b}_{4,\tilde{\uparrow}}$$,\tilde{b}_{3,\tilde{\downarrow}}$$,\tilde{b}^\dagger_{2,\tilde{\uparrow}}$$,\tilde{b}^\dagger_{1,\tilde{\downarrow}}$$,\tilde{b}^\dagger_{4,\tilde{\uparrow}}$$,\tilde{b}^\dagger_{3,\tilde{\downarrow}})^T$, where $\tilde{b}_{i,\tilde{s}}$ destroys a hole band with eigenenergy $E_i$ and spin $\tilde{s}$ aligned or anti-aligned with the spin-orbit interaction. Importantly, on this basis, the pairing block of the Hamiltonian takes the form
\begin{equation}
\label{eq:pairing-to-eigen}
\begin{aligned}
    H^{\Delta\text{ (d)}}&=UH^\Delta U'^\dagger=\Delta_{ij} \\ &=\left(\begin{array}{c|c}
        \tilde{\boldsymbol{\Delta}}_{l}(\mathbf{k}_1,\mathbf{k}_2)\cdot \boldsymbol{\sigma} & \tilde{\boldsymbol{\Delta}}_{lh}(\mathbf{k}_i)\cdot \boldsymbol{\sigma} \\ \hline
        \tilde{\boldsymbol{\Delta}}_{hl}(\mathbf{k}_i)\cdot \boldsymbol{\sigma} & \tilde{\boldsymbol{\Delta}}_{h}(\mathbf{k}_3,\mathbf{k}_4)\cdot \boldsymbol{\sigma}
    \end{array}\right) \,,
    \end{aligned}
\end{equation}
where $H^\Delta$ may include pairing amplitudes between any of the bands, and we introduce the block operators $\tilde{\boldsymbol{\Delta}}_i$ as the pairing components $\Delta_{ij}$ grouped in blocks and projected to the space of Pauli matrices in the Rashba basis, such that 
\begin{equation}
 \label{Pauli-Rashba} \tilde{\boldsymbol{\Delta}}_{i}\cdot\boldsymbol{\sigma}=\sum_{j=\{0,x,y,z\}} \tilde{\Delta}_{i,j}\sigma_j .
\end{equation}
The diagonal blocks $\tilde{\boldsymbol{\Delta}}_{l/h}$ are intraband pairing terms associated to the bands with LH/HH character at low $k$ values, and the off-diagonal blocks $\tilde{\boldsymbol{\Delta}}_{lh/hl}$ are interband pairing terms related to different anticrossings between hole and their time-reversed partners of different bands away from the Fermi level. In this basis, the pairings corresponding to the $\sigma_0$ and $\sigma_z$ components of intraband blocks introduce gaps at the Fermi level; we name these terms as longitudinal pairing terms. On the other hand, the $\sigma_x$ and $\sigma_y$ terms lead to anticrossings away from the Fermi level; we group these components as transverse pairing terms. Longitudinal and transverse terms can be related to singlet and triplet pairing correlations. Eq.~\eqref{eq:pairing-to-eigen} is one of the main results of our paper and allows us to obtain explicit pairing terms for the different hole bands. In what follows, we provide analytical expressions of these pairings terms expanded up to third order in momenta, and refer to Appendix~\ref{Appendix:exactdiag} for the exact expressions.

Focusing on the intraband blocks, and 
assuming that the proximity effect originates from the conduction band only, the expansions of the non-zero contributions coming from $H^\Delta=H^\Delta_{(cb)}$ are:
\begin{equation}
\label{eq:2DHGgaps}
\begin{aligned}
    \frac{\tilde{\Delta}_{l,0}^{(c)}}{\Delta_{CB}}&=\frac{\tilde{\Delta}_{h,0}^{(c)}}{\Delta_{CB}}\approx \frac{-3i\alpha_R\gamma_\Delta}{2(E_g-2\mu)m_0}p^2 e^{3i\phi_k} \,,
    \\
    \frac{\tilde{\Delta}_{l,x}^{(c)}}{\Delta_{CB}}&=\frac{\tilde{\Delta}_{h,x}^{(c)}}{\Delta_{CB}}\approx\frac{3i\alpha_R\gamma_\Delta\left(\gamma_+-i\gamma_-e^{4i\phi_k}\right)}{2(E_g-2\mu)m_0\gamma_3}p^2 e^{3i\phi_k} \,,
    \\
    \frac{\tilde{\Delta}_{l,z}^{(c)}}{\Delta_{CB}}&\approx\frac{-i\gamma_{\Delta}}{4(E_g-2\mu)}(p^2+4\langle p_z^2\rangle)e^{3i\phi_k} \,,
    \\
    \frac{\tilde{\Delta}_{h,z}^{(c)}}{\Delta_{CB}}&\approx\frac{3i\gamma_{\Delta}}{4(E_g-2\mu)}p^2e^{3i\phi_k} \,,
\end{aligned}
\end{equation}
where $\alpha_R$ is a linear-in-momentum adimensional Rashba coefficient $\alpha_R=\gamma_3\alpha_0p/(E_{hl}m_0)$,  $p=\hbar\sqrt{k_x^2+k_y^2}$, and $\phi_k=\arctan(k_x/k_y)$. These expressions feature both longitudinal and transverse types of pairing terms. 

If, on the other hand, we assume that direct proximity effect is possible in the  valence band with pairing terms $H^\Delta=H^\Delta_v$ including both $\Delta_\mathrm{HH}$ and $\Delta_\mathrm{LH}$, see Eq. \eqref{eq:VB_pairing-term-main}, the non-zero intraband components are
\begin{equation}
\label{eq:2DHGgaps-direct}
\begin{aligned}
    \frac{\tilde{\Delta}_{l,0}^{(v)}}{\bar{\Delta}_H}&=\frac{\tilde{\Delta}_{h,0}^{(v)}}{\bar{\Delta}_H}\approx -i\alpha_R\zeta e^{3i\phi_k} \,,
    \\
    \frac{\tilde{\Delta}_{l,z}^{(v)}}{\Delta_{\mathrm{LH}}}&\approx i\left(1-\frac{6\bar{\Delta}_H}{\Delta_{\mathrm{LH}}}\alpha_R^2\right)e^{3i\phi_k} \,,
    \\
    \frac{\tilde{\Delta}_{h,z}^{(v)}}{\Delta_{\mathrm{HH}}}&\approx i\left(1-\frac{6\bar{\Delta}_H}{\Delta_{\mathrm{HH}}}\alpha_R^2\right)e^{3i\phi_k} \,,
\end{aligned}
\end{equation}
where we have  defined an average pairing $\bar{\Delta}_H=(\Delta_{\mathrm{HH}}+\Delta_{\mathrm{LH}})/2$ and the adimensional quantity $\zeta=6(\gamma_+-\gamma_-e^{4i\phi_k})p^2/(E_{hl}m_0)$. Note that, in contrast to Eq.~\eqref{eq:2DHGgaps}, the effective pairings in Eq.~\eqref{eq:2DHGgaps-direct} only feature longitudinal components since the transverse components are identically zero up to any order, see Appendix~\ref{Appendix:exactdiag}.

\subsection{Conduction band contributions to the gap}
\label{subsec:directCB}
We now analyze some representative results corresponding to the case $\Delta_\mathrm{HH}=\Delta_\mathrm{LH}=0$ and $\Delta_\mathrm{CB}\neq 0$. We turn on the electric field in Fig.~\ref{fig:panel_4KP-2DHG}(a), where we show the band structure along the $k_x$ direction after projecting the vertical motion to the ground state. 
All four crossing points between bands may display a gap, as seen in Fig.~\ref{fig:panel_4KP-2DHG}(b-f). The value of the gap at these different crossing points is heavily dependent on the nature of the bands and, strikingly, exhibits a certain degree of gate-tunability through the vertical electric field $F$ and the chemical potential $\mu$. As can be seen in Fig.~\ref{fig:panel_4KP-2DHG}, this gap can go up to 10$\%$ of $\Delta_{CB}$ for the first band crossings in the large negative $\mu$ regime. Hence, assuming that the CB superconducting gap $\Delta_{\rm CB}$ comes entirely from Al, with $\Delta_{\rm Al}\approx 200\mu$eV, the induced superconducting gap through the conduction band can reach values of around 20$\mu$eV in the valence band. For $k_{3,4}$, the value of the gap is about five times smaller: for Al, it would be a few $\mu$eV at most. 

To further understand the nature of the induced superconductivity, we go to the expressions of the pairings in the eigenbasis in Eq.~\eqref{eq:2DHGgaps}. 
These expressions already provide an intuitive framework to analyze the pairing potentials in Fig.~\ref{fig:panel_4KP-2DHG}. The presence of $\sigma_0$ and $\sigma_z$ pairing terms as well as their intrinsic dependence on $k$ lead to each gap displaying differences in their dependence with $\mu$ and $F$. The gaps are mainly dominated by $\tilde{\Delta}_{l,z}(\mathbf{k}_1,\mathbf{k}_2)$ and $\tilde{\Delta}_{h,z}(\mathbf{k}_3,\mathbf{k}_4)$, which are $\propto p^2$, a kinetic  term. For $\tilde{\Delta}_{l,z}$ there is an extra $p_z^2$ dependence, which is responsible for the relative shift of $\Delta(k_1),\ \Delta(k_2)$ compared to $\Delta(k_3),\ \Delta(k_4)$ in Fig.~\ref{fig:panel_4KP-2DHG}(b). In Fig.~\ref{fig:panel_4KP-2DHG}(g), we show a non-trivial dependence with $\mu$. As $\mu$ goes from more negative values to zero, the gaps tend to decrease. This is a consequence of their proportionality with $k$: lower values of $|\mu|$ decrease the values of $k_i$, leading to lower kinetic pairing terms. 

The gap dependence on vertical electric field in Fig.~\ref{fig:panel_4KP-2DHG}(h) can be identified from the identity components $\tilde{\Delta}_{l/h,0}$ in Eq.~\eqref{eq:2DHGgaps}, which are directly related to the cubic Rashba interaction, that introduce a gate-dependence through the Rashba coefficient $\alpha_0(F_z)$. The direct dependence on $p_z^2$ induces a stronger field dependence in $\Delta(k_{1,2})$ than that of $\Delta(k_{3,4})$, as can be seen in Fig.~\ref{fig:panel_4KP-2DHG}(h). Finally, the transverse pairing contributions $\tilde{\Delta}_{i=l/h,x}$ also arise from the cubic Rashba interaction. Interestingly, this transverse pairing term appears even in the absence of a magnetic field and is proportional to the vertical electric field. Physically, the transverse pairing term does not influence the gap at zero magnetic field but induces anticrossings between hole and their time-reversed partners with opposite spin outside the Fermi level.  

\subsection{Direct heavy-hole and light-hole contributions to the gap}
\label{subsec:directHH-LH}
\begin{figure}[t]
    \centering
    \includegraphics[width=0.7\linewidth]{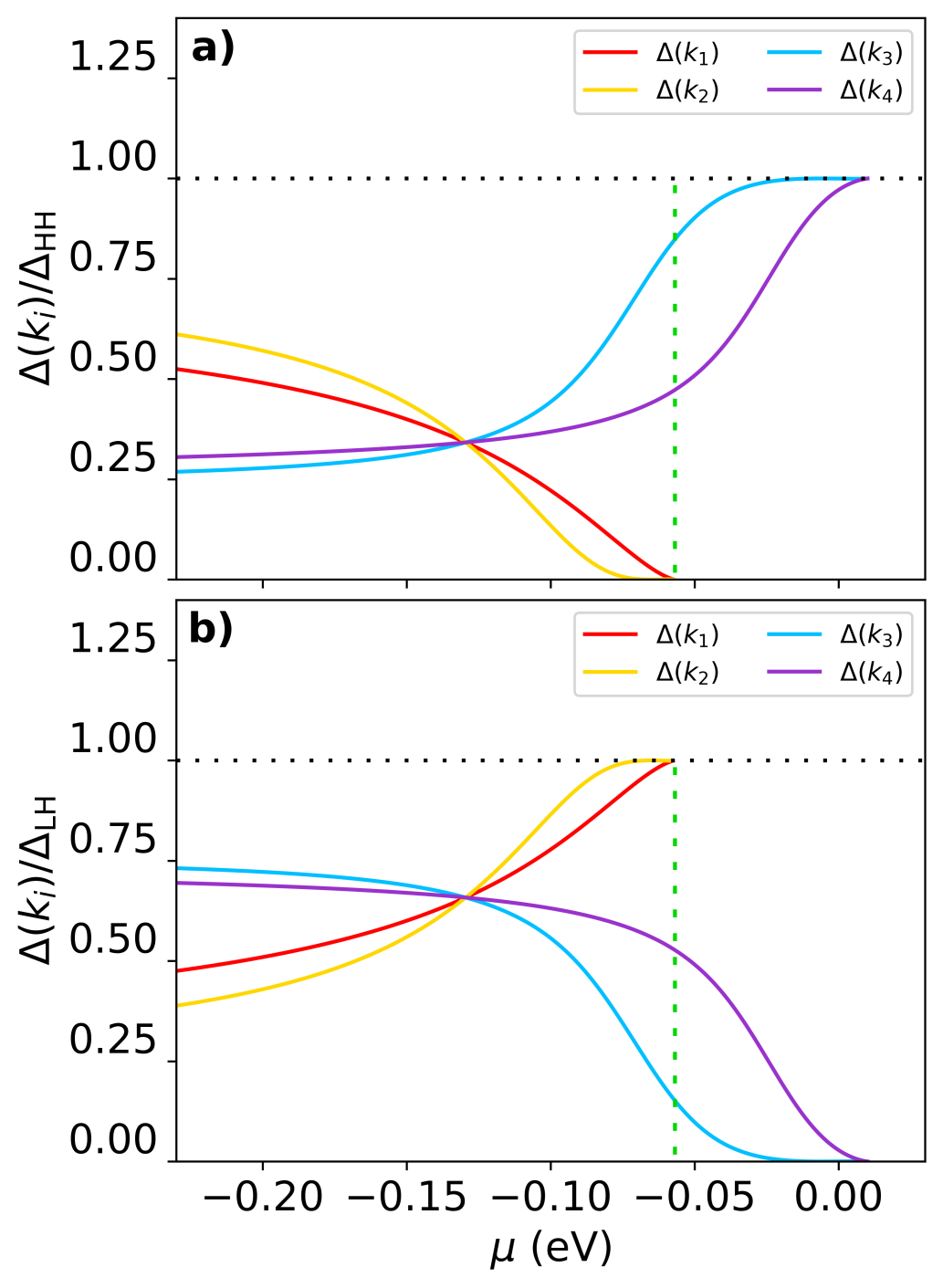}
    \caption{\textbf{Direct LH/HH pairing.} Gaps $\Delta(k_{1,2,3,4})$ along the $k_x$ direction, as a function of $\mu$ for (a) HH and (b) LH direct pairing. The vertical dashed line indicates $\mu=-E_{hl}$, where $\Delta(k_1,k_2)$ emerge. Parameters used: $\Delta_{\mathrm{LH/HH}} = 200$ $\mu$eV; $F=5$ MV/m; Table \ref{table:Ge-parameters}; Appendix \ref{sec:Appendix_confinement}.}
    \label{fig:4KP_2DHG_LH-HH-direct}
\end{figure}

The measured gaps in experiments can be close to the parent superconducting gap~\cite{Tosato2023, Sagi2024, Valentini2024, lakic2024} unlike our much smaller prediction (at most $\sim 10\% \Delta_{\rm CB}$) when assuming a CB-only proximity effect, see Sec.~\ref{subsec:directCB}. Even though symmetry arguments~\cite{moghaddam2014exporting} may imply that the only direct proximity effect can be induced by the CB, the superconductor-semiconductor interface reduces the symmetry and may introduce direct tunneling between the superconductor electrons and the semiconductor valence band~\cite{futterer2011band}. It is then relevant to introduce $\Delta_{\mathrm{HH}}$ and $\Delta_{\mathrm{LH}}$ (see Appendix~\ref{Appendix:approximation}).

In Fig.~\ref{fig:4KP_2DHG_LH-HH-direct} we show the dependence of the induced pairings on the chemical potential $\mu$.
At low values of $|\mu|$, in the perturbative regime, the bands crossing the Fermi level at $k_3$ and $k_4$ have strong HH character. Consequently, there is a strong contribution to the gaps in these bands coming directly from the pairing term $\Delta_{\mathrm{HH}}$, such that $\Delta(k_{3,4})\approx \Delta_{\mathrm{HH}}$, see Fig.~\ref{fig:4KP_2DHG_LH-HH-direct}(a). As $\mu$ gets more negative, LH and HH bands hybridize, reducing the relative weight of $\Delta_{\mathrm{HH}}$, hence $\Delta(k_{3,4})$ decrease.  $\Delta(k_{1,2})$, with a well-marked LH character at low values of $k$, exhibit the opposite behavior, starting from $\mu\approx -E_{hl}$. These trends are exchanged when $\Delta_{\mathrm{LH}}$ is switched on, Fig.~\ref{fig:4KP_2DHG_LH-HH-direct}(b).

This picture is consistent with Eq.~\eqref{eq:2DHGgaps-direct}, where we find longitudinal pairing contributions with a series of corrections that can be interpreted as the effect of mixing HH and LH states. Proportional to the identity, we find another correction that is linear with the electric field and comes from the cubic Rashba interaction. The competing $\sigma_0$ and $\sigma_z$ terms introduce a spin-dependent gap size for states within the same band that is tunable mostly through the value of $\mu$, which changes the values of $k_i$ at which the Fermi level crossings occur. 

\subsection{Mixed heavy-light contributions}
\label{subsec:disorder}
Interface mismatches and the existence of intermediate barriers, for example amorphous oxides, may introduce new terms in the Hamiltonian due to the reduced symmetry. This occurs even in quite compatible interfaces, such as Si/SiGe and GaAs/AlGaAs~\cite{ivchenko1996heavy, rodriguez2023linear}.
Such reduced symmetry, may allow tunneling events between the superconductor and semiconductor that mix heavy- and light-hole states~\cite{babkin2024superconducting}, see Appendix~\ref{Appendix:approximation}. These terms are analogous to the Bir-Pikus Hamiltonian~\cite{bir1974symmetry} for strain, where deformations breaking the symmetry along different directions lead to the general expression:
\begin{equation}
\label{eq:4KP-pairing-mixed}
    H_{m}^\Delta=\begin{pmatrix}
        0 & \Delta_Re^{i\chi_{R}} & \Delta_{S}e^{i\chi_{S}} & 0 \\
        -\Delta_Re^{i\chi_{R}} & 0 & 0 & -\Delta_{S}e^{-i\chi_{S}} \\
        -\Delta_{S}e^{i\chi_{S}} & 0 & 0 & \Delta_Re^{-i\chi_{R}} \\
        0 & \Delta_{S}e^{-i\chi_{S}} & -\Delta_Re^{-i\chi_{R}} & 0
    \end{pmatrix} \,.
\end{equation}
$\Delta_R$ and $\Delta_S$ can be written as $\Delta_Re^{i\chi_{R}}=(\Delta_{YY}-\Delta_{XX})+i\Delta_{XY}$ and $\Delta_Se^{i\chi_{S}}=\Delta_{XZ}+i\Delta_{YZ}$, illustrating broken symmetries along different directions in analogy to strain in the Bir-Pikus Hamiltonian.
We follow the same procedure as in subsection~\ref{sec:analytics-2DHG-4KP} and write this pairing matrix in the frame that diagonalizes the 4-bands Kohn-Luttinger Hamiltonian. For the mixed intraband corrections $\tilde{\boldsymbol{\Delta}}^{(m)}_{l/h}$ we get: 
\begin{equation}
\label{eq:2DHGgaps-mixed}
\begin{aligned}
    \frac{\tilde{\Delta}_{l,0}^{(m)}}{\Delta_R}&=\frac{\tilde{\Delta}_{h,0}^{(m)}}{\Delta_R} \approx  2i\sqrt{3}\alpha_Re^{3i\phi_k}\cos(2\phi_k+\chi_R) \\
    \frac{\tilde{\Delta}_{l,x}^{(m)}}{\Delta_R}&=-\frac{\tilde{\Delta}_{h,x}^{(m)}}{\Delta_R} \approx  2\sqrt{3}\alpha_Re^{3i\phi_k}\sin(2\phi_k+\chi_R)  \\
    \frac{\tilde{\Delta}_{l,y}^{(m)}}{\Delta_S}&=\frac{\tilde{\Delta}_{h,y}^{(m)}}{\Delta_S}\approx 2\sqrt{3}\alpha_Re^{3i\phi_k}\sin(\phi_k+\chi_S) \\
    \frac{\tilde{\Delta}_{l,z}^{(m)}}{\Delta_R}&=\frac{\tilde{\Delta}_{h,z}^{(m)}}{\Delta_R} \approx  \frac{-i\zeta e^{3i\phi_k}}{2\sqrt{3}}\cos(2\phi_k+\chi_R).
\end{aligned}
\end{equation}
Interestingly, mixed terms $\Delta_{S}$ introduce transverse terms, while those coming from $\Delta_R$ introduce both longitudinal and transverse pairing in the eigenbasis of the 4-bands Hamiltonian. All terms exhibit a p-type dependence in momentum, coming from the Rashba interaction, except the $z$-term which has a kinetic nature with cubic symmetry. Moreover, their low-symmetry nature leads to highly directional contributions, introducing rotations to the in-plane momentum angle. 

\begin{figure}[ht]
    \centering
    \includegraphics[width=0.7\linewidth]{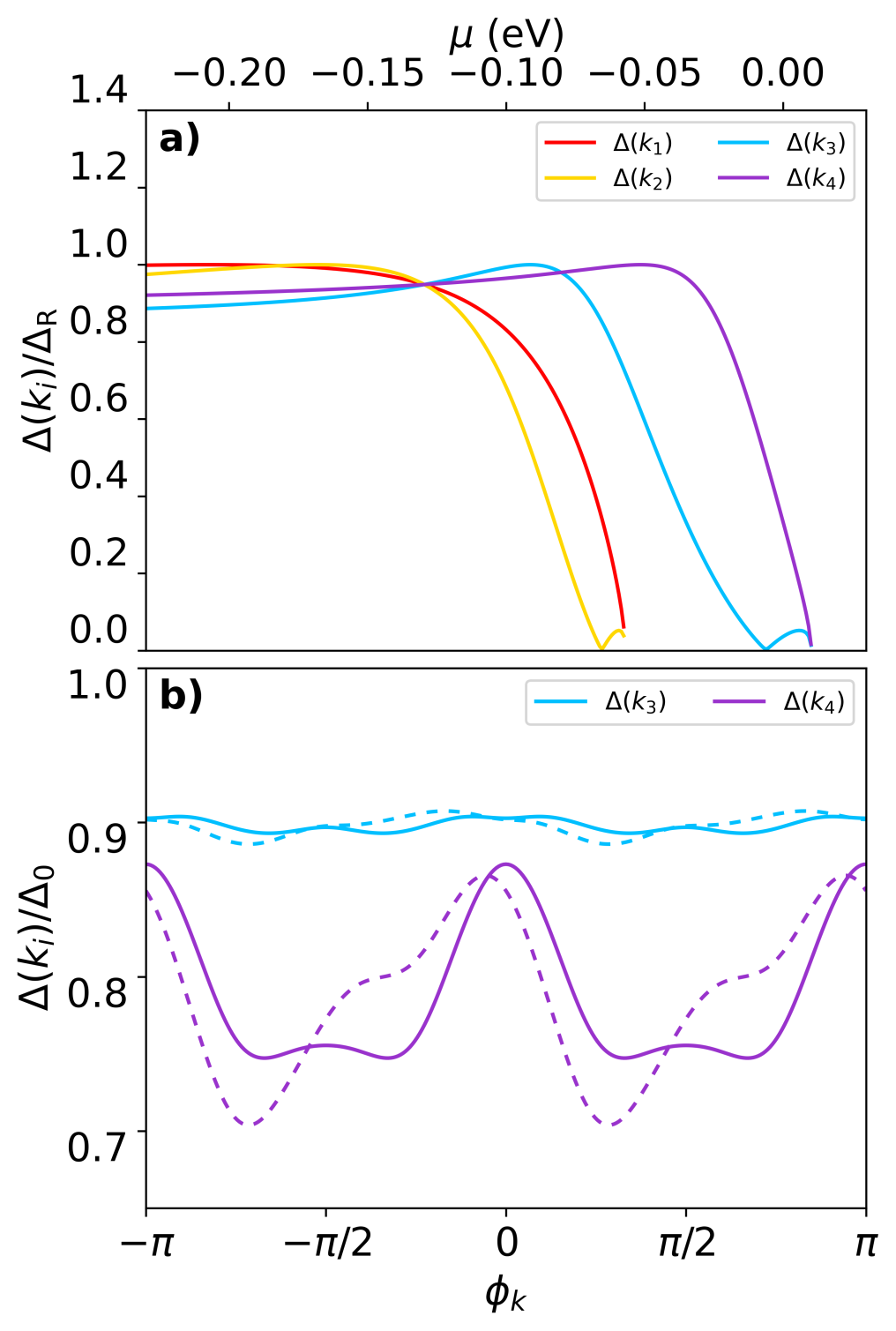}
    \caption{\textbf{Mixed LH-HH pairing.} (a) Gaps $\Delta(k_1,k_2,k_3,k_4)$ and their dependence on $\mu$ for mixed pairing $\Delta_R=200\,\mu$eV and $\chi_R=0$. (b) Gaps $\Delta(k_3,k_4)$ as a function of $\phi_k$ for $\mu=-0.01$ eV for $\Delta_{\mathrm{HH}}=180\,\mu$eV and $\Delta_{\mathrm{R}}=20\,\mu$eV with $\chi_R=0$ (solid) and $\chi_R=\pi/4$ (dashed).}
    \label{fig:disorder}
\end{figure}

In Fig.~\ref{fig:disorder}(a), we illustrate the effect of these pairing terms as a function of $\mu$. As $\mu$ decreases, the pairing terms coming from $\Delta_R$ increase their influence for all different gaps. However, this increase is not always monotonous, and there can be sign changes when the $0$-component exactly cancels out the $z$-component for a specific spin state leading to the gap closures in Fig.~\ref{fig:disorder}(a). Furthermore, the phase $\chi_R$ breaks the cubic symmetry and rotates the preferred directions in the different gaps, see Fig.~\ref{fig:disorder}(b). Taking $\chi_R=\pi/4$ (dashed line) induces a directional correction that reduces the symmetry of the gaps.

\section{Experimental signatures}
\label{sec:Signatures}

\begin{figure*}[ht]
    \includegraphics[width=\linewidth]{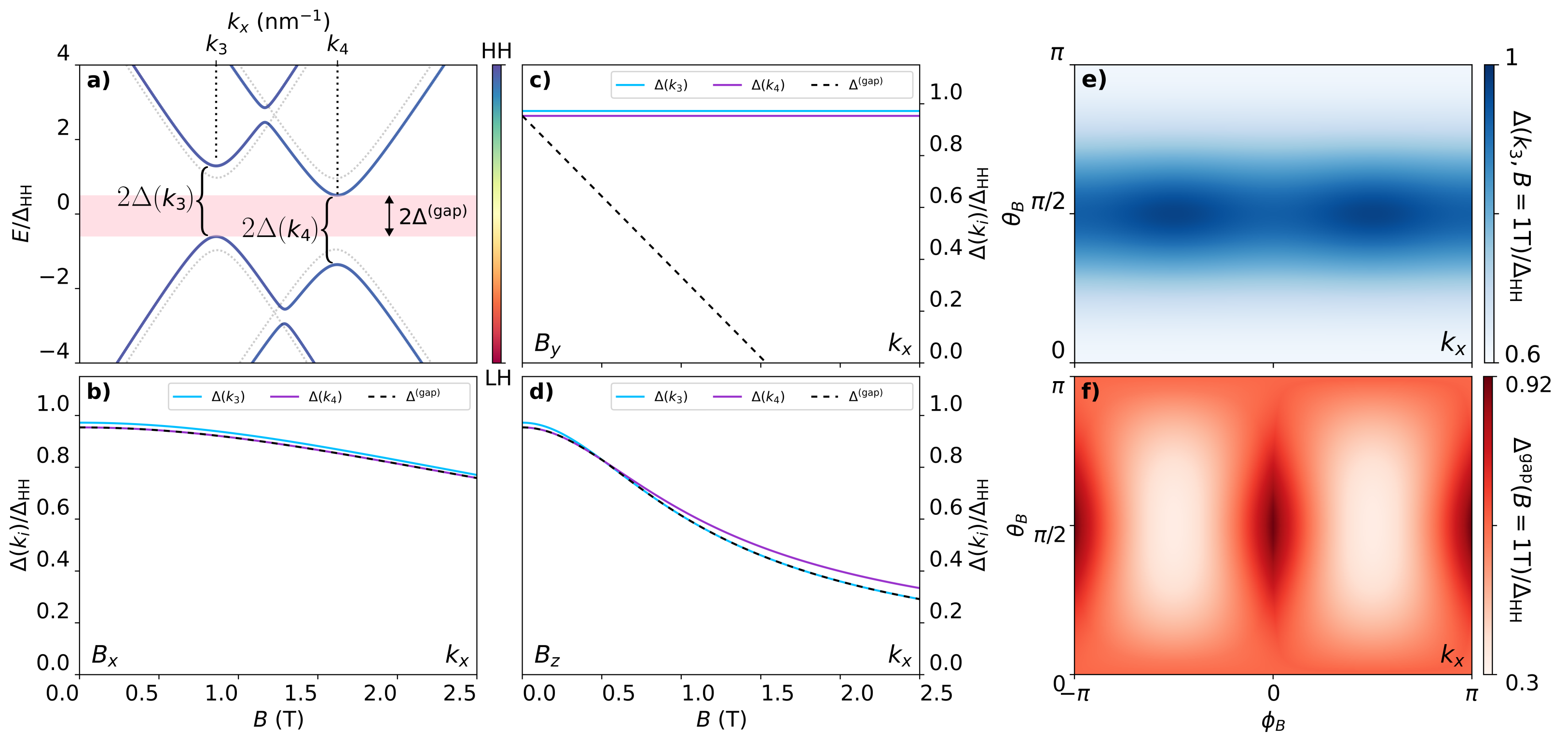}
    \caption{\textbf{Magnetic-field dependence of pairings with direct HH contribution.} (a) Tilted bands as a function of $k_x$ around the Fermi level crossings at $k_i$ for $\mathbf{B}=1/\sqrt{2}(1,1,0)$ (solid blue) and $||B||=0$ (dashed grey). The gap $\Delta^{\mathrm{(gap)}}$ is shown as a pink band around $E=0$ and the pairing terms $\Delta(k_i)$ are associated to the Fermi-level anticrossings. (b-d) Dependence of pairings $\Delta(k_3)$ and $\Delta(k_4)$ on the strength of a magnetic field for: (b) $B_x$, (c) $B_y$, and (d) $B_z$. The value of the spectral gap $\Delta^{\mathrm{(gap)}}$ is given as a dashed line. (e) Dependence of pairing $\Delta(k_3)$ on the orientation of a magnetic field with $\mathbf{B}=(\sin\theta_B\cos\phi_B,\sin\theta_B\sin\phi_B,\cos\theta_B)$ T, along the $k_x$ direction. (f) Dependence of spectral gap $\Delta^{\mathrm{(gap)}}$ on the orientation of a magnetic field with $|B|=1$ T, along the $k_x$ direction. Parameters used: $\Delta_\mathrm{HH}=200$ $\mu$eV; $\mu=-0.01$ eV; $F=0.5$ MV/m; Table \ref{table:Ge-parameters}; Appendix \ref{sec:Appendix_confinement}.}
    \label{fig:4KP_2DHG_Bfield}
\end{figure*}

We address now the question of how to distinguish among all the possible pairing mechanisms discussed above. Specifically, we focus on the intricate response of the proximitized 2DHG to external magnetic fields and discuss how they modify the band profiles and the induced gaps, giving rise to unique experimental signatures. They include distinct DOS (a quantity that can be directly linked to tunneling spectroscopy experiments) for both out-of-plane and in-plane magnetic fields as well as the emergence of strongly anisotropic Bogoliubov-Fermi surfaces.

\subsection{Magnetic field effects}
\label{sec:Magnetic}
In proximitized Rashba semiconductors, the Zeeman term competes with the SOI leading to magnetic-field tunable spin triplet and spin singlet pairing terms and, hence, magnetic-dependent gaps \cite{PhysRevLett.104.040502,alicea2010majorana, lutchyn2010majorana,PhysRevLett.105.177002}. Given the strong anisotropic response of holes to magnetic fields, we expect to find strong anisotropic behavior in the superconducting gaps as well. 
We focus now on the effect of Zeeman terms. Magneto-orbital corrections due to the vector potential will be discussed in Appendix~\ref{Appendix:vectorpotential}.

In the large negative $\mu$ regime, strong Rashba spin splitting appears near the crossings at the Fermi level $k_i$, which is expected since the Rashba spin-splitting is, effectively, cubic in momentum for Ge holes. In this regime, the spin is strongly polarized and the magnetic field required to modify this polarization and influence the nature of the gap is too large to be experimentally feasible. Consequently, magnetic signatures on the gap are expected to be more evident at small values of $|\mu|$, where $k_i$ are not too large and the Zeeman field can compete with the cubic-in-momentum spin splitting. In the low $|\mu|$ regime, the excitations crossing the Fermi level at $k_{3,4}$ have strong HH character, therefore, we focus on the case $\Delta_{\mathrm{HH}}\neq 0$. In this regime, we expand the spin splitting $\delta_{h}=E_4-E_3$ to lowest order in momentum in the eigenbasis, using Eq.~\eqref{eq:H-to-eigen}, to obtain:
\begin{equation}
\label{eq:h-spinsplitting}
    \delta_{h}=E_4-E_3\approx |\zeta\alpha_R| E_{hl}.
\end{equation}

Depending on the magnetic field orientation, there can be different types of signatures in the gap, as illustrated in Figs.~\ref{fig:4KP_2DHG_Bfield}(e,f). In particular, in-plane magnetic fields tilt the spectrum, potentially leading to the closing of the spectral gap~\cite{PhysRevB.97.115139, doi:10.1126/science.abf1077,10.21468/SciPostPhys.16.5.115} (even though each band still shows a finite pairing).
This distinction between the pairing terms and the spectral gaps is illustrated
in Fig~\ref{fig:4KP_2DHG_Bfield}(a), where a zoom near the Fermi level crossings around $k_3$ and $k_4$ shows the tilt in the spectrum (lack of mirror symmetry around $E=0$). The effective longitudinal pairings $\Delta(k_i)$ are defined as half the minimal energy distance in $k$ between hole and the time-reversed spectra, while the spectral gap $\Delta^{\mathrm{(gap)}}(\phi_k)$ is given by the energy range
where the bands never cross for any $k$.

We focus first on the effective pairings; in Fig.~\ref{fig:4KP_2DHG_Bfield}(b-e), we show the dependence of the size of the longitudinal pairings at $k_3$ and $k_4$ against magnetic field in different directions. These pairing amplitudes are tunable mainly using vertical magnetic fields, see Fig.~\ref{fig:4KP_2DHG_Bfield}(d,e) along $k_x$. In particular, Fig.~\ref{fig:4KP_2DHG_Bfield}(e) shows a map of the pairing as a function of magnetic field orientation taking $||\mathbf{B}||=1$T. In this scenario, the pairing is strongly normalized when $B||\hat{z}$ but in-plane magnetic fields cause only a small anisotropy.

The anisotropic behavior of the pairings in Fig.~\ref{fig:4KP_2DHG_Bfield} can be understood in terms of geometrical relationships. In the original basis, the Zeeman Hamiltonian is $H_Z=-2\mu_B\kappa\mathbf{J}\cdot\mathbf{B}-2q\mu_B\mathbf{J}^3\cdot\mathbf{B}$, where $\mathbf{J}$ and $\mathbf{J}^3=(J_x^3,J_y^3,J_z^3)$ are vectors of spin 3/2 matrices, $\kappa=3.41$, and $q=0.06$ in Ge. In the eigenbasis, the Zeeman term can be written as:
\begin{equation}
\label{eq:gmatrix}
    H_{Z,34}^{(d)}=\frac{1}{2}\mu_B\boldsymbol{\sigma}\cdot\begin{pmatrix}
        0 & 0 & g_{xz} \\
        g_{yx} & g_{yy} & 0 \\
        g_{zx} & g_{zy} & 0
    \end{pmatrix}\cdot\mathbf{B},
\end{equation}
where the $g_{zx}$ and $g_{zy}$ characterize the in-plane Zeeman terms that are parallel to the Rashba field and, hence, compete directly with the Rashba spin-splitting $\delta_{h}$, while $g_{yx}$ and $g_{yy}$ ($g_{xz}$) characterize the in-plane (out-of-plane) Zeeman terms that are perpendicular to the Rashba spin-splitting.
These g-factors can be approximated as:
\begin{equation}
\label{eq:gfactors-diag}
    \begin{aligned}
        g_{zx}&\approx 3q\sin3\phi_k-|\zeta|\kappa\sin\phi_k\,, \\
        g_{yx}&\approx 3q\cos3\phi_k-|\zeta|\kappa\cos\phi_k\,, \\
        g_{zy}&\approx 3q\cos3\phi_k + |\zeta|\kappa\cos\phi_k\,, \\
        g_{yy}&\approx -3q\sin 3\phi_k-|\zeta|\kappa\sin\phi_k\,, \\
        g_{xz}&\approx 6\kappa+\frac{27}{2}q.
    \end{aligned}
\end{equation}
In Appendix~\ref{Appendix:vectorpotential}, we estimate via perturbation theory the strong renormalizations of the g-factors~\cite{ares2013nature} caused by the vector potential, not included here.
Note that the in-plane terms $g_{ix}$ and $g_{iy}$ exhibit a strong in-plane anisotropic behavior through their dependence on $\phi_k$. Compared to the hole g-matrix $g_{ij}$, the g-matrix $\bar{g}_{ij}$ and spin-splitting $\bar{\delta}_h$ of the time-reversed states have opposite signs $\bar{g}_{ij}=-g_{ij}$ for $i\neq z$ and $\bar{\delta}_{h}=-\delta_{h}$. The only exception are the g-factors associated to Zeeman terms that are parallel to the Rashba interaction, in which case we find $\bar{g}_{zi}=g_{zi}$, introducing a direct competition with the Rashba splitting that tilts the spectrum \cite{PhysRevB.97.115139, 10.21468/SciPostPhys.16.5.115}.

When the Zeeman field points along the $y$ direction, we get Zeeman terms that are only parallel to the Rashba SOI for $k_y=0$ as in Fig.~\ref{fig:4KP_2DHG_Bfield}(c), leaving the pairing Hamiltonian invariant, yet introducing a tilt in the spectrum. As a result, the longitudinal pairings remain constant along $\mathbf{B}||\hat{y}$, Figs.~\ref{fig:4KP_2DHG_Bfield}(c, e). As can be deduced from the g-factor expressions, the magnetic field that leaves the pairing invariant heavily depends on the orientation in $k$-space. When $\mathbf{B}||\hat{x}$, as in Fig.~\ref{fig:4KP_2DHG_Bfield}(b), we get Zeeman contributions that are perpendicular to the Rashba field, however, the HH in-plane g-factors are quite small, suppressed by $E_{hl}$, and, while there is a visible renormalization of the longitudinal pairing, this is relatively small for $||\mathbf{B}||=1$T. Overall, the spin-splitting tends to dominate over the in-plane Zeeman terms as long as $\alpha_0 p/m_0 \gg \kappa\mu_B B_\parallel$. 

For a vertical magnetic field we get a perpendicular contribution to the Rashba field and the g-factor $g_{xz}$ is not suppressed, leading to a change in the spin polarization and a stronger longitudinal pairing renormalization. Geometrically, when $B_z\neq 0$, the pairing Hamiltonian in the spin space is rotated around the $y$ direction. At $B=0$, low $k$ and $|\mu|$, the pairing term in Eq.~\eqref{eq:2DHGgaps-direct} can be approximated to $\tilde{\Delta}_d^{(v)}(\mathbf{k}_3,\mathbf{k}_4)\cdot\boldsymbol{\sigma}\approx i\Delta_{\mathrm{HH}}\sigma_z$ while, when $B_z\neq 0$, it becomes
\begin{equation}
\label{eq:f-superconductivity}
    \frac{\tilde{\boldsymbol{\Delta}}_d^{(v)}(\mathbf{k}_3,\mathbf{k}_4)\cdot\boldsymbol{\sigma}}{\Delta_{\mathrm{HH}}}\approx \frac{i(\delta_{h}\sigma_z-g_{xz}\mu_B B_z\sigma_x)}{\sqrt{\delta_{h}^2+(g_{xz}\mu_B B_z)^2}}.
\end{equation}
Interestingly, Eq. \eqref{eq:f-superconductivity} resembles the familiar result of a proximitized Rashba semiconductor \cite{Alicea_RPP2012,LeijnseReview,AguadoReview,LutchynReview} with the main difference of the cubic-in-momentum dependence of $\delta_{h}$ suggesting an effective f-type superconducting pairing. Furthermore, we see clearly that, as $B_z$ grows, the longitudinal pairing becomes more a transverse contribution and vice-versa. For the rotation of the pairing terms under a general magnetic field, we refer to Appendix~\ref{Appendix:general-B}. The emergence of a transverse pairing contribution introduces another anticrossing in the spectrum between $k_3$ and $k_4$ at non-zero magnetic fields in Fig.~\ref{fig:4KP_2DHG_Bfield}(a).

The spectral gap exhibits a clearly distinct behavior to the pairing amplitude, as shown in Fig.~\ref{fig:4KP_2DHG_Bfield}(f). In the case where Zeeman contributions are purely transverse to the Rashba interaction, the spectral gap $\Delta^{\mathrm{(gap)}}$ is equivalent to the minimal value of the two pairings $\Delta(k_3)$ and $\Delta(k_4)$. This is exactly what happens along $k_x$ for finite fields along $B_x$ and $B_z$, Fig.~\ref{fig:4KP_2DHG_Bfield}(b,d). In contrast, any Zeeman component parallel to the Rashba field, introduces a tilt in the spectrum. Along $k_x$, any $B$-field component parallel to $\hat{y}$ introduces this tilt and, hence, a reduction in the gap occurs without a concomitant reduction of the pairing, see Fig.~\ref{fig:4KP_2DHG_Bfield}(c). As a result, in Fig.~\ref{fig:4KP_2DHG_Bfield}(f), we see a strong suppression of the gap as long as there is any $B_y$ component in the magnetic field.

\subsection{Density of states and Bogoliubov
Fermi surfaces}

\begin{figure*}[ht]
    \includegraphics[width=\linewidth]{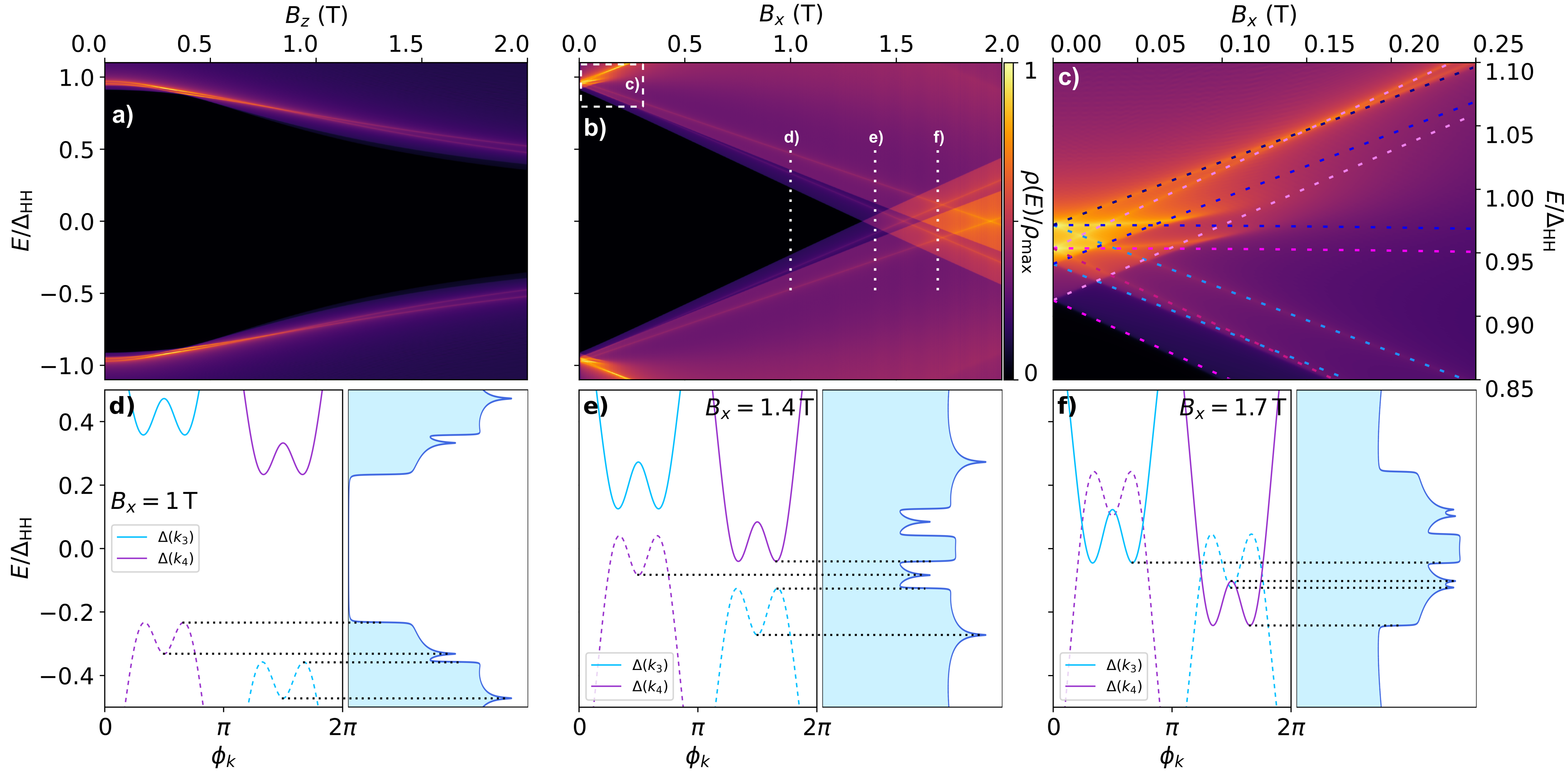}
    \caption{\textbf{DOS as a function of magnetic field.} (a) DOS centered around $E=0$ as a function of the vertical magnetic field $B_z$ with $\mu=-0.01$ eV. (b) DOS centered around $E=0$ as a function of the in-plane magnetic field $B_x$. (c) DOS at low in-plane magnetic field, in the corresponding energy window marked in panel (b). Dashed lines mark the expected van Hove singularity positions, see Fig.~\ref{fig:phis_Bx}. (d-f) Hole band minima as a function of $\phi_k$ associated to $\Delta(k_3)$ and $\Delta(k_4)$ for the in-plane magnetic field marked with the corresponding dashed lines in panel (b), at (d) $B_x=1$ T, (e) $B_x=1.4$ T, and (f) $B_x=1.7$ T. At their right, their corresponding cuts of the DOS. The black dashed lines relate the local minima and maxima with the different singularities in the DOS. Parameters: $\Delta_\mathrm{HH}=200$ $\mu$eV; $\mu=-0.01$ eV; $F=0.5$ MV/m; Table \ref{table:Ge-parameters}; Appendix \ref{sec:Appendix_confinement}.}
    \label{fig:4KP_2DHG_DOS}
\end{figure*}

\begin{figure}[ht]
\includegraphics[width=\linewidth]{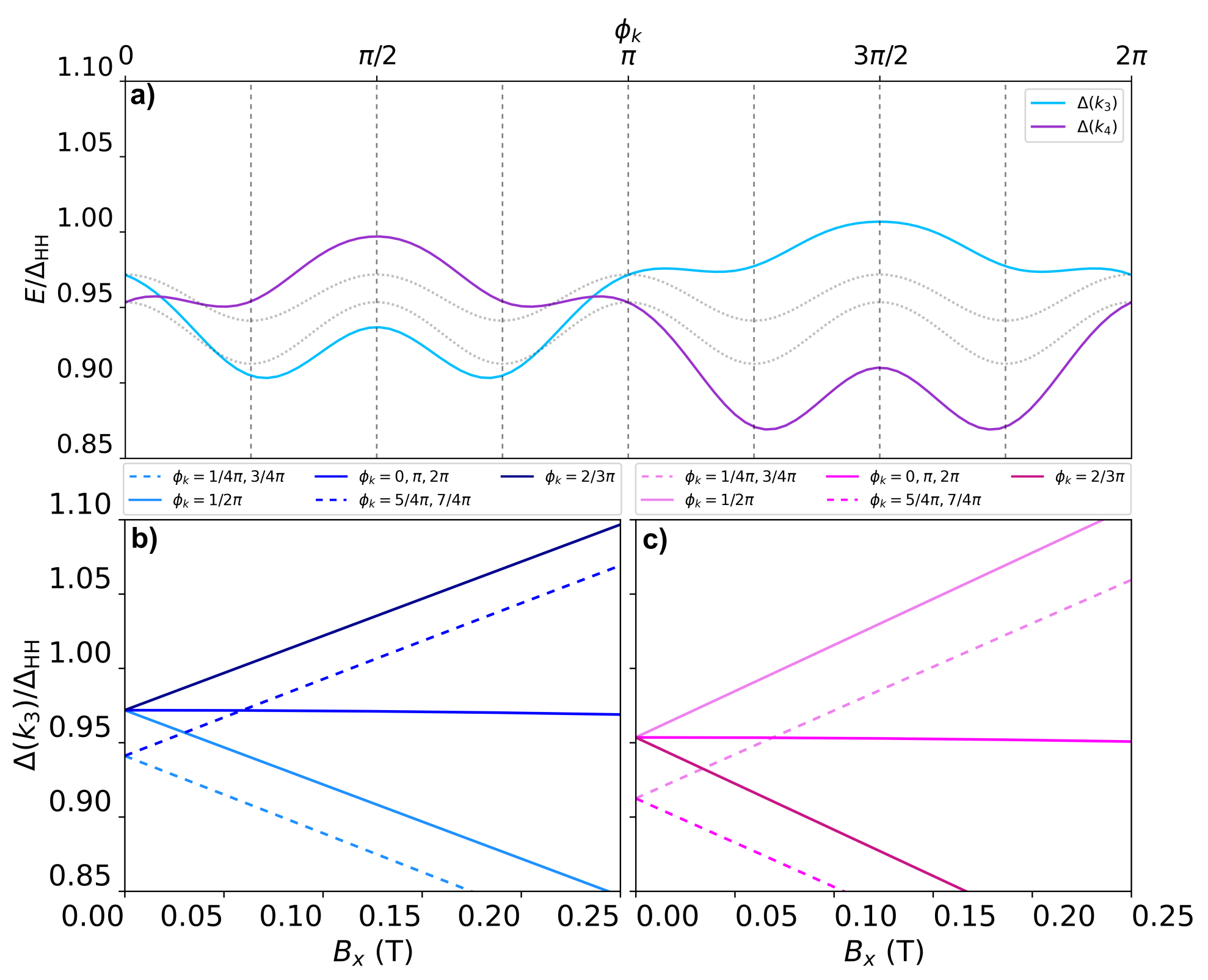}
\caption{\textbf{Maxima, minima and van Hove singularities}. (a) Energy at $k_3$ and $k_4$ as a function of $\phi_k$. While maxima and minima of the oscillations in both bands coincide at the same $\phi_k$ for $B=0$ (grey dotted lines), this is no longer the case at finite in-plane magnetic field ($B_x=0.07$ T in the plot) owing to the anistropies discussed in the main text. This gives rise to the splittings in the energy positions of the van Hove singularities as a function of $B_x$ (lower panels b and c) that are clearly seen in the DOS in Fig. \ref{fig:4KP_2DHG_DOS}c. Parameters: $\Delta_\mathrm{HH}=200$ $\mu$eV; $\mu=-0.01$ eV; $F=0.5$ MV/m; Table \ref{table:Ge-parameters}; Appendix \ref{sec:Appendix_confinement}}.
\label{fig:phis_Bx}
\end{figure}

\begin{figure}[ht]
    \includegraphics[width=\linewidth]{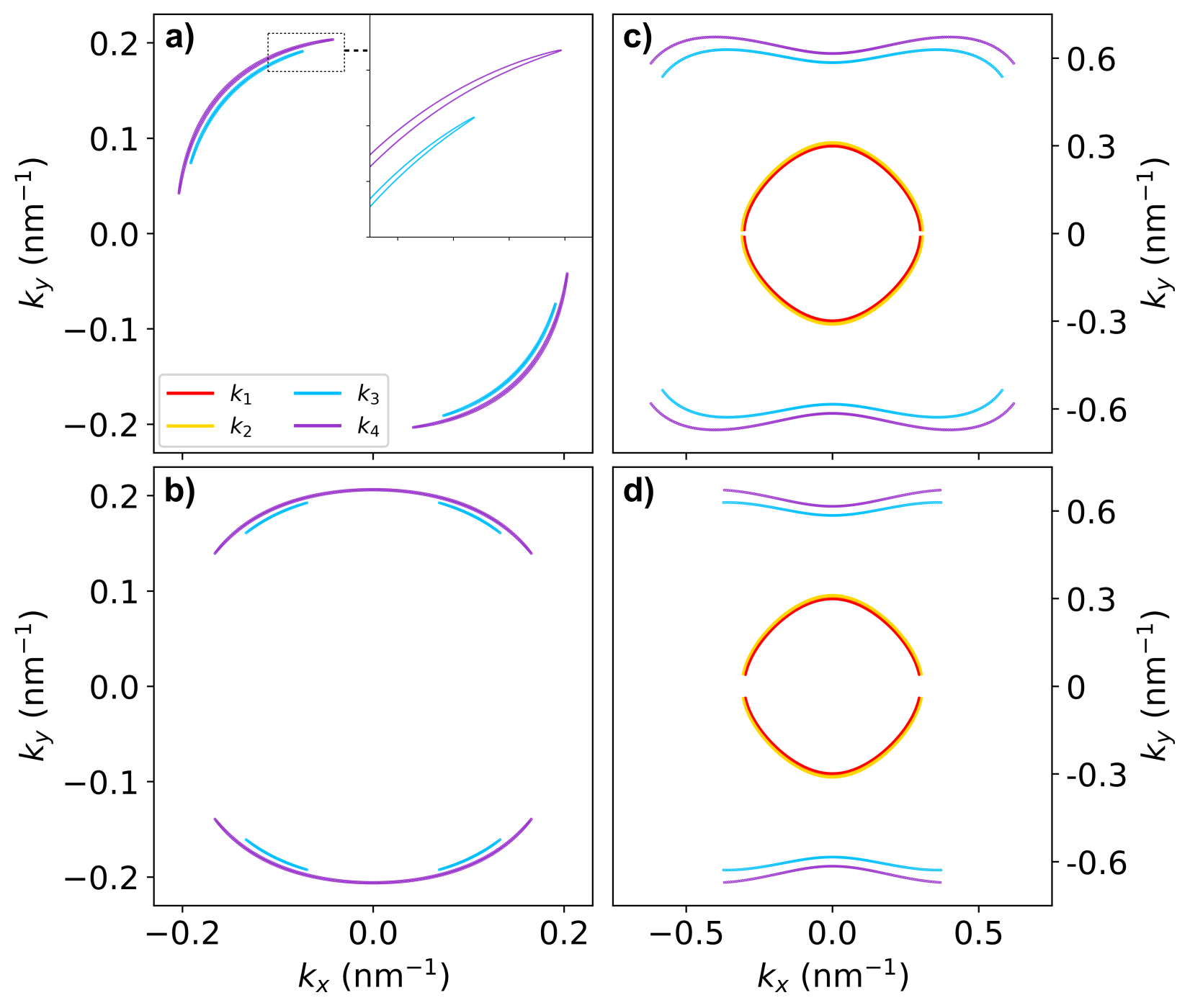}
    \caption{\textbf{Bogoliubov-Fermi bands.} (a-b) Bogoliubov-Fermi surfaces for $\mu=-0.01$ eV, $||B||=1.7$ T, $\Delta_\mathrm{HH}=200$ $\mu$eV, with (a) $\phi_B=0$, and (b) $\phi_B=\pi/4$. (c-d) Bogoliubov-Fermi surfaces for $\mu=-0.1$ eV, $B_x=1.5$ T; with (c) $\Delta_{\mathrm{LH}}=0$ $\mu$eV, $\Delta_{\mathrm{HH}}=200$ $\mu$eV, and with (d) $\Delta_{\mathrm{LH}}=200$ $\mu$eV, $\Delta_{\mathrm{HH}}=0$ $\mu$eV. Parameters used: $F=0.5$ MV/m (rest of parameters in Table \ref{table:Ge-parameters}; Appendix \ref{sec:Appendix_confinement}). }
    \label{fig:4KP_2DHG_Bogobands}
\end{figure}

\label{BdGFermi-DOS}
The DOS of proximitized holes exhibits a strong magnetic field anisotropy due to the very distinct behavior of hole spins under vertical or in-plane magnetic fields. For a given magnetic field, the DOS can be calculated as
\begin{equation}
    \rho(E) = -\frac{1}{\pi}\int \Im \Tr \frac{1}{E+i0^+ - H_\mathrm{BdG}} k\, dk\,d\phi_k \;,
\end{equation}
integrating over the in-plane momentum. As previously noted, the in-plane magnetic field induces tilts to the spectrum depending on whether the Zeeman term is parallel to the Rashba field or not. By integrating the momentum orientation, we average out this behavior. Hence, under an in-plane magnetic field, the DOS shall give a qualitatively similar picture irrespective of the specific in-plane Zeeman orientation $\phi_B$. In contrast, a vertical magnetic field is always perpendicular to the Rashba field for any value of $\phi_k$. Therefore, the DOS provides different signatures for vertical and in-plane magnetic fields. 

In Fig.~\ref{fig:4KP_2DHG_DOS}(a,b), we show the DOS as a function of vertical $B_z$ and in-plane $B_x$ magnetic fields at low $|\mu|$. As a function of $B_z$, Fig.~\ref{fig:4KP_2DHG_DOS}(a), the DOS exhibits a series of peaks which are the van Hove singularities associated to the maxima in $\phi_k$ of the pairings $\Delta(k_3)$ and $\Delta(k_4)$. The black region in these plots illustrates the spectral gap, which decreases with increasing $B_z$, as expected from the previous analysis with magnetic fields. The
different dependence on $k$ of the two gaps leads to a non-trivial behavior of the van Hove singularities, which cross and swap places as a function of energy.

In contrast, in Fig.~\ref{fig:4KP_2DHG_DOS}(b) we see many more features for the in-plane behavior of the DOS due to the spectrum tilting. In this case, multiple van Hove singularities appear. The van Hove singularities are associated to newly emerging local maxima of the gap against $\phi_k$ for non-zero magnetic field, Figs.~\ref{fig:4KP_2DHG_DOS}(d-f). Interestingly, as shown in Fig.~\ref{fig:4KP_2DHG_DOS}(d), the anisotropic behavior with $\phi_k$ of the spin splitting $\Delta_\text{HH}$ in Eq.~\eqref{eq:h-spinsplitting} introduce two local maxima at each anticrossing for $B_\parallel=0$, leading to a couple of the van Hove singularities. These van Hove singularities split when $B_\parallel\neq 0$. A superconductor-semiconductor interface breaking cubic symmetry would introduce highly directional pairing terms treated in  Sec.~\ref{subsec:disorder}
and, therefore, break the symmetry between the two minima in each band, leading to measurable signatures in the DOS. 

Importantly, as the in-plane Zeeman energy grows, the spectrum tilts linearly with $B$, reducing the gap as illustrated in Fig.~\ref{fig:4KP_2DHG_Bfield}(a) and Fig.~\ref{fig:4KP_2DHG_DOS}(b). When the Zeeman energy parallel to the Rashba field $g_{zi}\mu_B B_\parallel$ surpasses the pairing term at a given $k_i$, the gap closes at that value of $k$. An intermediate region emerges where only one of the two bands can get occupied since $\Delta(k_3)$ and $\Delta(k_4)$ exhibit different behavior with $B$ as well. For sufficiently large in-plane magnetic fields, the gap closes and a finite DOS is observed at $E=0$, see Figs.~\ref{fig:4KP_2DHG_DOS}(e) and (f). In this case, the van Hove singularities from hole and time-reversed states converge to approximately the same energy regions, giving rise to a more complex DOS with a diamond-like structure as a function of $E$ and $B_x$ in Fig.~\ref{fig:4KP_2DHG_DOS}(b).  
Interestingly, this strong anisotropic behavior is already evident at low in-plane magnetic fields, see Fig.~\ref{fig:4KP_2DHG_DOS}(c), where the van Hove singularities show an intricate behavior. 

In Fig.~\ref{fig:phis_Bx} we focus on the behavior of these van Hove singularities. Despite the apparent complexity, the position of the peaks can be deduced from the competition between the Rashba-induced spin splitting $\Delta_\text{HH}$ and the Zeeman spin splitting related to the parallel g-factors $g_{zx}$ and $g_{zy}$. $\Delta_\text{HH}$ has a $\cos 4\phi_k$ dependence coming from the cubic symmetry, leading to the two van Hove peaks per band at zero magnetic field in Fig.~\ref{fig:phis_Bx}(b,c). When the magnetic field is turned on, the Zeeman field breaks the mirror symmetry around $\phi_k=n\pi$, see Fig.~\ref{fig:phis_Bx}(a), leading to a splitting of the van Hove singularities with increasing $B$. An intriguing case occurs when $\phi_k$ is such that the Zeeman energy for a given $B_\parallel$ does not induce a tilt --the Zeeman term is perpendicular to the Rashba term-- leaving a trace in the DOS as a van Hove singularity that remains invariant as a function of the amplitude $||B||$. This invariant singularity gets washed away for larger magnetic fields due to the deformation of the gap dependence with $\phi_k$.

All the above predictions should constitute strong experimental signatures of a proximitized hole gas with large anisotropies and expected to appear in tunneling spectroscopy experiments, where the tunneling conductance is directly proportional to the DOS.

The anisotropic behavior of the gap against $\phi_k$ for in-plane magnetic fields is illustrated in Fig.~\ref{fig:4KP_2DHG_Bogobands} through Bogoliubov-Fermi surfaces~\cite{PhysRevB.97.115139, 10.21468/SciPostPhys.16.5.115}. The points in $k$-space where the BdG bands cross the Fermi level form closed anisotropic surfaces, due to the anisotropic Rashba and Zeeman fields for holes. 
These Bogoliubov-Fermi surfaces are strongly dependent on the different parameters of the system. The orientation of the magnetic field $\phi_B$ leads to rotations of the surfaces in $k$-space. The distance between surfaces associated to the closure of different HH bands is related to the Rashba spin splitting, which is tunable through the electric field. Furthermore, by tuning the amplitude of the magnetic field, the shape and topology of the surfaces can be changed. These features are not unique to the low $|\mu|$ regime, and it is possible to show similar surfaces associated to four different bands for states with hybridized LH components. In this hybridized scenario, the value of $\Delta_{\mathrm{LH}}$ further tunes the different surfaces, as can be seen in Fig.~\ref{fig:4KP_2DHG_Bogobands}(c-d). Moreover, these surfaces illustrate the cubic symmetry as $\mu$ gets more negative, becoming less symmetric than the surfaces in Fig.~\ref{fig:4KP_2DHG_Bogobands}(a-b) cases.

Experimentally, the emergence of such anisotropic Bogoliubov-Fermi surfaces could be probed in a circuit QED setup. Specifically, the frequency of a resonant microwave circuit is expected to be modified by  the contribution of the 2DHG to the kinetic inductance, which is inversely proportional to the superfluid
density. This method has recently been demonstrated in a two-dimensional Al-InAs hybrid heterostructure \cite{PhysRevLett.128.107701}.

\section{Conclusions}
\label{sec:Conclusions} 
In this work, we have studied the proximity-induced superconductivity in a two-dimensional hole gas. We have found very rich anisotropic behavior in the inherited superconducting correlations for hole spins. 
We have distinguished two main regimes: at low $|\mu|$, only the bands with strong HH character at low $k$ cross the Fermi level leading to a dominant pairing in the HH band; at $\mu$ below $-E_{hl}$ four different bands cross the Fermi level, including two bands with strong LH character at low $k$. We have considered pairing terms coming directly from the conduction band, HH band, and LH band, all of which exhibiting different dependencies with $\mu$, which potentially allows to infer their contributions. The electron pairing through the conduction band induces both superconducting longitudinal and transverse pairing correlations in the hole subspace. These contributions decrease as $\mu$ goes from the four-level regime to the low $|\mu|$ two-level and HH-dominated regime. In contrast, direct HH and LH pairing mechanisms induce longitudinal pairing correlations at zero magnetic field. In this case, a direct HH pairing mechanism leads to a maximal gap for low $|\mu|$ in the HH-like bands, while it increases with more negative $\mu$ for the LH-like bands. The opposite behavior occurs for a direct LH pairing amplitude. We have also considered disorder effects allowing mixed HH-LH pairing mechanisms. These introduce both longitudinal and transverse superconducting pairing correlations in the hole subspace that decrease with $\mu$, and rotate the pairings in $k$-space, reducing the overall symmetry of the gaps, potentially leading to extra singularities in DOS measurements. Overall, all different pairing mechanisms induce a non-trivial anisotropic $k$-dependence, in particular due to the cubic symmetry of the Ge crystal, which influences the singularities that are observed in DOS measurements.

When turning the magnetic field on, we have found strongly anisotropic behavior that can be experimentally tested. For vertical magnetic fields and low $|\mu|$ values, we find analogous behavior to proximitized Rashba nanowires where longitudinal and transverse pairing correlations are exchanged as $B_z$ grows. The main difference with Rashba nanowires is the f-type superconductivity arising due to the cubic Rashba of holes. We have related the magnetic dependence to geometrical relationships with the g-factors of the hole spins. In contrast, when the magnetic field is in-plane, the component perpendicular to the Rashba field induces an identical behavior as with vertical magnetic fields in the pairing correlations with a reduced g-factor, while the component parallel to the Rashba field competes directly with the spin splitting inducing a tilt to the spectrum.
Consequently, for sufficiently large in-plane magnetic fields, the gap may close at different points in $k$-space, forming non-trivial Bogoliubov-Fermi surfaces. In DOS measurements, this leads to a magnetic field value at which the gap closes and does not reopen, and where van Hove singularities coming from hole and time-reversed states converge in energy, giving rise to a diamond-like structure in the density of states. 
Note that the orbital corrections considered in Appendix~\ref{Appendix:vectorpotential} introduce strong g-factor renormalizations, particularly for vertical magnetic fields, which do not qualitatively change the analyzed behavior but quantitatively affect the values of magnetic field at which all these experimental signatures occur. Further g-factor corrections may directly arise from the interaction with the superconductor \cite{luethi2023planar, babkin2024superconducting}.

To conclude, after analyzing the wealth of possible pairings and anisotropies that can arise in a proximitized 2DHG, we believe that our model provides an excellent starting point for theoretical modeling of novel concepts and devices based on proximitized hole gases. In this context, systems of great current interest include superconducting spin qubits, subgap physics including Andreev and Shiba, minimal Kitaev chains and Majorana-based devices in general; as well as novel Josephson junctions including superconducting diodes. Given that so far most of the experiments with hybrid devices have been carried out on electron-based systems and that much of the experimental efforts are now directed towards holes, we are confident that the new physics introduced in the latter, as yet almost unexplored, will undoubtedly enhance the palette of new phenomenologies and functionalities.

\acknowledgments
Work supported by the Horizon Europe Framework Program of the European Commission through the European Innovation Council Pathfinder Grant No. 101115315 (QuKiT), the Spanish Comunidad de Madrid (CM) “Talento Program” (Project No. 2022-T1/IND-24070), the Fundaci\'on General CSIC’s ComFuturo programme under the Marie Sklodowska-Curie grant agreement No. 101034263 for EU Horizon 2020, the State Research Agency through the pre-
doctoral Grant No. PRE2022-103741 under the Program “State Program to Develop, Attract and Retain Talent”, the Spanish Ministry of Science, innovation, and Universities through Grants PID2021-125343NB-I00, RYC2022-037527-I, PID2022-
140552NA-I00, PID2023-148257NA-I00, and TED2021-130292B-C43 funded by MCIN/AEI/10.13039/501100011033, “ERDF A way of making Europe” and European Union Next Generation EU/PRTR. We acknowledge the support of the CSIC’s Quantum Technologies Platform (QTEP).

{\bf Note added.} Recently, another work~\cite{babkin2024superconducting} also investigates the theory of the superconducting proximity effect in two-dimensional hole gases. This study partially overlaps with our own work.

\textbf{Data availability}. Our code can be found in \footnote{Github repository for the code:~\href{https://github.com/dmichelpino/superholes}{https://github.com/dmichelpino/superholes}}.

\appendix
\renewcommand{\thefigure}{\Alph{section}.\arabic{figure}}
\setcounter{figure}{0}

\section{8KP Hamiltonian} \label{sec:Appendix_8KP-Hamiltonian}

Below we detail the explicit form of the different blocks appearing in the Hamiltonian (\ref{8kpBdG}), using the spinor basis
$\Psi=$($c_{1/2,\mathbf{k}}$, $c_{-1/2,\mathbf{k}}$, $b_{3/2,3/2,\mathbf{k}}$, $b_{3/2,1/2,\mathbf{k}}$, 
$b_{3/2,-1/2,\mathbf{k}}$, $b_{3/2,-3/2,\mathbf{k}}$, $b_{1/2,1/2,\mathbf{k}}$, $b_{1/2,-1/2,\mathbf{k}}$, $c_{1/2,-\mathbf{k}}^\dagger$, $c_{-1/2,-\mathbf{k}}^\dagger$, $b_{3/2,3/2,-\mathbf{k}}^\dagger$, $b_{3/2,1/2,-\mathbf{k}}^\dagger$, $b_{3/2,-1/2,-\mathbf{k}}^\dagger$, $b_{3/2,-3/2,-\mathbf{k}}^\dagger$, $b_{1/2,1/2,-\mathbf{k}}^\dagger$, $b_{1/2,-1/2,-\mathbf{k}}^\dagger$)$^T$, where $c_{s_z,\mathbf{k}}$ ($b_{j,j_z\mathbf{k}}$) destroys an electron (hole) with angular momentum $j$, spin projection $s_z$ (angular momentum projection $j_z$), and momentum $\mathbf{k}$.
The standard blocks of conduction and valence bands and their coupling terms are:
\begin{equation}
\begin{aligned}
   H_{cb} =& E_g + \frac{\hbar^2}{2m'} (k_x^2 + k_y^2 + k_z^2)
   \\
   \\
   H_{cb-v} &= \frac{P_\mathrm{Ge}}{\sqrt{3}} \times 
   \\
   &\left( \begin{matrix}
       -\sqrt{\frac{3}{2}}k_+ & \sqrt{2}k_z & \frac{1}{\sqrt{2}}k_- & 0 & -k_z & -k_-
       \\
       0 & -\frac{1}{\sqrt{2}}k_+ & \sqrt{2}k_z & \sqrt{\frac{3}{2}}k_- & -k_+ & k_z
   \end{matrix}\right)
   \\
   \\
   H_{v} =& \left( \begin{matrix}
       H_\mathrm{4KP} & H_{8v-7v} \\
       H_{8v-7v}^\dagger & H_{7v}
   \end{matrix} \right) \;,
\end{aligned}
\end{equation}
where $H_\mathrm{4KP}$ is the Kohn-Luttinger Hamiltonian for LH and HH,
\begin{equation} \label{eq:Kohn-Luttinger_Hamiltonian}
\begin{aligned}
   H_\mathrm{4KP} =& -\left( \begin{matrix}
       P+Q & -S & R & 0 \\
       -S^\dagger & P-Q & 0 & R \\
       R^\dagger & 0 & P-Q & S \\
       0 & R^\dagger & S^\dagger & P+Q
   \end{matrix} \right)
   \\
   \\
   P =& \frac{\hbar^2}{2m_0}\gamma_1'(k_x^2 + k_y^2 + k_z^2)
   \\
   Q =& \frac{\hbar^2}{2m_0}\gamma_2'(k_x^2 + k_y^2 - 2k_z^2)
   \\
   R =& \frac{\sqrt{3}\hbar^2}{2m_0}[-\gamma_2'(k_x^2-k_y^2) + 2i\gamma_3'k_xk_y]
   \\
   S =& \frac{2\sqrt{3}\hbar^2}{2m_0}\gamma_3'(k_x-ik_y)k_z \;,
\end{aligned}
\end{equation}
and $H_{7v}$ comprises the split-off valence band,
\begin{equation}
\begin{aligned}
   H_{7v} =& -P - E_\mathrm{so}
   \\
   H_{8v-7v} =& \left( \begin{matrix}
        -\frac{1}{\sqrt{2}}S & \sqrt{2}R \\
        -\sqrt{2}Q & \sqrt{\frac{3}{2}}S \\
        \sqrt{\frac{3}{2}}S^* & \sqrt{2}Q \\
        -\sqrt{2}R^* & -\frac{1}{\sqrt{2}}S^*
   \end{matrix}\right) \;.
\end{aligned}
\end{equation}
All the parameters needed for these Hamiltonians are tabulated and explained in Appendix \ref{sec:Appendix_Ge-parameters}. We have also included a direct pairing potential induced to each band ($\Delta_\mathrm{CB}$, $\Delta_\mathrm{LH}$ and $\Delta_\mathrm{HH}$) via contact with a superconductor, being
\begin{equation} \label{eq:VB_pairing-term}
\begin{aligned}
    H^\Delta_{cb} =& -i\sigma_y \Delta_\mathrm{CB}
    \\
    H^\Delta_v =& \left(\begin{matrix}
        0 & 0 & 0 & \Delta_\mathrm{HH} \\
        0 & 0 & \Delta_\mathrm{LH} & 0 \\
        0 & -\Delta_\mathrm{LH} & 0 & 0 \\
        -\Delta_\mathrm{HH} & 0 & 0 & 0
    \end{matrix}\right)\oplus 0_{2\times 2}^{7v} \;.
\end{aligned}
\end{equation}
These pairings are assumed to be real without loss of generality and constant as discussed in Appendix \ref{Appendix:approximation}.

\section{Ge parameters for 8KP and 4KP models}\label{sec:Appendix_Ge-parameters}

Below we list the Ge band structure parameters for the 4KP model and their corrections for the 8KP model, extracted from~\cite{paul20168}.

\begin{table}[ht]
\begin{tabular}{|c|c|}
    \hline
    $E_g$ (eV) & 0.8981 \\
    $E_\mathrm{so}$ (eV) & 0.289 \\
    $m_e^*$ ($m_0$) & 0.041 \\
    $\gamma_1$ & 13.37 \\
    $\gamma_2$ & 4.23 \\
    $\gamma_3$ & 5.68 \\
    $\kappa$ & 3.41 \\
    $q$ & 0.06 \\
    $P_\mathrm{Ge}$ ($\mathrm{eV}\cdot\mathrm{\AA}$) & $9.19 $ \\ \hline
\end{tabular}
\caption{Band structure parameters for Ge.}
\label{table:Ge-parameters}
\end{table}

Note that these band parameters contain remote-band contributions of second order in $k$. Thus, for the 8KP model, we need to use reduced band parameters, subtracting the contributions of remote bands which are explicitly taken into account:
\begin{equation}
\begin{aligned}
    \frac{m_0}{m'} =& \frac{m_0}{m_e^*} - \underbrace{\frac{2}{3}\frac{2m_0}{\hbar^2}\frac{P_\mathrm{Ge}^2}{E_g}}_{\Gamma_{8v}} - \underbrace{\frac{1}{3}\frac{2m_0}{\hbar^2}\frac{P_\mathrm{Ge}^2}{E_g+\Delta_0}}_{\Gamma_{7v}}
    \\
    \gamma_1' =& \gamma_1 - \frac{1}{3}\frac{2m_0}{\hbar^2}\frac{P_\mathrm{Ge}^2}{E_g}
    \\
    \gamma_{2,3}' =& \gamma_{2,3} - \frac{1}{6}\frac{2m_0}{\hbar^2}\frac{P_\mathrm{Ge}^2}{E_g}
    \\
    \kappa' =& \kappa - \frac{1}{6}\frac{2m_0}{\hbar^2}\frac{P_\mathrm{Ge}^2}{E_g} \\
    \gamma_\Delta&=\frac{2m_0P_\mathrm{Ge}^2}{3\hbar^2E_g}=7.38
\end{aligned}
\end{equation}

\section{Validity of the constant pairing terms}
\label{Appendix:approximation}
Throughout this work, we consider effectively constant pairing terms along the different bands and, also, connecting the heavy-hole and light-hole bands (in Sec. \ref{subsec:disorder}) as a starting point. More formally, the constant pairing occurs at the s-type superconducting region instead of the semiconducting region, with the superconducting correlations being induced in the semiconductor region only through the direct contact between semiconductor and superconducting regions \cite{luethi2023planar}. Given our aim towards an effective theory of hole states with proximity-induced superconductivity, we have directly assumed that integrating out the superconducting region leads to the approximate effective constant pairings we propose. In this Appendix, we discuss the validity of such approximation. 

For this discussion, we follow the reasoning given in Ref.~\cite{babkin2024superconducting} to perform a similar derivation and extend it to also justify the constant conduction band pairing terms. First, we consider the Hamiltonian of a superconductor as $H_\text{super}=\frac{1}{2}\sum_\mathbf{k}\Psi_{\sigma,\mathbf{k}}^{s\dagger} H_\text{super}^\text{BdG}\Psi^s_{\sigma,\mathbf{k}'}$, where $\Psi^{s}_{\sigma,\mathbf{k}}$ is the Nambu spinor for superconductor states with momentum $\mathbf{k}$, and spin $\sigma$, and  $H_\text{super}^\text{BdG}$ is given by
\begin{equation}
\begin{aligned}
    H_\text{super}^\text{BdG}&= \\ \begin{pmatrix}
        \frac{p^2}{2m_s}-\mu_s+\frac{1}{2}g_s\mu_B \mathbf{B}\cdot\boldsymbol{\sigma} & i\sigma_y\Delta_s  \\
        -i\sigma_y\Delta_s & -\frac{p^2}{2m_s}+\mu_s-\frac{1}{2}g_s\mu_B \mathbf{B}\cdot\boldsymbol{\sigma}^*
    \end{pmatrix},
\end{aligned}
\label{eq:superconductingH}
\end{equation}
where $m_s$, $\mu_s$, $g_s$, and $\Delta_s$ are the effective mass, chemical potential, g-factor, and parent pairing potential, as given in the main text.

In the most general form, we can assume the interface between semiconductor and superconductor may break the symmetry along any spatial direction but it must preserve spin. In this general scenario, an electron from the superconductor may tunnel to the semiconductor region right to the $s$-orbitals from the conduction band with tunnel coupling $t_s$, or to the $p$-valence orbitals $p_{x,y,z}$ with tunnel coupling $t_{px,py,pz}$, respectively. The effective tunneling Hamiltonian is then:
\begin{equation}
    \begin{aligned}
        H_\text{tunnel}=\sum_{\sigma,\mathbf{k}} t_s d^\dagger_{s,\sigma,\mathbf{k}}c_{\sigma,\mathbf{k}}+\sum_{i=x,y,z}t_{pi}d^\dagger_{s,\sigma,\mathbf{k}}b_{\sigma,p_i,\mathbf{k}}+\text{H.c.}
    \end{aligned},
\end{equation}
where $d_{s,\sigma,\mathbf{k}}$ ($b_{\sigma,p_i,\mathbf{k}}$) destroys a superconducting electron ($p_i$ hole) with momentum $\mathbf{k}$ and spin $\sigma$. Given that our basis is not written in terms of $p$-orbitals but HHs and LHs, a unitary transformation must be applied to write down the tunneling in our basis. We skip the details of this rotation, which are given in Ref.~\cite{babkin2024superconducting}.

To obtain the effective pairing terms in the semiconductor region, we integrate out the superconducting states. We perform a Schrieffer-Wolff transformation~\cite{Winkler:684956} to extract the effective pairings in the different bands, finding to third-order in a 6KP theory with CB, HH, and LH bands:
\begin{equation}
\begin{aligned}
    H^\Delta&=\\
    &\left(\begin{array}{cc|ccccc}
        0 & \Delta_\text{CB} & 0 & \frac{\Delta_\text{CH1}}{\sqrt{3}} & -\Delta_\text{CH2} & -\Delta_\text{CH1}^* \\
        -\Delta_\text{CB} & 0 & -\Delta_\text{CH1} & \Delta_\text{CH2} & \frac{\Delta_\text{CH1}^*}{\sqrt{3}} & 0 \\ \hline
        0 & \Delta_\text{CH1} & 0 & \Delta_{R'} & \Delta_{S'} & \Delta_\mathrm{HH} \\
        -\frac{\Delta_\text{CH1}}{\sqrt{3}} & -\Delta_\text{CH2} & -\Delta_{R'} & 0 & \Delta_\mathrm{LH} & -\Delta_{S'}^* \\
        \Delta_\text{CH2} & -\frac{\Delta_\text{CH1}^*}{\sqrt{3}} & -\Delta_{S'} & -\Delta_\mathrm{LH} & 0 & \Delta_{R'}^* \\
        \Delta_\text{CH1}^* & 0 & -\Delta_\mathrm{HH} & \Delta_{S'}^* & -\Delta_{R'}^* & 0
    \end{array}\right),
\end{aligned}
\end{equation}
where the effective gaps we find can be directly related to the different gaps $\Delta_\text{CB},\Delta_\text{HH},\Delta_\text{LH}, \Delta_\text{R}, \Delta_\text{S}$ we assume in the main text. In terms of the parent superconducting gap $\Delta_s$, we get:
\begin{equation}
    \begin{aligned}
        \frac{\Delta_{CB}}{\Delta_s}&\approx \frac{t_s^2}{E_s^2-E_g^2} \\
        \frac{\Delta_{HH}}{\Delta_s}&\approx -\frac{t_{px}^2+t_{py}^2}{2E_s^2} \\
        \frac{\Delta_{LH}}{\Delta_s}&\approx \frac{t_{px}^2+t_{py}^2+4t_{pz}^2}{6E_s^2} \\
        \frac{\Delta_{R'}}{\Delta_s}&=\frac{\Delta_{R}e^{i\chi_R}}{\Delta_s}\approx \frac{(t_{px}-it_{py})^2}{2\sqrt{3}E_s^2} \\
        \frac{\Delta_{S'}}{\Delta_s}&=\frac{\Delta_{S}e^{i\chi_S}}{\Delta_s}\approx -\frac{(t_{px}-it_{py})t_{pz}}{\sqrt{3}E_s^2},
    \end{aligned}
\end{equation}
where $E_s$ is the energy of the superconducting states measured from the top of the valence band, which are given by the eigenvalues of the diagonal blocks from Eq.~(\ref{eq:superconductingH}) with the proper boundary conditions for the problem (e.g. vertically confined superconductor). The energy of the superconducting states has an intrinsic dependence on the magnetic field $E_s=E_s(\mu_s, \mathbf{B})$, which effectively introduces extra magnetic dependence on the induced pairing terms. We neglect these effects since one expects the main energy scale to be given by the chemical potential difference between the superconductor states and the semiconductor states. Besides, we find hybrid conduction-valence pairing terms $\Delta_\text{CH1,2}$, whose expressions are
\begin{equation}
    \begin{aligned}
        \frac{\Delta_\text{CH1}}{\Delta_s}\approx -\frac{t_s(t_{px}-it_{py})}{2\sqrt{2}}\left(\frac{1}{E_s^2}+\frac{1}{E_s^2-E_g^2}\right) \\
        \frac{\Delta_\text{CH2}}{\Delta_s}\approx -\frac{t_st_{pz}}{\sqrt{6}}\left(\frac{1}{E_s^2}+\frac{1}{E_s^2-E_g^2}\right).
    \end{aligned}
\end{equation}
We neglect these mixed conduction-valence pairing terms since they effectively have a higher order within the hole subspace compared to the constant pairing terms. However, it can be shown that they produce linear-in-k corrections to $\Delta_\text{HH}$, $\Delta_\text{LH}$, $\Delta_\text{R}$, and $\Delta_\text{S}$ which may be necessary in more quantitative treatments of the problem.

\section{Comparison of effective 4KP theory and the parent 8KP Hamiltonian}
\label{Appendix:benchmark}
\begin{figure}[ht]
   \centering
    \includegraphics[width=\linewidth]{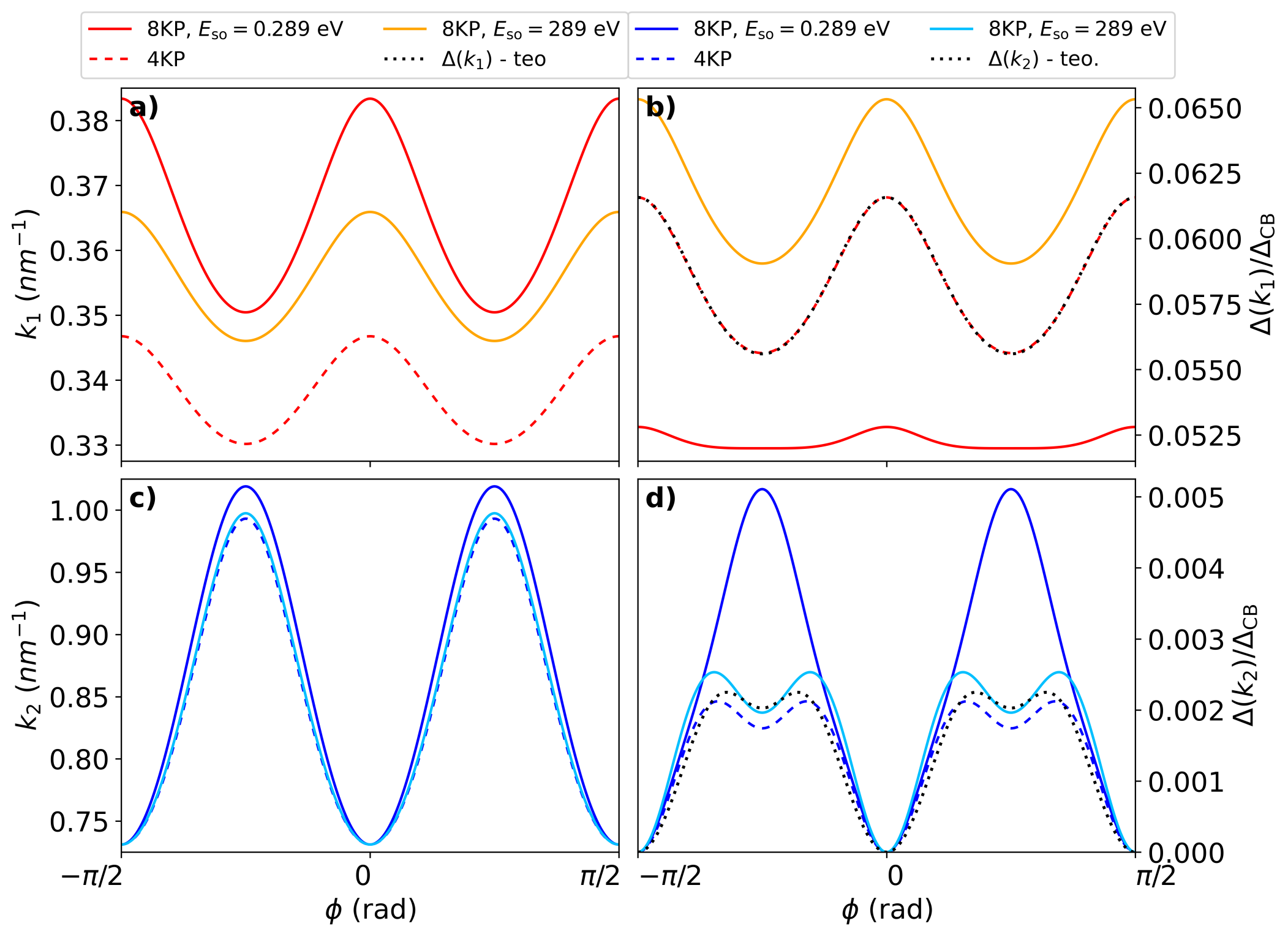}    \caption{\textbf{Comparison between 8KP and 4KP models.} (b,d) Gaps and (a,c) their positions in $k$, as a function of the spherical coordinate $\phi_k$ with $\theta_k=\pi/2$; at (a,b) $k_1(\phi_k,\pi/2,\mu)$ and (c,d) $k_2(\phi_k,\pi/2,\mu)$. Parameters used: $\Delta_\mathrm{CB}=200$ $\mu$eV; $\mu=-0.1$ eV; Table \ref{table:Ge-parameters}.}
    \label{fig:4KP_compare_8KP}
\end{figure}

In Fig. \ref{fig:4KP_compare_8KP}, we show a comparison between the effective 4KP and the original 8KP model for $H_v^\Delta=0$, $H_c^\Delta = i\sigma_y\Delta_\mathrm{CB}$ at $\theta_k=\pi/2$. In Fig. \ref{fig:4KP_compare_8KP}(a,c) we show the position of the superconducting anticrossings $k_1(\phi_k)$ and $k_2(\phi_k)$. The effective model qualitatively captures the correct orientation dependence and only deviates from the parent Hamiltonian results in, roughly, a $15\%$. In Fig. \ref{fig:4KP_compare_8KP}(b,d), we compare the magnitude of the anticrossings, resulting in an overall good agreement, except around $\phi_k=\pi/4$, which shows a factor 2 disagreement. The resulting error is mainly due to the lack of the split-off band, as illustrated when comparing the 4KP results with the 8KP for a decoupled split-off band (see light blue in Fig. \ref{fig:4KP_compare_8KP}(d)). As noted previously, differences coming from the (lack of) split-off band lose relevance at the 2DHG regime for strained Ge, where most experiments take place.

\section{Confinement along $z$ direction} \label{sec:Appendix_confinement}

Assuming a vertical electric field $F$ and hard infinite wall boundary conditions at $z=\pm L_W/2$, we find the following trial wave function for the ground-state Bastard wavefunction \cite{npj2021}
\begin{equation}
\begin{aligned}
    \psi_z(z,\beta) =& \sqrt{\frac{4\beta(\pi^2+\beta^2)}{\pi^2L_W(1-e^{-2\beta})}} \cos\left(\frac{\pi z}{L_W}\right)
    \\
    &\times \exp\left[-\beta\left(\frac{z}{L_W}+\frac{1}{2}\right)\right]
\end{aligned}
\end{equation}

The variational parameter $\beta$ minimizes the ground-state energy of the vertical Hamiltonian
\begin{equation}
    H_z = \frac{p_z^2}{2m_\perp} + eFz + V_\mathrm{barrier}\theta_k\left(|z|-\frac{L_W}{2}\right)
\end{equation}
where the barrier potential is approximated to a hard-wall condition $V_\mathrm{barrier}\rightarrow \infty$, and $m_\perp$ is the confinement mass along $z$, which is different for light and heavy holes,
\begin{equation}
    m_\perp^\mathrm{L} = \frac{m_0}{\gamma_1+2\gamma_2} \quad,\quad m_\perp^\mathrm{H} = \frac{m_0}{\gamma_1-2\gamma_2} \;.
\end{equation}
Then, we compute the ground-state energy
\begin{equation}
\begin{aligned}
    E &= \int_{-L_W/2}^{L_W/2} \psi_z(z,\beta) H_z \psi_z(z,\beta) \, dz
    \\
    & = FL_W\left(\frac{1}{2\beta} + \frac{\beta}{\pi^2+\beta^2} - \frac{1}{2}\coth\beta\right) + \frac{\hbar^2(\pi^2+\beta^2)}{2m_\perp L_W^2}
\end{aligned}
\end{equation}
and we minimize this quantity to (numerically) obtain a $\beta_{L,H}$ such that $\partial_\beta E = 0$ for each $m_\perp^L$ and $m_\perp^H$. Now, we extract the following quantities, 
\begin{equation}
\begin{aligned}
\label{eq:vertical-quantities}
    \langle p_z^2\rangle_H &= \bra{0_H} p_z^2\ket{0_H} = \frac{\hbar^2}{L_W^2}(\pi^2+\beta_H^2) \\
    \langle p_z^2\rangle_L &= \bra{0_L} p_z^2\ket{0_L} = \frac{\hbar^2}{L_W^2}(\pi^2+\beta_L^2) \\
    E_{hl} &= E_{hl}^{\mathrm{(strain)}}+\frac{\hbar^2(\pi^2+\beta_L^2)}{2m_\perp^LL_W^2}-\frac{\hbar^2(\pi^2+\beta_H^2)}{2m_\perp^HL_W^2} \\
    \mathcal{O}_0 &= \langle 0_H | 0_L \rangle = \frac{4(e^{\bar{\beta}}-1)\tilde{K}}{\bar{\beta}(4\pi^2+\bar{\beta}^2)}
    \\
    \alpha_0 &= -i\bra{0_H}p_z\ket{0_L} = \frac{2\hbar(\beta_L-\beta_H)(e^{\bar{\beta}}-1)\tilde{K}}{L_W\bar{\beta}(4\pi^2+\bar{\beta}^2)}
    \\
    z_0 &= \bra{0_H}z\ket{0_L} = -\frac{4L_W\tilde{K}e^{\bar{\beta}/2}}{\bar{\beta}^2(4\pi^2+\bar{\beta}^2)^2}
    \\
    &\times \left[\bar{\beta}(\bar{\beta}^2+4\pi^2)\cosh(\bar{\beta}/2)-2(3\bar{\beta}^2+4\pi^2)\sinh(\beta/2) \right]
    \\
    \tilde{K} &= \prod_{\alpha=H,L}\sqrt{\beta_\alpha(\pi^2+\beta_\alpha^2)(\mathrm{coth}\beta_\alpha-1)}
    \\
    \bar{\beta} &= \beta_H+\beta_L
\end{aligned}
\end{equation}
where the expectation values are integrated in the interval $z\in[-L_W/2,L_W/2]$.

\section{Exact diagonalization of the 4-bands Hamiltonian}
\label{Appendix:exactdiag}
By projecting to the ground state of the vertical motion in the 2DHG and assuming $B=0$, it is possible to analytically diagonalize the 4-band Kohn-Luttinger Hamiltonian in Eq.~\eqref{eq:Kohn-Luttinger_Hamiltonian} and obtain exact analytical expressions for the pairing terms. When projecting to the vertical ground state, the expected value of $\bra{0_H}p_z\ket{0_L}$ is an imaginary quantity, see Eq.~\eqref{eq:vertical-quantities}, transforming the terms that go with $S$ in Eq.~\eqref{eq:Kohn-Luttinger_Hamiltonian} into an effective k-dependent Zeeman interaction. The diagonalization procedure consists on the concatenation of different block-diagonalizations. The first step is to define a mixing angle $\theta_R$ between states $\ket{3/2,\pm 3/2}$ and $\ket{3/2,\mp 1/2}$, respectively:
\begin{equation}
    \begin{aligned}
        \Omega_{RQ}=\sqrt{(-E_{hl}+Q)^2+|R|^2} \\
        \theta_R=-\arcsin{\left(\frac{R\exp(-i\phi_{R})}{\Omega_{RQ}}\right)},
    \end{aligned}
\end{equation}
where $\phi_R$ is the phase of the $R$ term, such that $R=\Omega_{RQ}\sin\theta_Re^{i\phi_R}$ and $Q=\frac{Ehl}{2}+\Omega_{RQ}\cos\theta_R$. From here, we diagonalize the $\ket{3/2,\pm 3/2}$-$\ket{3/2,\mp 1/2}$ blocks with an unitary $U_{1}$, arriving at
\begin{equation}
\begin{aligned}
    H_{4KP}^{(1)}=U_1H_{4KP}U_1^\dagger=-P-\mu+\\
    \begin{pmatrix}
        -\Omega_{RQ} & S^*\sin\theta_R & 0 & S^*\cos\theta_R \\
        S\sin\theta_R & -\Omega_{RQ} & -S\cos\theta_R & 0 \\
        0 & -S^*\cos\theta_R & \Omega_{RQ} & S^*\sin\theta_R \\
        S\cos\theta_R & 0 & S\sin\theta_R & \Omega_{RQ}
    \end{pmatrix}.
    \end{aligned}
\end{equation}
The next step is to diagonalize the blocks connected by $\cos\theta_R$. We define here the mixing angle $\theta_S$, such that
\begin{equation}
    \begin{aligned}
        \Omega_{SRQ}=\sqrt{(\Omega_{RQ}^2+|S|^2\cos^2\theta_R} \\
        \theta_{S}=-\arcsin{\left(\frac{S\cos\theta_R\exp(-i\phi_{S})}{\Omega_{SRQ}}\right)},
    \end{aligned}
\end{equation}
where $\phi_S$ is the imaginary phase of $S$, analogously to $\phi_R$ and $R$. Note that $S$ has been projected to the vertical ground state. We end up with
\begin{equation}
\begin{aligned}
    H_{4KP}^{(2)}=U_2H_{4KP}^{(1)}U_2^\dagger= -P-\mu+\Omega_{SRQ}\times \\\begin{pmatrix}
        -1 & -h_{1} e^{i\phi_S} & 0 & h_2e^{i\phi_S} \\
        -h_{1} e^{-i\phi_S} & -1 & h_2e^{-i\phi_S} & 0 \\
        0 & h_2e^{i\phi_S} & 1 & h_{1}e^{i\phi_S} \\
        h_2e^{-i\phi_S} & 0 & h_{1}e^{-i\phi_S} & 1
    \end{pmatrix},
    \end{aligned}
\end{equation}
where $h_1=\cos\theta_S\sin\theta_S\tan\theta_R$ and $h_2=\sin^2\theta_S\tan\theta_R$. The next step is to rotate the blocks with same diagonal terms with unitary $U_3$. This leaves only the anti-diagonal that goes with $h_2$ as the only non-diagonal terms. To finish the diagonalization, we define the following terms
\begin{equation}
\begin{aligned}
    \Omega_{\pm}&=\Omega_{SRQ}\sqrt{\sin^4\theta_S\tan\theta^2_{RQ}+(\cos\theta_S\sin\theta_S\tan\theta_R\pm 1)^2} \\
    \theta_\pm &= \arcsin{\left(\frac{\Omega_{SRQ}\sin^2\theta_S\tan\theta_R}{\Omega_\pm}\right)}
\end{aligned},
\end{equation}
where the mixing angles $\theta_\pm$ are not fully independent. However, using these mixing angles to perform the final rotation $U_4$, leads to a full diagonal form:
\begin{equation}
\begin{aligned}
    H_{4KP}^{(diag)}&=U_4U_3U_2U_1 H_{4KP} U_1^\dagger U_2^\dagger U_3^\dagger U_4^\dagger =\\ -P-\mu &+ \mathrm{diag}(+\Omega_-,-\Omega_-,+\Omega_+,-\Omega_+).
\end{aligned}
\end{equation}
\subsection{Exact expressions for the pairing terms}
The exact diagonalization of the time-reversed terms follows an identical reasoning. Let $U=U_4U_3U_2U_1$ and $U'=U'_4U'_3U'_2U'_1$, with $U_i'$ being the time-reversed equivalent of $U_i$, any pairing Hamiltonian within the valence band $H_{v,j}^\Delta$ becomes
\begin{equation}
    H_{v,j}^{\Delta\ (d)}=UH_{v,j}^{\Delta}U'^\dagger,
\end{equation}
this equation allows us to extract exact analytical expressions of the pairings, valid at all orders. For the non-zero conduction band pairing terms, we get
\begin{widetext}
\begin{equation}
\label{eq:2DHGgaps-analytical}
\begin{aligned}
    \frac{\tilde{\Delta}_{l,0}^{(c)}}{\Delta_{CB}}
    &=\frac{\tilde{\Delta}_{h,0}^{(c)}}{\Delta_{CB}}=
    \frac{e^{-i \left(2 \phi_R+\phi_S\right)}}{E_g-2\mu}\left[\sin \left(\frac{\theta_-}{2}-\frac{\theta_+}{2}+\theta_S\right)\Big(R_{\Delta } \cos \left(\frac{\theta_-}{2}+\frac{\theta_+}{2}-\theta_R\right)+ Q_{\Delta } e^{i \phi_R} \sin \left(\frac{\theta_-}{2}+\frac{\theta_+}{2}-\theta_R\right)\Big)\right] \\
    \frac{\tilde{\Delta}_{l,x}^{(c)}}{\Delta_{CB}}&=\frac{\tilde{\Delta}_{h,x}^{(c)}}{\Delta_{CB}}=\frac{S_{\Delta } e^{-i \left(\phi_R+2 \phi_S\right)} \sin \left(\frac{\theta_-}{2}+\frac{\theta_+}{2}-\theta_R\right)}{\text{E}_g-2 \mu } \\
    \frac{\tilde{\Delta}_{l,z}^{(c)}}{\Delta_{CB}}&=\frac{e^{-i \left(2 \phi_R+\phi_S\right)}}{E_g-2\mu} \Big[\cos \left(\frac{\theta_-}{2}-\frac{\theta_+}{2}+\theta_S\right)
    \Big(-R_{\Delta} \sin \left(\theta_-/2+\theta_+/2-\theta_R\right)+Q_{\Delta } e^{i \phi_R} \cos \left(\theta_-/2+\theta_+/2-\theta_R\right)\Big)-P_{\Delta } e^{i \phi_R}\Big] \\
   \frac{\tilde{\Delta}_{h,z}^{(c)}}{\Delta_{CB}}&=-\frac{e^{-i \left(2 \phi_R+\phi_S\right)}}{E_g-2\mu} \Big[\cos \left(\frac{\theta_-}{2}-\frac{\theta_+}{2}+\theta_S\right)\Big(-R_{\Delta} \sin \left(\theta_-/2+\theta_+/2-\theta_R\right)+Q_{\Delta } e^{i \phi_R} \cos \left(\theta_-/2+\theta_+/2-\theta_R\right)\Big)+P_{\Delta } e^{i \phi_R}\Big].
\end{aligned}
\end{equation}
For the direct HH and LH we define $\bar{\Delta}_H=(\Delta_{\mathrm{HH}}+\Delta_{\mathrm{LH}})/2$ and $\Delta_{H-}=\Delta_{\mathrm{HH}}-\Delta_{\mathrm{LH}}$, and the non-zero pairing terms are
\begin{equation}
    \begin{aligned}
    \frac{\tilde{\Delta}_{l,0}^{(v)}}{\bar{\Delta}_H}&=\frac{\tilde{\Delta}_{h,0}^{(v)}}{\bar{\Delta}_H}=\frac{1}{2} e^{-i \left(\phi_R+\phi_S\right)} \big(\cos \left(\theta_+-\theta_R-\theta_S\right)-\cos \left(\theta_--\theta_R+\theta_S\right)\big) \\
    \tilde{\Delta}_{l,z}^{(v)}&= \frac{1}{2} e^{-i \left(\phi_R+\phi_S\right)} \big[\Delta _{H-}- \bar{\Delta}_H \left(\cos \left(\theta_--\theta_R+\theta_S\right)+\cos \left(\theta_+-\theta_R-\theta_S\right)\right)\big]
   \\
    \tilde{\Delta}_{h,z}^{(v)}&= \frac{1}{2} e^{-i \left(\phi_R+\phi_S\right)} \big[\Delta _{H-}+ \bar{\Delta}_H \left(\cos \left(\theta_--\theta_R+\theta_S\right)+\cos \left(\theta_+-\theta_R-\theta_S\right)\right)\big].
    \end{aligned}
\end{equation}
Finally, the mixed terms are
\begin{equation}
    \begin{aligned}
        \frac{\tilde{\Delta}_{l,0}^{(m)}}{\Delta_R}&=\frac{\tilde{\Delta}_{h,0}^{(m)}}{\Delta_R}=e^{-i \left(\phi_R+\phi_S\right)}\cos\left(\theta_R-\frac{\theta_-}{2}-\frac{\theta_+}{2}\right)\sin\left(\theta_S+\frac{\theta_-}{2}-\frac{\theta_+}{2}\right)\cos(\phi_R-\chi_R) \\
   \frac{\tilde{\Delta}_{l,x}^{(m)}}{\Delta_R}&=-\frac{\tilde{\Delta}_{h,x}^{(m)}}{\Delta_R}=-i e^{-i \left(\phi_R+\phi_S\right)} \sin \left(\frac{\theta_-}{2}-\frac{\theta_+}{2}+\theta_S\right) \sin(\phi_R-\chi_R) \\
   \frac{\tilde{\Delta}_{l,y}^{(m)}}{\Delta_S}&=-\frac{\tilde{\Delta}_{h,y}^{(m)}}{\Delta_S}=i e^{-i \left(\phi_R+\phi_S\right)} \sin \left(\frac{\theta_-}{2}-\frac{\theta_+}{2}+\theta_S\right)\cos(\phi_S-\chi_S) \\
   \frac{\tilde{\Delta}_{l,z}^{(m)}}{\Delta_R}&=-\frac{\tilde{\Delta}_{h,z}^{(m)}}{\Delta_R}=-e^{-i \left(\phi_R+\phi_S\right)}\sin\left(\theta_R-\frac{\theta_-}{2}-\frac{\theta_+}{2}\right)\cos\left(\theta_S+\frac{\theta_-}{2}-\frac{\theta_+}{2}\right)\cos(\phi_R-\chi_R).
    \end{aligned}
\end{equation}
\end{widetext}

\section{Rotation of the longitudinal and transverse pairing terms for a general magnetic field}
\label{Appendix:general-B}
Given the Zeeman Hamiltonian given in Eq.~\eqref{eq:gmatrix}, the Larmor vectors of hole and their time-reversed partners are given by:
\begin{equation}
    \begin{aligned}
        \boldsymbol{\omega}_L&=(g_{xz}B_z, g_{yx}B_x+g_{yy}B_y,\frac{\delta_{h}}{\mu_B}+g_{zx}B_x+g_{zy}B_y) \\
        \bar{\boldsymbol{\omega}}_L&=-(g_{xz}B_z, g_{yx}B_x+g_{yy}B_y,\frac{\delta_{h}}{\mu_B}-g_{zx}B_x-g_{zy}B_y),
    \end{aligned}
\end{equation}
such that $(\mu_B/2)\boldsymbol{\omega}_L\cdot\boldsymbol{\sigma}$ and $(\mu_B/2)\bar{\boldsymbol{\omega}}_L\cdot\boldsymbol{\sigma}$ are the effective Zeeman terms in each subspace. To understand how the pairing terms transform in the eigenbasis, we need to diagonalize the spin-splitting terms. For this purpose, we define the angles $\Theta$ and $\bar{\Theta}$ as the deviation angles from the $z$ and $-z$ axis, respectively:
\begin{equation}
    \begin{aligned}
    \Theta&=\arccos\left({\frac{\omega_L^{(z)}}{||\boldsymbol{\omega}_L||}}\right) \\
        \bar{\Theta}&=\arccos\left({-\frac{\bar{\omega}_L^{(z)}}{||\bar{\boldsymbol{\omega}}_L||}}\right).
    \end{aligned}
\end{equation}
Furthermore, we define a normalized rotation axis given by the perpendicular between $\boldsymbol{\omega}_L$ and the $z$ axis:
\begin{equation}
    \mathbf{n}=\frac{\boldsymbol{\omega}_L\times \hat{z}}{||\boldsymbol{\omega}_L||},
\end{equation}
while generally $\mathbf{n}\neq \mathbf{\bar{n}}$, the choice of $\bar{\Theta_k}$ and basis set leads to the same rotation axis for the time-reversed sector $\mathbf{n}=\mathbf{\bar{n}}$. The rotations that bring hole and time-reversed Hamiltonians to diagonal form are given by $R=\exp(i\Theta\mathbf{n}\cdot\boldsymbol{\sigma}/2)$ and $\bar{R}=\exp(i\bar{\Theta}\mathbf{n}\cdot\boldsymbol{\sigma}/2)$. Defining $\Theta_\pm=(\Theta\pm\bar{\Theta})/2$ and given a general pairing term of the form $\Delta_0\sigma_0+\boldsymbol{\Delta}\cdot\boldsymbol{\sigma}$, the rotated term becomes
\begin{equation}
    \begin{aligned}
        R(\Delta_0\sigma_0)\bar{R}^\dagger&=\Delta_0\left[\cos\Theta_-\sigma_0+i\sin\Theta_-\mathbf{n}\cdot\boldsymbol{\sigma}\right] \\
        R(\mathbf{\Delta}\cdot\boldsymbol{\sigma})\bar{R}^\dagger&=\cos\Theta_+\mathbf{n}\times(\boldsymbol{\Delta}\times\mathbf{n})\cdot\boldsymbol{\sigma}-\sin\Theta_+(\mathbf{n}\times \mathbf{\Delta})\cdot\boldsymbol{\sigma} \\
        &+ (\mathbf{n}\cdot\boldsymbol{\Delta})\left[\cos\Theta_-(\mathbf{n}\cdot\boldsymbol{\sigma})+i\sin\Theta_-\sigma_0\right],
    \end{aligned}
\end{equation}
which indicates in a general manner the geometrical relationships between longitudinal and transverse pairings with the imbalance of the Zeeman and Rashba spin-splittings.

\section{Magneto-orbital corrections}
\label{Appendix:vectorpotential}
The magnetic response of hole spins is known to be heavily influenced by the vector potential~\cite{ares2013nature}. We estimate corrections to the proximitized hole Hamiltonian arising due to the vector potential. In the Kohn-Luttinger Hamiltonian, the vector potential $\mathbf{A}$ can be included through the minimal coupling $\mathbf{p}\rightarrow \boldsymbol{\Pi}=\mathbf{p}+e\mathbf{A}$. Given a choice of gauge, the vector potential acquires a position dependence, and the momenta can no longer be treated as a good quantum number anymore. In this scenario, exact diagonalization of the Kohn-Luttinger Hamiltonian is not possible, hence, we focus on the perturbative regime of HHs at low $|\mu|$. For considering magneto-orbital effects we perform quasi-degenerate perturbation theory~\cite{Winkler:684956, AbadilloPRL2023} to integrate out the contributions from the conduction and light-hole bands to the heavy-hole manifold.

The effective HH 2DHG Hamiltonian in the presence of orbital effects can be estimated to be:
\begin{equation}
\label{eq:perturbative-H}
\begin{aligned}
    H_{HH}^{(0)}&=\frac{1}{2m_\parallel}(\Pi_x^2+\Pi_y^2)+\frac{\mu_B}{2}\boldsymbol{\sigma}\cdot g(\boldsymbol{\Pi})\cdot\mathbf{B}-i\alpha_{\bigcirc}\Pi_-^3\sigma_- \\
        &-i\alpha_\square\Pi_+\Pi_-\Pi_+\sigma_--\mu+h.c.,
\end{aligned}
\end{equation}
where $m_\parallel$ is the effective in-plane effective mass of heavy-holes, $\Pi_\pm=(\Pi_x\pm i\Pi_y)$ and $\alpha_{\bigcirc}$, and $\alpha_\square$ are spherical and cubic Rashba coefficients:
\begin{equation}
    \begin{aligned}
        \alpha_\bigcirc&= \frac{3\alpha_0\gamma_3(\gamma_2+\gamma_3)\mathcal{O}_0}{2E_{hl}m_0^2} \\
        \alpha_\square&= \frac{3\alpha_0\gamma_3(\gamma_2-\gamma_3)\mathcal{O}_0}{2E_{hl}m_0^2}. 
    \end{aligned}
\end{equation}
It is important to note that perturbative results in Eq.~\eqref{eq:perturbative-H} are not given in the Rashba eigenbasis used for prior results. 
The main corrections are a renormalization of the g-factors:
\begin{equation}
\label{eq:gfactors-perturb}
    \begin{aligned}
        g_{xx}&=3q+\left(\lambda \Pi_x^2-\lambda'\Pi_y^2\right)    \\ 
        g_{yy}&=-3q-\left(\lambda \Pi_y^2-\lambda'\Pi_x^2\right)      \\ 
        g_{xy}&=-g_{yx}=\lambda_{xy}\{\Pi_x,\Pi_y\} \\
        g_{zz}&=6\kappa+\frac{27}{2}q- 2\gamma_{h},
    \end{aligned}
\end{equation}
where $\{A,B\}=(AB+BA)/2$, and $\lambda,\ \lambda',\ \lambda_{xy},\ \text{and } \gamma_h$ are parameters characterizing the perturbative corrections to the g-factors.
\begin{equation}
    \begin{aligned}
        \lambda&=\frac{6}{\hbar m_0E_{hl}}\left(4\gamma_3^2\alpha_0z_0+\hbar\gamma_2\kappa\mathcal{O}^2_0\right) \\ 
        \lambda'&=\frac{6\gamma_2}{\hbar m_0E_{hl}}\left(4\gamma_3\alpha_0z_0+\hbar\kappa\mathcal{O}^2_0\right). \\
        \lambda_{xy}&=\frac{6\gamma_3}{\hbar m_0E_{hl}}\left(2(\gamma_2+\gamma_3)\alpha_0z_0+\hbar\kappa\mathcal{O}^2_0\right).
    \end{aligned}
\end{equation}
In particular, we get a strong correction to the vertical g factor, which can go from $g_{zz}\approx 21$ without $\mathbf{A}$ to $g_{zz}\approx 13$ when considering $\mathbf{A}$. Comparing with the g-factors in the Rashba eigenbasis from Eq.~\eqref{eq:gfactors-diag}, we expect the $g_{xz}$ term to acquire such correction. In addition, we get a correction $\lambda\neq\lambda'$, which prevents cancellation of the perturbative correction for $k_x=k_y$ of the in-plane g-factors $g_{xx}$ and $g_{yy}$, which is equivalent to a small renormalization of the parameters and angular dependence in the in-plane g-factors of Eq.~\eqref{eq:gfactors-diag}. Therefore, we expect quantitative corrections to the observables, such as a small renormalization of how fast the bands tilt with in-plane magnetic fields and a lower dependence on the vertical magnetic field for the longitudinal pairing, as given in Fig.~\ref{fig:4KP_2DHG_Bfield}(c,f). 

Strikingly, orbital corrections to the pairing terms are negligible to linear order in the magnetic field, with only quadratic corrections becoming relevant. Perturbatively, the contribution to the pairings coming from holes with $\mathbf{k}$ gets canceled out exactly with the contribution from their time-reversed states with $-\mathbf{k}$. Hence, only at large magnetic fields the pairing terms get noticeable corrections. For instance, using $\Pi_i^*=\Pi_i(-k)$, the pairing terms from the conduction band in Eq.~\eqref{eq:4KP_pairing} become:
\begin{equation}
\begin{aligned}
P_\Delta&=\frac{1}{2m_0}\gamma_\Delta(\Pi_x\Pi_x^*+\Pi_y\Pi_y^*+\Pi_z\Pi_z^*+i\Pi_x\Pi_y^*-\Pi_y\Pi_x^*)  \\
Q_\Delta&=\frac{1}{4m_0}\gamma_\Delta(\Pi_x\Pi_x^*+\Pi_y\Pi_y^*-2\Pi_z\Pi_z^*+i\Pi_x\Pi_y^*-\Pi_y\Pi_x^*) \\
R_\Delta&=\frac{1}{4m_0}\sqrt{3}\left[-\gamma_\Delta(\Pi_x\Pi_x^*-\Pi_y\Pi_y^*)+i\gamma_\Delta\left(\Pi_x\Pi_y^*+\Pi_x^*\Pi_y\right)\right]\\
S_\Delta&=\frac{1}{2m_0}\sqrt{3}\gamma_\Delta\Pi_-\Pi_z^*.
\end{aligned}
\end{equation}

\bibliography{bibliography}

\begin{thebibliography}{73}%
\makeatletter
\providecommand \@ifxundefined [1]{%
 \@ifx{#1\undefined}
}%
\providecommand \@ifnum [1]{%
 \ifnum #1\expandafter \@firstoftwo
 \else \expandafter \@secondoftwo
 \fi
}%
\providecommand \@ifx [1]{%
 \ifx #1\expandafter \@firstoftwo
 \else \expandafter \@secondoftwo
 \fi
}%
\providecommand \natexlab [1]{#1}%
\providecommand \enquote  [1]{``#1''}%
\providecommand \bibnamefont  [1]{#1}%
\providecommand \bibfnamefont [1]{#1}%
\providecommand \citenamefont [1]{#1}%
\providecommand \href@noop [0]{\@secondoftwo}%
\providecommand \href [0]{\begingroup \@sanitize@url \@href}%
\providecommand \@href[1]{\@@startlink{#1}\@@href}%
\providecommand \@@href[1]{\endgroup#1\@@endlink}%
\providecommand \@sanitize@url [0]{\catcode `\\12\catcode `\$12\catcode `\&12\catcode `\#12\catcode `\^12\catcode `\_12\catcode `\%12\relax}%
\providecommand \@@startlink[1]{}%
\providecommand \@@endlink[0]{}%
\providecommand \url  [0]{\begingroup\@sanitize@url \@url }%
\providecommand \@url [1]{\endgroup\@href {#1}{\urlprefix }}%
\providecommand \urlprefix  [0]{URL }%
\providecommand \Eprint [0]{\href }%
\providecommand \doibase [0]{https://doi.org/}%
\providecommand \selectlanguage [0]{\@gobble}%
\providecommand \bibinfo  [0]{\@secondoftwo}%
\providecommand \bibfield  [0]{\@secondoftwo}%
\providecommand \translation [1]{[#1]}%
\providecommand \BibitemOpen [0]{}%
\providecommand \bibitemStop [0]{}%
\providecommand \bibitemNoStop [0]{.\EOS\space}%
\providecommand \EOS [0]{\spacefactor3000\relax}%
\providecommand \BibitemShut  [1]{\csname bibitem#1\endcsname}%
\let\auto@bib@innerbib\@empty
\bibitem [{\citenamefont {Alicea}(2012)}]{Alicea_RPP2012}%
  \BibitemOpen
  \bibfield  {author} {\bibinfo {author} {\bibfnamefont {J.}~\bibnamefont {Alicea}},\ }\bibfield  {title} {\bibinfo {title} {New directions in the pursuit of majorana fermions in solid state systems},\ }\href {https://doi.org/10.1088/0034-4885/75/7/076501} {\bibfield  {journal} {\bibinfo  {journal} {Rep. Prog. Phys.}\ }\textbf {\bibinfo {volume} {75}},\ \bibinfo {pages} {076501} (\bibinfo {year} {2012})}\BibitemShut {NoStop}%
\bibitem [{\citenamefont {Leijnse}\ and\ \citenamefont {Flensberg}(2012)}]{LeijnseReview}%
  \BibitemOpen
  \bibfield  {author} {\bibinfo {author} {\bibfnamefont {M.}~\bibnamefont {Leijnse}}\ and\ \bibinfo {author} {\bibfnamefont {K.}~\bibnamefont {Flensberg}},\ }\bibfield  {title} {\bibinfo {title} {{Introduction to topological superconductivity and Majorana fermions}},\ }\href {https://doi.org/10.1088/0268-1242/27/12/124003} {\ \textbf {\bibinfo {volume} {27}},\ \bibinfo {pages} {124003} (\bibinfo {year} {2012})}\BibitemShut {NoStop}%
\bibitem [{\citenamefont {Aguado}(2017)}]{AguadoReview}%
  \BibitemOpen
  \bibfield  {author} {\bibinfo {author} {\bibfnamefont {R.}~\bibnamefont {Aguado}},\ }\href {https://doi.org/10.1393/ncr/i2017-10141-9} {\bibfield  {journal} {\bibinfo  {journal} {La Rivista del Nuovo Cimento}\ }\textbf {\bibinfo {volume} {40}},\ \bibinfo {pages} {523} (\bibinfo {year} {2017})}\BibitemShut {NoStop}%
\bibitem [{\citenamefont {Lutchyn}\ \emph {et~al.}(2018)\citenamefont {Lutchyn}, \citenamefont {Bakkers}, \citenamefont {Kouwenhoven}, \citenamefont {Krogstrup}, \citenamefont {Marcus},\ and\ \citenamefont {Oreg}}]{LutchynReview}%
  \BibitemOpen
  \bibfield  {author} {\bibinfo {author} {\bibfnamefont {R.~M.}\ \bibnamefont {Lutchyn}}, \bibinfo {author} {\bibfnamefont {E.~P. A.~M.}\ \bibnamefont {Bakkers}}, \bibinfo {author} {\bibfnamefont {L.~P.}\ \bibnamefont {Kouwenhoven}}, \bibinfo {author} {\bibfnamefont {P.}~\bibnamefont {Krogstrup}}, \bibinfo {author} {\bibfnamefont {C.~M.}\ \bibnamefont {Marcus}},\ and\ \bibinfo {author} {\bibfnamefont {Y.}~\bibnamefont {Oreg}},\ }\href {https://doi.org/10.1038/s41578-018-0003-1} {\bibfield  {journal} {\bibinfo  {journal} {Nature Review Materials}\ }\textbf {\bibinfo {volume} {3}},\ \bibinfo {pages} {52} (\bibinfo {year} {2018})}\BibitemShut {NoStop}%
\bibitem [{\citenamefont {Prada}\ \emph {et~al.}(2020)\citenamefont {Prada}, \citenamefont {San-Jose}, \citenamefont {de~Moor}, \citenamefont {Geresdi}, \citenamefont {Lee}, \citenamefont {Klinovaja}, \citenamefont {Loss}, \citenamefont {Nyg{\aa}rd}, \citenamefont {Aguado},\ and\ \citenamefont {Kouwenhoven}}]{Prada_review}%
  \BibitemOpen
  \bibfield  {author} {\bibinfo {author} {\bibfnamefont {E.}~\bibnamefont {Prada}}, \bibinfo {author} {\bibfnamefont {P.}~\bibnamefont {San-Jose}}, \bibinfo {author} {\bibfnamefont {M.~W.~A.}\ \bibnamefont {de~Moor}}, \bibinfo {author} {\bibfnamefont {A.}~\bibnamefont {Geresdi}}, \bibinfo {author} {\bibfnamefont {E.~J.~H.}\ \bibnamefont {Lee}}, \bibinfo {author} {\bibfnamefont {J.}~\bibnamefont {Klinovaja}}, \bibinfo {author} {\bibfnamefont {D.}~\bibnamefont {Loss}}, \bibinfo {author} {\bibfnamefont {J.}~\bibnamefont {Nyg{\aa}rd}}, \bibinfo {author} {\bibfnamefont {R.}~\bibnamefont {Aguado}},\ and\ \bibinfo {author} {\bibfnamefont {L.~P.}\ \bibnamefont {Kouwenhoven}},\ }\bibfield  {title} {\bibinfo {title} {From andreev to majorana bound states in hybrid superconductor--semiconductor nanowires},\ }\href {https://doi.org/10.1038/s42254-020-0228-y} {\bibfield  {journal} {\bibinfo  {journal} {Nat. Rev. Phys.}\ }\textbf {\bibinfo {volume} {2}},\ \bibinfo {pages} {575} (\bibinfo {year} {2020})}\BibitemShut
  {NoStop}%
\bibitem [{\citenamefont {Aguado}\ and\ \citenamefont {Kouwenhoven}(2020)}]{10.1063/PT.3.4499}%
  \BibitemOpen
  \bibfield  {author} {\bibinfo {author} {\bibfnamefont {R.}~\bibnamefont {Aguado}}\ and\ \bibinfo {author} {\bibfnamefont {L.~P.}\ \bibnamefont {Kouwenhoven}},\ }\bibfield  {title} {\bibinfo {title} {{Majorana qubits for topological quantum computing}},\ }\href {https://doi.org/https://doi.org/10.1063/PT.3.4499} {\bibfield  {journal} {\bibinfo  {journal} {Physics Today}\ }\textbf {\bibinfo {volume} {73}},\ \bibinfo {pages} {44} (\bibinfo {year} {2020})}\BibitemShut {NoStop}%
\bibitem [{\citenamefont {Flensberg}\ \emph {et~al.}(2021)\citenamefont {Flensberg}, \citenamefont {von Oppen},\ and\ \citenamefont {Stern}}]{flensberg2021engineered}%
  \BibitemOpen
  \bibfield  {author} {\bibinfo {author} {\bibfnamefont {K.}~\bibnamefont {Flensberg}}, \bibinfo {author} {\bibfnamefont {F.}~\bibnamefont {von Oppen}},\ and\ \bibinfo {author} {\bibfnamefont {A.}~\bibnamefont {Stern}},\ }\bibfield  {title} {\bibinfo {title} {Engineered platforms for topological superconductivity and majorana zero modes},\ }\href {https://doi.org/https://doi.org/10.1038/s41578-021-00336-6} {\bibfield  {journal} {\bibinfo  {journal} {Nat. Rev. Mat.}\ }\textbf {\bibinfo {volume} {6}},\ \bibinfo {pages} {944} (\bibinfo {year} {2021})}\BibitemShut {NoStop}%
\bibitem [{\citenamefont {Aguado}(2020)}]{APL-Aguado}%
  \BibitemOpen
  \bibfield  {author} {\bibinfo {author} {\bibfnamefont {R.}~\bibnamefont {Aguado}},\ }\bibfield  {title} {\bibinfo {title} {A perspective on semiconductor-based superconducting qubits},\ }\href {https://doi.org/10.1063/5.0024124} {\bibfield  {journal} {\bibinfo  {journal} {Applied Physics Letters}\ }\textbf {\bibinfo {volume} {117}},\ \bibinfo {pages} {240501} (\bibinfo {year} {2020})}\BibitemShut {NoStop}%
\bibitem [{\citenamefont {Seoane~Souto}\ and\ \citenamefont {Aguado}(2024)}]{Seoane-Aguado2024}%
  \BibitemOpen
  \bibfield  {author} {\bibinfo {author} {\bibfnamefont {R.}~\bibnamefont {Seoane~Souto}}\ and\ \bibinfo {author} {\bibfnamefont {R.}~\bibnamefont {Aguado}},\ }\href {https://link.springer.com/chapter/10.1007/978-3-031-55657-9_3} {\emph {\bibinfo {title} {Subgap States in Semiconductor-Superconductor Devices for Quantum Technologies: Andreev Qubits and Minimal Majorana Chains, chapter 3 in New Trends and Platforms for Quantum Technologies}}},\ edited by\ \bibinfo {editor} {\bibfnamefont {R.}~\bibnamefont {Aguado}}, \bibinfo {editor} {\bibfnamefont {R.}~\bibnamefont {Citro}}, \bibinfo {editor} {\bibfnamefont {M.}~\bibnamefont {Lewenstein}},\ and\ \bibinfo {editor} {\bibfnamefont {M.}~\bibnamefont {Stern}}\ (\bibinfo  {publisher} {Springer Nature Switzerland},\ \bibinfo {address} {Cham},\ \bibinfo {year} {2024})\ pp.\ \bibinfo {pages} {133--223}\BibitemShut {NoStop}%
\bibitem [{\citenamefont {Burkard}\ \emph {et~al.}(2023)\citenamefont {Burkard}, \citenamefont {Ladd}, \citenamefont {Pan}, \citenamefont {Nichol},\ and\ \citenamefont {Petta}}]{burkard2023semiconductor}%
  \BibitemOpen
  \bibfield  {author} {\bibinfo {author} {\bibfnamefont {G.}~\bibnamefont {Burkard}}, \bibinfo {author} {\bibfnamefont {T.~D.}\ \bibnamefont {Ladd}}, \bibinfo {author} {\bibfnamefont {A.}~\bibnamefont {Pan}}, \bibinfo {author} {\bibfnamefont {J.~M.}\ \bibnamefont {Nichol}},\ and\ \bibinfo {author} {\bibfnamefont {J.~R.}\ \bibnamefont {Petta}},\ }\bibfield  {title} {\bibinfo {title} {Semiconductor spin qubits},\ }\href {https://doi.org/10.1103/RevModPhys.95.025003} {\bibfield  {journal} {\bibinfo  {journal} {Reviews of Modern Physics}\ }\textbf {\bibinfo {volume} {95}},\ \bibinfo {pages} {025003} (\bibinfo {year} {2023})}\BibitemShut {NoStop}%
\bibitem [{\citenamefont {Huang}\ \emph {et~al.}(2019)\citenamefont {Huang}, \citenamefont {Yang}, \citenamefont {Chan}, \citenamefont {Tanttu}, \citenamefont {Hensen}, \citenamefont {Leon}, \citenamefont {Fogarty}, \citenamefont {Hwang}, \citenamefont {Hudson}, \citenamefont {Itoh} \emph {et~al.}}]{huang2019fidelity}%
  \BibitemOpen
  \bibfield  {author} {\bibinfo {author} {\bibfnamefont {W.}~\bibnamefont {Huang}}, \bibinfo {author} {\bibfnamefont {C.}~\bibnamefont {Yang}}, \bibinfo {author} {\bibfnamefont {K.}~\bibnamefont {Chan}}, \bibinfo {author} {\bibfnamefont {T.}~\bibnamefont {Tanttu}}, \bibinfo {author} {\bibfnamefont {B.}~\bibnamefont {Hensen}}, \bibinfo {author} {\bibfnamefont {R.}~\bibnamefont {Leon}}, \bibinfo {author} {\bibfnamefont {M.}~\bibnamefont {Fogarty}}, \bibinfo {author} {\bibfnamefont {J.}~\bibnamefont {Hwang}}, \bibinfo {author} {\bibfnamefont {F.}~\bibnamefont {Hudson}}, \bibinfo {author} {\bibfnamefont {K.~M.}\ \bibnamefont {Itoh}}, \emph {et~al.},\ }\bibfield  {title} {\bibinfo {title} {Fidelity benchmarks for two-qubit gates in silicon},\ }\href {https://doi.org/10.1038/s41586-019-1197-0} {\bibfield  {journal} {\bibinfo  {journal} {Nature}\ }\textbf {\bibinfo {volume} {569}},\ \bibinfo {pages} {532} (\bibinfo {year} {2019})}\BibitemShut {NoStop}%
\bibitem [{\citenamefont {Noiri}\ \emph {et~al.}(2022)\citenamefont {Noiri}, \citenamefont {Takeda}, \citenamefont {Nakajima}, \citenamefont {Kobayashi}, \citenamefont {Sammak}, \citenamefont {Scappucci},\ and\ \citenamefont {Tarucha}}]{noiri2022fast}%
  \BibitemOpen
  \bibfield  {author} {\bibinfo {author} {\bibfnamefont {A.}~\bibnamefont {Noiri}}, \bibinfo {author} {\bibfnamefont {K.}~\bibnamefont {Takeda}}, \bibinfo {author} {\bibfnamefont {T.}~\bibnamefont {Nakajima}}, \bibinfo {author} {\bibfnamefont {T.}~\bibnamefont {Kobayashi}}, \bibinfo {author} {\bibfnamefont {A.}~\bibnamefont {Sammak}}, \bibinfo {author} {\bibfnamefont {G.}~\bibnamefont {Scappucci}},\ and\ \bibinfo {author} {\bibfnamefont {S.}~\bibnamefont {Tarucha}},\ }\bibfield  {title} {\bibinfo {title} {Fast universal quantum gate above the fault-tolerance threshold in silicon},\ }\href {https://doi.org/10.1038/s41586-021-04182-y} {\bibfield  {journal} {\bibinfo  {journal} {Nature}\ }\textbf {\bibinfo {volume} {601}},\ \bibinfo {pages} {338} (\bibinfo {year} {2022})}\BibitemShut {NoStop}%
\bibitem [{\citenamefont {Mills}\ \emph {et~al.}(2022)\citenamefont {Mills}, \citenamefont {Guinn}, \citenamefont {Gullans}, \citenamefont {Sigillito}, \citenamefont {Feldman}, \citenamefont {Nielsen},\ and\ \citenamefont {Petta}}]{mills2022two}%
  \BibitemOpen
  \bibfield  {author} {\bibinfo {author} {\bibfnamefont {A.~R.}\ \bibnamefont {Mills}}, \bibinfo {author} {\bibfnamefont {C.~R.}\ \bibnamefont {Guinn}}, \bibinfo {author} {\bibfnamefont {M.~J.}\ \bibnamefont {Gullans}}, \bibinfo {author} {\bibfnamefont {A.~J.}\ \bibnamefont {Sigillito}}, \bibinfo {author} {\bibfnamefont {M.~M.}\ \bibnamefont {Feldman}}, \bibinfo {author} {\bibfnamefont {E.}~\bibnamefont {Nielsen}},\ and\ \bibinfo {author} {\bibfnamefont {J.~R.}\ \bibnamefont {Petta}},\ }\bibfield  {title} {\bibinfo {title} {Two-qubit silicon quantum processor with operation fidelity exceeding 99\%},\ }\href {https://doi.org/10.1126/sciadv.abn513} {\bibfield  {journal} {\bibinfo  {journal} {Science Advances}\ }\textbf {\bibinfo {volume} {8}},\ \bibinfo {pages} {eabn5130} (\bibinfo {year} {2022})}\BibitemShut {NoStop}%
\bibitem [{\citenamefont {Xue}\ \emph {et~al.}(2019)\citenamefont {Xue}, \citenamefont {Watson}, \citenamefont {Helsen}, \citenamefont {Ward}, \citenamefont {Savage}, \citenamefont {Lagally}, \citenamefont {Coppersmith}, \citenamefont {Eriksson}, \citenamefont {Wehner},\ and\ \citenamefont {Vandersypen}}]{xue2019benchmarking}%
  \BibitemOpen
  \bibfield  {author} {\bibinfo {author} {\bibfnamefont {X.}~\bibnamefont {Xue}}, \bibinfo {author} {\bibfnamefont {T.}~\bibnamefont {Watson}}, \bibinfo {author} {\bibfnamefont {J.}~\bibnamefont {Helsen}}, \bibinfo {author} {\bibfnamefont {D.~R.}\ \bibnamefont {Ward}}, \bibinfo {author} {\bibfnamefont {D.~E.}\ \bibnamefont {Savage}}, \bibinfo {author} {\bibfnamefont {M.~G.}\ \bibnamefont {Lagally}}, \bibinfo {author} {\bibfnamefont {S.~N.}\ \bibnamefont {Coppersmith}}, \bibinfo {author} {\bibfnamefont {M.}~\bibnamefont {Eriksson}}, \bibinfo {author} {\bibfnamefont {S.}~\bibnamefont {Wehner}},\ and\ \bibinfo {author} {\bibfnamefont {L.~M.}\ \bibnamefont {Vandersypen}},\ }\bibfield  {title} {\bibinfo {title} {Benchmarking gate fidelities in a si/sige two-qubit device},\ }\href {https://doi.org/10.1103/PhysRevX.9.021011} {\bibfield  {journal} {\bibinfo  {journal} {Physical Review X}\ }\textbf {\bibinfo {volume} {9}},\ \bibinfo {pages} {021011} (\bibinfo {year} {2019})}\BibitemShut {NoStop}%
\bibitem [{\citenamefont {Blais}\ \emph {et~al.}(2021)\citenamefont {Blais}, \citenamefont {Grimsmo}, \citenamefont {Girvin},\ and\ \citenamefont {Wallraff}}]{blais2021circuit}%
  \BibitemOpen
  \bibfield  {author} {\bibinfo {author} {\bibfnamefont {A.}~\bibnamefont {Blais}}, \bibinfo {author} {\bibfnamefont {A.~L.}\ \bibnamefont {Grimsmo}}, \bibinfo {author} {\bibfnamefont {S.~M.}\ \bibnamefont {Girvin}},\ and\ \bibinfo {author} {\bibfnamefont {A.}~\bibnamefont {Wallraff}},\ }\bibfield  {title} {\bibinfo {title} {Circuit quantum electrodynamics},\ }\href {https://doi.org/10.1103/RevModPhys.93.025005} {\bibfield  {journal} {\bibinfo  {journal} {Reviews of Modern Physics}\ }\textbf {\bibinfo {volume} {93}},\ \bibinfo {pages} {025005} (\bibinfo {year} {2021})}\BibitemShut {NoStop}%
\bibitem [{\citenamefont {Hays}\ \emph {et~al.}(2021)\citenamefont {Hays}, \citenamefont {Fatemi}, \citenamefont {Bouman}, \citenamefont {Cerrillo}, \citenamefont {Diamond}, \citenamefont {Serniak}, \citenamefont {Connolly}, \citenamefont {Krogstrup}, \citenamefont {Nygård}, \citenamefont {Yeyati}, \citenamefont {Geresdi},\ and\ \citenamefont {Devoret}}]{Hays_Science2021}%
  \BibitemOpen
  \bibfield  {author} {\bibinfo {author} {\bibfnamefont {M.}~\bibnamefont {Hays}}, \bibinfo {author} {\bibfnamefont {V.}~\bibnamefont {Fatemi}}, \bibinfo {author} {\bibfnamefont {D.}~\bibnamefont {Bouman}}, \bibinfo {author} {\bibfnamefont {J.}~\bibnamefont {Cerrillo}}, \bibinfo {author} {\bibfnamefont {S.}~\bibnamefont {Diamond}}, \bibinfo {author} {\bibfnamefont {K.}~\bibnamefont {Serniak}}, \bibinfo {author} {\bibfnamefont {T.}~\bibnamefont {Connolly}}, \bibinfo {author} {\bibfnamefont {P.}~\bibnamefont {Krogstrup}}, \bibinfo {author} {\bibfnamefont {J.}~\bibnamefont {Nygård}}, \bibinfo {author} {\bibfnamefont {A.~L.}\ \bibnamefont {Yeyati}}, \bibinfo {author} {\bibfnamefont {A.}~\bibnamefont {Geresdi}},\ and\ \bibinfo {author} {\bibfnamefont {M.~H.}\ \bibnamefont {Devoret}},\ }\bibfield  {title} {\bibinfo {title} {Coherent manipulation of an andreev spin qubit},\ }\href {https://doi.org/10.1126/science.abf0345} {\bibfield  {journal} {\bibinfo  {journal} {Science}\ }\textbf {\bibinfo {volume} {373}},\
  \bibinfo {pages} {430} (\bibinfo {year} {2021})}\BibitemShut {NoStop}%
\bibitem [{\citenamefont {Pita-Vidal}\ \emph {et~al.}(2023)\citenamefont {Pita-Vidal}, \citenamefont {Bargerbos}, \citenamefont {{\v{Z}}itko}, \citenamefont {Splitthoff}, \citenamefont {Gr{\"u}nhaupt}, \citenamefont {Wesdorp}, \citenamefont {Liu}, \citenamefont {Kouwenhoven}, \citenamefont {Aguado}, \citenamefont {van Heck}, \citenamefont {Kou},\ and\ \citenamefont {Andersen}}]{Pita-Vidal_NatPhys2023}%
  \BibitemOpen
  \bibfield  {author} {\bibinfo {author} {\bibfnamefont {M.}~\bibnamefont {Pita-Vidal}}, \bibinfo {author} {\bibfnamefont {A.}~\bibnamefont {Bargerbos}}, \bibinfo {author} {\bibfnamefont {R.}~\bibnamefont {{\v{Z}}itko}}, \bibinfo {author} {\bibfnamefont {L.~J.}\ \bibnamefont {Splitthoff}}, \bibinfo {author} {\bibfnamefont {L.}~\bibnamefont {Gr{\"u}nhaupt}}, \bibinfo {author} {\bibfnamefont {J.~J.}\ \bibnamefont {Wesdorp}}, \bibinfo {author} {\bibfnamefont {Y.}~\bibnamefont {Liu}}, \bibinfo {author} {\bibfnamefont {L.~P.}\ \bibnamefont {Kouwenhoven}}, \bibinfo {author} {\bibfnamefont {R.}~\bibnamefont {Aguado}}, \bibinfo {author} {\bibfnamefont {B.}~\bibnamefont {van Heck}}, \bibinfo {author} {\bibfnamefont {A.}~\bibnamefont {Kou}},\ and\ \bibinfo {author} {\bibfnamefont {C.~K.}\ \bibnamefont {Andersen}},\ }\bibfield  {title} {\bibinfo {title} {Direct manipulation of a superconducting spin qubit strongly coupled to a transmon qubit},\ }\href {https://doi.org/10.1038/s41567-023-02071-x} {\bibfield  {journal}
  {\bibinfo  {journal} {Nature Physics}\ }\textbf {\bibinfo {volume} {19}},\ \bibinfo {pages} {1110} (\bibinfo {year} {2023})}\BibitemShut {NoStop}%
\bibitem [{\citenamefont {Lee}\ \emph {et~al.}(2014)\citenamefont {Lee}, \citenamefont {Jiang}, \citenamefont {Houzet}, \citenamefont {Aguado}, \citenamefont {Lieber},\ and\ \citenamefont {De~Franceschi}}]{Lee-NatureNano14}%
  \BibitemOpen
  \bibfield  {author} {\bibinfo {author} {\bibfnamefont {E.~J.~H.}\ \bibnamefont {Lee}}, \bibinfo {author} {\bibfnamefont {X.}~\bibnamefont {Jiang}}, \bibinfo {author} {\bibfnamefont {M.}~\bibnamefont {Houzet}}, \bibinfo {author} {\bibfnamefont {R.}~\bibnamefont {Aguado}}, \bibinfo {author} {\bibfnamefont {C.~M.}\ \bibnamefont {Lieber}},\ and\ \bibinfo {author} {\bibfnamefont {S.}~\bibnamefont {De~Franceschi}},\ }\bibfield  {title} {\bibinfo {title} {Spin-resolved andreev levels and parity crossings in hybrid superconductor--semiconductor nanostructures},\ }\href {https://doi.org/10.1038/nnano.2013.267} {\bibfield  {journal} {\bibinfo  {journal} {Nature Nanotechnology}\ }\textbf {\bibinfo {volume} {9}},\ \bibinfo {pages} {79} (\bibinfo {year} {2014})}\BibitemShut {NoStop}%
\bibitem [{\citenamefont {Janvier}\ \emph {et~al.}(2015)\citenamefont {Janvier}, \citenamefont {Tosi}, \citenamefont {Bretheau}, \citenamefont {\c{C}. \"{O}.~Girit}, \citenamefont {Stern}, \citenamefont {Bertet}, \citenamefont {Joyez}, \citenamefont {Vion}, \citenamefont {Esteve}, \citenamefont {Goffman}, \citenamefont {Pothier},\ and\ \citenamefont {Urbina}}]{Janvier_Science2015}%
  \BibitemOpen
  \bibfield  {author} {\bibinfo {author} {\bibfnamefont {C.}~\bibnamefont {Janvier}}, \bibinfo {author} {\bibfnamefont {L.}~\bibnamefont {Tosi}}, \bibinfo {author} {\bibfnamefont {L.}~\bibnamefont {Bretheau}}, \bibinfo {author} {\bibnamefont {\c{C}. \"{O}.~Girit}}, \bibinfo {author} {\bibfnamefont {M.}~\bibnamefont {Stern}}, \bibinfo {author} {\bibfnamefont {P.}~\bibnamefont {Bertet}}, \bibinfo {author} {\bibfnamefont {P.}~\bibnamefont {Joyez}}, \bibinfo {author} {\bibfnamefont {D.}~\bibnamefont {Vion}}, \bibinfo {author} {\bibfnamefont {D.}~\bibnamefont {Esteve}}, \bibinfo {author} {\bibfnamefont {M.~F.}\ \bibnamefont {Goffman}}, \bibinfo {author} {\bibfnamefont {H.}~\bibnamefont {Pothier}},\ and\ \bibinfo {author} {\bibfnamefont {C.}~\bibnamefont {Urbina}},\ }\bibfield  {title} {\bibinfo {title} {Coherent manipulation of andreev states in superconducting atomic contacts},\ }\href {https://doi.org/10.1126/science.aab2179} {\bibfield  {journal} {\bibinfo  {journal} {Science}\ }\textbf {\bibinfo {volume}
  {349}},\ \bibinfo {pages} {1199} (\bibinfo {year} {2015})}\BibitemShut {NoStop}%
\bibitem [{\citenamefont {Hays}\ \emph {et~al.}(2020)\citenamefont {Hays}, \citenamefont {Fatemi}, \citenamefont {Serniak}, \citenamefont {Bouman}, \citenamefont {Diamond}, \citenamefont {de~Lange}, \citenamefont {Krogstrup}, \citenamefont {Nyg{\aa}rd}, \citenamefont {Geresdi},\ and\ \citenamefont {Devoret}}]{HaysNaturePhysics2020}%
  \BibitemOpen
  \bibfield  {author} {\bibinfo {author} {\bibfnamefont {M.}~\bibnamefont {Hays}}, \bibinfo {author} {\bibfnamefont {V.}~\bibnamefont {Fatemi}}, \bibinfo {author} {\bibfnamefont {K.}~\bibnamefont {Serniak}}, \bibinfo {author} {\bibfnamefont {D.}~\bibnamefont {Bouman}}, \bibinfo {author} {\bibfnamefont {S.}~\bibnamefont {Diamond}}, \bibinfo {author} {\bibfnamefont {G.}~\bibnamefont {de~Lange}}, \bibinfo {author} {\bibfnamefont {P.}~\bibnamefont {Krogstrup}}, \bibinfo {author} {\bibfnamefont {J.}~\bibnamefont {Nyg{\aa}rd}}, \bibinfo {author} {\bibfnamefont {A.}~\bibnamefont {Geresdi}},\ and\ \bibinfo {author} {\bibfnamefont {M.~H.}\ \bibnamefont {Devoret}},\ }\bibfield  {title} {\bibinfo {title} {Continuous monitoring of a trapped superconducting spin},\ }\href {https://www.nature.com/articles/s41567-020-0952-3} {\bibfield  {journal} {\bibinfo  {journal} {Nature Physics}\ }\textbf {\bibinfo {volume} {16}},\ \bibinfo {pages} {1103} (\bibinfo {year} {2020})}\BibitemShut {NoStop}%
\bibitem [{\citenamefont {Bargerbos}\ \emph {et~al.}(2022)\citenamefont {Bargerbos}, \citenamefont {Pita-Vidal}, \citenamefont {\ifmmode~\check{Z}\else \v{Z}\fi{}itko}, \citenamefont {\'Avila}, \citenamefont {Splitthoff}, \citenamefont {Gr\"unhaupt}, \citenamefont {Wesdorp}, \citenamefont {Andersen}, \citenamefont {Liu}, \citenamefont {Kouwenhoven}, \citenamefont {Aguado}, \citenamefont {Kou},\ and\ \citenamefont {van Heck}}]{PRXQuantum.3.030311}%
  \BibitemOpen
  \bibfield  {author} {\bibinfo {author} {\bibfnamefont {A.}~\bibnamefont {Bargerbos}}, \bibinfo {author} {\bibfnamefont {M.}~\bibnamefont {Pita-Vidal}}, \bibinfo {author} {\bibfnamefont {R.}~\bibnamefont {\ifmmode~\check{Z}\else \v{Z}\fi{}itko}}, \bibinfo {author} {\bibfnamefont {J.}~\bibnamefont {\'Avila}}, \bibinfo {author} {\bibfnamefont {L.~J.}\ \bibnamefont {Splitthoff}}, \bibinfo {author} {\bibfnamefont {L.}~\bibnamefont {Gr\"unhaupt}}, \bibinfo {author} {\bibfnamefont {J.~J.}\ \bibnamefont {Wesdorp}}, \bibinfo {author} {\bibfnamefont {C.~K.}\ \bibnamefont {Andersen}}, \bibinfo {author} {\bibfnamefont {Y.}~\bibnamefont {Liu}}, \bibinfo {author} {\bibfnamefont {L.~P.}\ \bibnamefont {Kouwenhoven}}, \bibinfo {author} {\bibfnamefont {R.}~\bibnamefont {Aguado}}, \bibinfo {author} {\bibfnamefont {A.}~\bibnamefont {Kou}},\ and\ \bibinfo {author} {\bibfnamefont {B.}~\bibnamefont {van Heck}},\ }\bibfield  {title} {\bibinfo {title} {Singlet-doublet transitions of a quantum dot josephson junction detected in a
  transmon circuit},\ }\href {https://doi.org/10.1103/PRXQuantum.3.030311} {\bibfield  {journal} {\bibinfo  {journal} {PRX Quantum}\ }\textbf {\bibinfo {volume} {3}},\ \bibinfo {pages} {030311} (\bibinfo {year} {2022})}\BibitemShut {NoStop}%
\bibitem [{\citenamefont {Bargerbos}\ \emph {et~al.}(2023)\citenamefont {Bargerbos}, \citenamefont {Pita-Vidal}, \citenamefont {\ifmmode~\check{Z}\else \v{Z}\fi{}itko}, \citenamefont {Splitthoff}, \citenamefont {Gr\"unhaupt}, \citenamefont {Wesdorp}, \citenamefont {Liu}, \citenamefont {Kouwenhoven}, \citenamefont {Aguado}, \citenamefont {Andersen}, \citenamefont {Kou},\ and\ \citenamefont {van Heck}}]{PhysRevLett.131.097001}%
  \BibitemOpen
  \bibfield  {author} {\bibinfo {author} {\bibfnamefont {A.}~\bibnamefont {Bargerbos}}, \bibinfo {author} {\bibfnamefont {M.}~\bibnamefont {Pita-Vidal}}, \bibinfo {author} {\bibfnamefont {R.}~\bibnamefont {\ifmmode~\check{Z}\else \v{Z}\fi{}itko}}, \bibinfo {author} {\bibfnamefont {L.~J.}\ \bibnamefont {Splitthoff}}, \bibinfo {author} {\bibfnamefont {L.}~\bibnamefont {Gr\"unhaupt}}, \bibinfo {author} {\bibfnamefont {J.~J.}\ \bibnamefont {Wesdorp}}, \bibinfo {author} {\bibfnamefont {Y.}~\bibnamefont {Liu}}, \bibinfo {author} {\bibfnamefont {L.~P.}\ \bibnamefont {Kouwenhoven}}, \bibinfo {author} {\bibfnamefont {R.}~\bibnamefont {Aguado}}, \bibinfo {author} {\bibfnamefont {C.~K.}\ \bibnamefont {Andersen}}, \bibinfo {author} {\bibfnamefont {A.}~\bibnamefont {Kou}},\ and\ \bibinfo {author} {\bibfnamefont {B.}~\bibnamefont {van Heck}},\ }\bibfield  {title} {\bibinfo {title} {Spectroscopy of spin-split andreev levels in a quantum dot with superconducting leads},\ }\href
  {https://doi.org/10.1103/PhysRevLett.131.097001} {\bibfield  {journal} {\bibinfo  {journal} {Phys. Rev. Lett.}\ }\textbf {\bibinfo {volume} {131}},\ \bibinfo {pages} {097001} (\bibinfo {year} {2023})}\BibitemShut {NoStop}%
\bibitem [{\citenamefont {Dimoulas}\ \emph {et~al.}(2006)\citenamefont {Dimoulas}, \citenamefont {Tsipas}, \citenamefont {Sotiropoulos},\ and\ \citenamefont {Evangelou}}]{dimoulas2006fermi}%
  \BibitemOpen
  \bibfield  {author} {\bibinfo {author} {\bibfnamefont {A.}~\bibnamefont {Dimoulas}}, \bibinfo {author} {\bibfnamefont {P.}~\bibnamefont {Tsipas}}, \bibinfo {author} {\bibfnamefont {A.}~\bibnamefont {Sotiropoulos}},\ and\ \bibinfo {author} {\bibfnamefont {E.}~\bibnamefont {Evangelou}},\ }\bibfield  {title} {\bibinfo {title} {Fermi-level pinning and charge neutrality level in germanium},\ }\bibfield  {journal} {\bibinfo  {journal} {Applied physics letters}\ }\textbf {\bibinfo {volume} {89}},\ \href {https://doi.org/10.1063/1.2410241} {10.1063/1.2410241} (\bibinfo {year} {2006})\BibitemShut {NoStop}%
\bibitem [{\citenamefont {Winkler}(2003)}]{Winkler:684956}%
  \BibitemOpen
  \bibfield  {author} {\bibinfo {author} {\bibfnamefont {R.}~\bibnamefont {Winkler}},\ }\href {https://doi.org/10.1007/b13586} {\emph {\bibinfo {title} {{Spin-orbit coupling effects in two-dimensional electron and hole systems}}}},\ Springer tracts in modern physics\ (\bibinfo  {publisher} {Springer},\ \bibinfo {address} {Berlin},\ \bibinfo {year} {2003})\BibitemShut {NoStop}%
\bibitem [{\citenamefont {Philippopoulos}\ \emph {et~al.}(2020)\citenamefont {Philippopoulos}, \citenamefont {Chesi}, \citenamefont {Culcer},\ and\ \citenamefont {Coish}}]{philippopoulos2020pseudospin}%
  \BibitemOpen
  \bibfield  {author} {\bibinfo {author} {\bibfnamefont {P.}~\bibnamefont {Philippopoulos}}, \bibinfo {author} {\bibfnamefont {S.}~\bibnamefont {Chesi}}, \bibinfo {author} {\bibfnamefont {D.}~\bibnamefont {Culcer}},\ and\ \bibinfo {author} {\bibfnamefont {W.}~\bibnamefont {Coish}},\ }\bibfield  {title} {\bibinfo {title} {Pseudospin-electric coupling for holes beyond the envelope-function approximation},\ }\href {https://doi.org/10.1103/PhysRevB.102.075310} {\bibfield  {journal} {\bibinfo  {journal} {Physical Review B}\ }\textbf {\bibinfo {volume} {102}},\ \bibinfo {pages} {075310} (\bibinfo {year} {2020})}\BibitemShut {NoStop}%
\bibitem [{\citenamefont {Liu}\ \emph {et~al.}(2022)\citenamefont {Liu}, \citenamefont {Xiong}, \citenamefont {Wang}, \citenamefont {Ma}, \citenamefont {Guan}, \citenamefont {Luo},\ and\ \citenamefont {Li}}]{liu2022emergent}%
  \BibitemOpen
  \bibfield  {author} {\bibinfo {author} {\bibfnamefont {Y.}~\bibnamefont {Liu}}, \bibinfo {author} {\bibfnamefont {J.-X.}\ \bibnamefont {Xiong}}, \bibinfo {author} {\bibfnamefont {Z.}~\bibnamefont {Wang}}, \bibinfo {author} {\bibfnamefont {W.-L.}\ \bibnamefont {Ma}}, \bibinfo {author} {\bibfnamefont {S.}~\bibnamefont {Guan}}, \bibinfo {author} {\bibfnamefont {J.-W.}\ \bibnamefont {Luo}},\ and\ \bibinfo {author} {\bibfnamefont {S.-S.}\ \bibnamefont {Li}},\ }\bibfield  {title} {\bibinfo {title} {Emergent linear rashba spin-orbit coupling offers fast manipulation of hole-spin qubits in germanium},\ }\href {https://doi.org/10.1103/PhysRevB.105.075313} {\bibfield  {journal} {\bibinfo  {journal} {Physical Review B}\ }\textbf {\bibinfo {volume} {105}},\ \bibinfo {pages} {075313} (\bibinfo {year} {2022})}\BibitemShut {NoStop}%
\bibitem [{\citenamefont {Rodr{\'\i}guez-Mena}\ \emph {et~al.}(2023)\citenamefont {Rodr{\'\i}guez-Mena}, \citenamefont {Abadillo-Uriel}, \citenamefont {Veste}, \citenamefont {Martinez}, \citenamefont {Li}, \citenamefont {Skl{\'e}nard},\ and\ \citenamefont {Niquet}}]{rodriguez2023linear}%
  \BibitemOpen
  \bibfield  {author} {\bibinfo {author} {\bibfnamefont {E.~A.}\ \bibnamefont {Rodr{\'\i}guez-Mena}}, \bibinfo {author} {\bibfnamefont {J.~C.}\ \bibnamefont {Abadillo-Uriel}}, \bibinfo {author} {\bibfnamefont {G.}~\bibnamefont {Veste}}, \bibinfo {author} {\bibfnamefont {B.}~\bibnamefont {Martinez}}, \bibinfo {author} {\bibfnamefont {J.}~\bibnamefont {Li}}, \bibinfo {author} {\bibfnamefont {B.}~\bibnamefont {Skl{\'e}nard}},\ and\ \bibinfo {author} {\bibfnamefont {Y.-M.}\ \bibnamefont {Niquet}},\ }\bibfield  {title} {\bibinfo {title} {Linear-in-momentum spin orbit interactions in planar ge/gesi heterostructures and spin qubits},\ }\href {https://doi.org/10.1103/PhysRevB.108.205416} {\bibfield  {journal} {\bibinfo  {journal} {Physical Review B}\ }\textbf {\bibinfo {volume} {108}},\ \bibinfo {pages} {205416} (\bibinfo {year} {2023})}\BibitemShut {NoStop}%
\bibitem [{\citenamefont {Crippa}\ \emph {et~al.}(2018)\citenamefont {Crippa}, \citenamefont {Maurand}, \citenamefont {Bourdet}, \citenamefont {Kotekar-Patil}, \citenamefont {Amisse}, \citenamefont {Jehl}, \citenamefont {Sanquer}, \citenamefont {Lavi{\'e}ville}, \citenamefont {Bohuslavskyi}, \citenamefont {Hutin} \emph {et~al.}}]{crippa2018electrical}%
  \BibitemOpen
  \bibfield  {author} {\bibinfo {author} {\bibfnamefont {A.}~\bibnamefont {Crippa}}, \bibinfo {author} {\bibfnamefont {R.}~\bibnamefont {Maurand}}, \bibinfo {author} {\bibfnamefont {L.}~\bibnamefont {Bourdet}}, \bibinfo {author} {\bibfnamefont {D.}~\bibnamefont {Kotekar-Patil}}, \bibinfo {author} {\bibfnamefont {A.}~\bibnamefont {Amisse}}, \bibinfo {author} {\bibfnamefont {X.}~\bibnamefont {Jehl}}, \bibinfo {author} {\bibfnamefont {M.}~\bibnamefont {Sanquer}}, \bibinfo {author} {\bibfnamefont {R.}~\bibnamefont {Lavi{\'e}ville}}, \bibinfo {author} {\bibfnamefont {H.}~\bibnamefont {Bohuslavskyi}}, \bibinfo {author} {\bibfnamefont {L.}~\bibnamefont {Hutin}}, \emph {et~al.},\ }\bibfield  {title} {\bibinfo {title} {Electrical spin driving by g-matrix modulation in spin-orbit qubits},\ }\href {https://doi.org/10.1103/PhysRevLett.120.137702} {\bibfield  {journal} {\bibinfo  {journal} {Physical review letters}\ }\textbf {\bibinfo {volume} {120}},\ \bibinfo {pages} {137702} (\bibinfo {year} {2018})}\BibitemShut
  {NoStop}%
\bibitem [{\citenamefont {Michal}\ \emph {et~al.}(2021)\citenamefont {Michal}, \citenamefont {Venitucci},\ and\ \citenamefont {Niquet}}]{michal2021longitudinal}%
  \BibitemOpen
  \bibfield  {author} {\bibinfo {author} {\bibfnamefont {V.~P.}\ \bibnamefont {Michal}}, \bibinfo {author} {\bibfnamefont {B.}~\bibnamefont {Venitucci}},\ and\ \bibinfo {author} {\bibfnamefont {Y.-M.}\ \bibnamefont {Niquet}},\ }\bibfield  {title} {\bibinfo {title} {Longitudinal and transverse electric field manipulation of hole spin-orbit qubits in one-dimensional channels},\ }\href {https://doi.org/10.1103/PhysRevB.103.045305} {\bibfield  {journal} {\bibinfo  {journal} {Physical Review B}\ }\textbf {\bibinfo {volume} {103}},\ \bibinfo {pages} {045305} (\bibinfo {year} {2021})}\BibitemShut {NoStop}%
\bibitem [{\citenamefont {Terrazos}\ \emph {et~al.}(2021)\citenamefont {Terrazos}, \citenamefont {Marcellina}, \citenamefont {Wang}, \citenamefont {Coppersmith}, \citenamefont {Friesen}, \citenamefont {Hamilton}, \citenamefont {Hu}, \citenamefont {Koiller}, \citenamefont {Saraiva}, \citenamefont {Culcer} \emph {et~al.}}]{terrazos2021theory}%
  \BibitemOpen
  \bibfield  {author} {\bibinfo {author} {\bibfnamefont {L.}~\bibnamefont {Terrazos}}, \bibinfo {author} {\bibfnamefont {E.}~\bibnamefont {Marcellina}}, \bibinfo {author} {\bibfnamefont {Z.}~\bibnamefont {Wang}}, \bibinfo {author} {\bibfnamefont {S.}~\bibnamefont {Coppersmith}}, \bibinfo {author} {\bibfnamefont {M.}~\bibnamefont {Friesen}}, \bibinfo {author} {\bibfnamefont {A.}~\bibnamefont {Hamilton}}, \bibinfo {author} {\bibfnamefont {X.}~\bibnamefont {Hu}}, \bibinfo {author} {\bibfnamefont {B.}~\bibnamefont {Koiller}}, \bibinfo {author} {\bibfnamefont {A.}~\bibnamefont {Saraiva}}, \bibinfo {author} {\bibfnamefont {D.}~\bibnamefont {Culcer}}, \emph {et~al.},\ }\bibfield  {title} {\bibinfo {title} {Theory of hole-spin qubits in strained germanium quantum dots},\ }\href@noop {} {\bibfield  {journal} {\bibinfo  {journal} {Physical Review B}\ }\textbf {\bibinfo {volume} {103}},\ \bibinfo {pages} {125201} (\bibinfo {year} {2021})}\BibitemShut {NoStop}%
\bibitem [{\citenamefont {Zhang}\ \emph {et~al.}(2021)\citenamefont {Zhang}, \citenamefont {Liu}, \citenamefont {Gao}, \citenamefont {Xu}, \citenamefont {Wang}, \citenamefont {Zhang}, \citenamefont {Cao}, \citenamefont {Wang}, \citenamefont {Zhang}, \citenamefont {Hu} \emph {et~al.}}]{zhang2021anisotropic}%
  \BibitemOpen
  \bibfield  {author} {\bibinfo {author} {\bibfnamefont {T.}~\bibnamefont {Zhang}}, \bibinfo {author} {\bibfnamefont {H.}~\bibnamefont {Liu}}, \bibinfo {author} {\bibfnamefont {F.}~\bibnamefont {Gao}}, \bibinfo {author} {\bibfnamefont {G.}~\bibnamefont {Xu}}, \bibinfo {author} {\bibfnamefont {K.}~\bibnamefont {Wang}}, \bibinfo {author} {\bibfnamefont {X.}~\bibnamefont {Zhang}}, \bibinfo {author} {\bibfnamefont {G.}~\bibnamefont {Cao}}, \bibinfo {author} {\bibfnamefont {T.}~\bibnamefont {Wang}}, \bibinfo {author} {\bibfnamefont {J.}~\bibnamefont {Zhang}}, \bibinfo {author} {\bibfnamefont {X.}~\bibnamefont {Hu}}, \emph {et~al.},\ }\bibfield  {title} {\bibinfo {title} {Anisotropic g-factor and spin--orbit field in a germanium hut wire double quantum dot},\ }\href {https://doi.org/10.1021/acs.nanolett.1c00263} {\bibfield  {journal} {\bibinfo  {journal} {Nano Letters}\ }\textbf {\bibinfo {volume} {21}},\ \bibinfo {pages} {3835} (\bibinfo {year} {2021})}\BibitemShut {NoStop}%
\bibitem [{\citenamefont {Froning}\ \emph {et~al.}(2021)\citenamefont {Froning}, \citenamefont {Camenzind}, \citenamefont {van~der Molen}, \citenamefont {Li}, \citenamefont {Bakkers}, \citenamefont {Zumb{\"u}hl},\ and\ \citenamefont {Braakman}}]{froning2021ultrafast}%
  \BibitemOpen
  \bibfield  {author} {\bibinfo {author} {\bibfnamefont {F.~N.}\ \bibnamefont {Froning}}, \bibinfo {author} {\bibfnamefont {L.~C.}\ \bibnamefont {Camenzind}}, \bibinfo {author} {\bibfnamefont {O.~A.}\ \bibnamefont {van~der Molen}}, \bibinfo {author} {\bibfnamefont {A.}~\bibnamefont {Li}}, \bibinfo {author} {\bibfnamefont {E.~P.}\ \bibnamefont {Bakkers}}, \bibinfo {author} {\bibfnamefont {D.~M.}\ \bibnamefont {Zumb{\"u}hl}},\ and\ \bibinfo {author} {\bibfnamefont {F.~R.}\ \bibnamefont {Braakman}},\ }\bibfield  {title} {\bibinfo {title} {Ultrafast hole spin qubit with gate-tunable spin--orbit switch functionality},\ }\href {https://www.nature.com/articles/s41565-020-00828-6} {\bibfield  {journal} {\bibinfo  {journal} {Nature Nanotechnology}\ }\textbf {\bibinfo {volume} {16}},\ \bibinfo {pages} {308} (\bibinfo {year} {2021})}\BibitemShut {NoStop}%
\bibitem [{\citenamefont {Liles}\ \emph {et~al.}(2021)\citenamefont {Liles}, \citenamefont {Martins}, \citenamefont {Miserev}, \citenamefont {Kiselev}, \citenamefont {Thorvaldson}, \citenamefont {Rendell}, \citenamefont {Jin}, \citenamefont {Hudson}, \citenamefont {Veldhorst}, \citenamefont {Itoh} \emph {et~al.}}]{liles2021electrical}%
  \BibitemOpen
  \bibfield  {author} {\bibinfo {author} {\bibfnamefont {S.}~\bibnamefont {Liles}}, \bibinfo {author} {\bibfnamefont {F.}~\bibnamefont {Martins}}, \bibinfo {author} {\bibfnamefont {D.}~\bibnamefont {Miserev}}, \bibinfo {author} {\bibfnamefont {A.}~\bibnamefont {Kiselev}}, \bibinfo {author} {\bibfnamefont {I.}~\bibnamefont {Thorvaldson}}, \bibinfo {author} {\bibfnamefont {M.}~\bibnamefont {Rendell}}, \bibinfo {author} {\bibfnamefont {I.}~\bibnamefont {Jin}}, \bibinfo {author} {\bibfnamefont {F.}~\bibnamefont {Hudson}}, \bibinfo {author} {\bibfnamefont {M.}~\bibnamefont {Veldhorst}}, \bibinfo {author} {\bibfnamefont {K.}~\bibnamefont {Itoh}}, \emph {et~al.},\ }\bibfield  {title} {\bibinfo {title} {Electrical control of the g tensor of the first hole in a silicon mos quantum dot},\ }\href {https://doi.org/10.1103/PhysRevB.104.235303} {\bibfield  {journal} {\bibinfo  {journal} {Physical Review B}\ }\textbf {\bibinfo {volume} {104}},\ \bibinfo {pages} {235303} (\bibinfo {year} {2021})}\BibitemShut {NoStop}%
\bibitem [{\citenamefont {Martinez}\ \emph {et~al.}(2022)\citenamefont {Martinez}, \citenamefont {Abadillo-Uriel}, \citenamefont {Rodr{\'\i}guez-Mena},\ and\ \citenamefont {Niquet}}]{martinez2022hole}%
  \BibitemOpen
  \bibfield  {author} {\bibinfo {author} {\bibfnamefont {B.}~\bibnamefont {Martinez}}, \bibinfo {author} {\bibfnamefont {J.~C.}\ \bibnamefont {Abadillo-Uriel}}, \bibinfo {author} {\bibfnamefont {E.~A.}\ \bibnamefont {Rodr{\'\i}guez-Mena}},\ and\ \bibinfo {author} {\bibfnamefont {Y.-M.}\ \bibnamefont {Niquet}},\ }\bibfield  {title} {\bibinfo {title} {Hole spin manipulation in inhomogeneous and nonseparable electric fields},\ }\href {https://doi.org/10.1103/PhysRevB.106.235426} {\bibfield  {journal} {\bibinfo  {journal} {Physical Review B}\ }\textbf {\bibinfo {volume} {106}},\ \bibinfo {pages} {235426} (\bibinfo {year} {2022})}\BibitemShut {NoStop}%
\bibitem [{\citenamefont {Jirovec}\ \emph {et~al.}(2022)\citenamefont {Jirovec}, \citenamefont {Mutter}, \citenamefont {Hofmann}, \citenamefont {Crippa}, \citenamefont {Rychetsky}, \citenamefont {Craig}, \citenamefont {Kukucka}, \citenamefont {Martins}, \citenamefont {Ballabio}, \citenamefont {Ares} \emph {et~al.}}]{jirovec2022dynamics}%
  \BibitemOpen
  \bibfield  {author} {\bibinfo {author} {\bibfnamefont {D.}~\bibnamefont {Jirovec}}, \bibinfo {author} {\bibfnamefont {P.~M.}\ \bibnamefont {Mutter}}, \bibinfo {author} {\bibfnamefont {A.}~\bibnamefont {Hofmann}}, \bibinfo {author} {\bibfnamefont {A.}~\bibnamefont {Crippa}}, \bibinfo {author} {\bibfnamefont {M.}~\bibnamefont {Rychetsky}}, \bibinfo {author} {\bibfnamefont {D.~L.}\ \bibnamefont {Craig}}, \bibinfo {author} {\bibfnamefont {J.}~\bibnamefont {Kukucka}}, \bibinfo {author} {\bibfnamefont {F.}~\bibnamefont {Martins}}, \bibinfo {author} {\bibfnamefont {A.}~\bibnamefont {Ballabio}}, \bibinfo {author} {\bibfnamefont {N.}~\bibnamefont {Ares}}, \emph {et~al.},\ }\bibfield  {title} {\bibinfo {title} {Dynamics of hole singlet-triplet qubits with large g-factor differences},\ }\href {https://doi.org/10.1103/PhysRevLett.128.126803} {\bibfield  {journal} {\bibinfo  {journal} {Physical Review Letters}\ }\textbf {\bibinfo {volume} {128}},\ \bibinfo {pages} {126803} (\bibinfo {year} {2022})}\BibitemShut {NoStop}%
\bibitem [{\citenamefont {Abadillo-Uriel}\ \emph {et~al.}(2023)\citenamefont {Abadillo-Uriel}, \citenamefont {Rodr{\'\i}guez-Mena}, \citenamefont {Martinez},\ and\ \citenamefont {Niquet}}]{AbadilloPRL2023}%
  \BibitemOpen
  \bibfield  {author} {\bibinfo {author} {\bibfnamefont {J.~C.}\ \bibnamefont {Abadillo-Uriel}}, \bibinfo {author} {\bibfnamefont {E.~A.}\ \bibnamefont {Rodr{\'\i}guez-Mena}}, \bibinfo {author} {\bibfnamefont {B.}~\bibnamefont {Martinez}},\ and\ \bibinfo {author} {\bibfnamefont {Y.-M.}\ \bibnamefont {Niquet}},\ }\bibfield  {title} {\bibinfo {title} {Hole-spin driving by strain-induced spin-orbit interactions},\ }\href {https://doi.org/10.1103/PhysRevLett.131.097002} {\bibfield  {journal} {\bibinfo  {journal} {Physical Review Letters}\ }\textbf {\bibinfo {volume} {131}},\ \bibinfo {pages} {097002} (\bibinfo {year} {2023})}\BibitemShut {NoStop}%
\bibitem [{\citenamefont {Maurand}\ \emph {et~al.}(2016)\citenamefont {Maurand}, \citenamefont {Jehl}, \citenamefont {Kotekar-Patil}, \citenamefont {Corna}, \citenamefont {Bohuslavskyi}, \citenamefont {Lavi{\'e}ville}, \citenamefont {Hutin}, \citenamefont {Barraud}, \citenamefont {Vinet}, \citenamefont {Sanquer} \emph {et~al.}}]{maurand2016cmos}%
  \BibitemOpen
  \bibfield  {author} {\bibinfo {author} {\bibfnamefont {R.}~\bibnamefont {Maurand}}, \bibinfo {author} {\bibfnamefont {X.}~\bibnamefont {Jehl}}, \bibinfo {author} {\bibfnamefont {D.}~\bibnamefont {Kotekar-Patil}}, \bibinfo {author} {\bibfnamefont {A.}~\bibnamefont {Corna}}, \bibinfo {author} {\bibfnamefont {H.}~\bibnamefont {Bohuslavskyi}}, \bibinfo {author} {\bibfnamefont {R.}~\bibnamefont {Lavi{\'e}ville}}, \bibinfo {author} {\bibfnamefont {L.}~\bibnamefont {Hutin}}, \bibinfo {author} {\bibfnamefont {S.}~\bibnamefont {Barraud}}, \bibinfo {author} {\bibfnamefont {M.}~\bibnamefont {Vinet}}, \bibinfo {author} {\bibfnamefont {M.}~\bibnamefont {Sanquer}}, \emph {et~al.},\ }\bibfield  {title} {\bibinfo {title} {A cmos silicon spin qubit},\ }\href {https://doi.org/10.1038/ncomms13575} {\bibfield  {journal} {\bibinfo  {journal} {Nature communications}\ }\textbf {\bibinfo {volume} {7}},\ \bibinfo {pages} {13575} (\bibinfo {year} {2016})}\BibitemShut {NoStop}%
\bibitem [{\citenamefont {Hendrickx}\ \emph {et~al.}(2021)\citenamefont {Hendrickx}, \citenamefont {Lawrie}, \citenamefont {Russ}, \citenamefont {van Riggelen}, \citenamefont {de~Snoo}, \citenamefont {Schouten}, \citenamefont {Sammak}, \citenamefont {Scappucci},\ and\ \citenamefont {Veldhorst}}]{hendrickx2021four}%
  \BibitemOpen
  \bibfield  {author} {\bibinfo {author} {\bibfnamefont {N.~W.}\ \bibnamefont {Hendrickx}}, \bibinfo {author} {\bibfnamefont {W.~I.}\ \bibnamefont {Lawrie}}, \bibinfo {author} {\bibfnamefont {M.}~\bibnamefont {Russ}}, \bibinfo {author} {\bibfnamefont {F.}~\bibnamefont {van Riggelen}}, \bibinfo {author} {\bibfnamefont {S.~L.}\ \bibnamefont {de~Snoo}}, \bibinfo {author} {\bibfnamefont {R.~N.}\ \bibnamefont {Schouten}}, \bibinfo {author} {\bibfnamefont {A.}~\bibnamefont {Sammak}}, \bibinfo {author} {\bibfnamefont {G.}~\bibnamefont {Scappucci}},\ and\ \bibinfo {author} {\bibfnamefont {M.}~\bibnamefont {Veldhorst}},\ }\bibfield  {title} {\bibinfo {title} {A four-qubit germanium quantum processor},\ }\href {https://doi.org/10.4121/13663442} {\bibfield  {journal} {\bibinfo  {journal} {Nature}\ }\textbf {\bibinfo {volume} {591}},\ \bibinfo {pages} {580} (\bibinfo {year} {2021})}\BibitemShut {NoStop}%
\bibitem [{\citenamefont {Yu}\ \emph {et~al.}(2023)\citenamefont {Yu}, \citenamefont {Zihlmann}, \citenamefont {Abadillo-Uriel}, \citenamefont {Michal}, \citenamefont {Rambal}, \citenamefont {Niebojewski}, \citenamefont {Bedecarrats}, \citenamefont {Vinet}, \citenamefont {Dumur}, \citenamefont {Filippone} \emph {et~al.}}]{yu2023strong}%
  \BibitemOpen
  \bibfield  {author} {\bibinfo {author} {\bibfnamefont {C.~X.}\ \bibnamefont {Yu}}, \bibinfo {author} {\bibfnamefont {S.}~\bibnamefont {Zihlmann}}, \bibinfo {author} {\bibfnamefont {J.~C.}\ \bibnamefont {Abadillo-Uriel}}, \bibinfo {author} {\bibfnamefont {V.~P.}\ \bibnamefont {Michal}}, \bibinfo {author} {\bibfnamefont {N.}~\bibnamefont {Rambal}}, \bibinfo {author} {\bibfnamefont {H.}~\bibnamefont {Niebojewski}}, \bibinfo {author} {\bibfnamefont {T.}~\bibnamefont {Bedecarrats}}, \bibinfo {author} {\bibfnamefont {M.}~\bibnamefont {Vinet}}, \bibinfo {author} {\bibfnamefont {{\'E}.}~\bibnamefont {Dumur}}, \bibinfo {author} {\bibfnamefont {M.}~\bibnamefont {Filippone}}, \emph {et~al.},\ }\bibfield  {title} {\bibinfo {title} {Strong coupling between a photon and a hole spin in silicon},\ }\href {https://doi.org/10.5281/zenodo.7533669} {\bibfield  {journal} {\bibinfo  {journal} {Nature Nanotechnology}\ }\textbf {\bibinfo {volume} {18}},\ \bibinfo {pages} {741} (\bibinfo {year} {2023})}\BibitemShut {NoStop}%
\bibitem [{\citenamefont {De~Palma}\ \emph {et~al.}(2024)\citenamefont {De~Palma}, \citenamefont {Oppliger}, \citenamefont {Jang}, \citenamefont {Bosco}, \citenamefont {Jan{\'\i}k}, \citenamefont {Calcaterra}, \citenamefont {Katsaros}, \citenamefont {Isella}, \citenamefont {Loss},\ and\ \citenamefont {Scarlino}}]{de2023strong}%
  \BibitemOpen
  \bibfield  {author} {\bibinfo {author} {\bibfnamefont {F.}~\bibnamefont {De~Palma}}, \bibinfo {author} {\bibfnamefont {F.}~\bibnamefont {Oppliger}}, \bibinfo {author} {\bibfnamefont {W.}~\bibnamefont {Jang}}, \bibinfo {author} {\bibfnamefont {S.}~\bibnamefont {Bosco}}, \bibinfo {author} {\bibfnamefont {M.}~\bibnamefont {Jan{\'\i}k}}, \bibinfo {author} {\bibfnamefont {S.}~\bibnamefont {Calcaterra}}, \bibinfo {author} {\bibfnamefont {G.}~\bibnamefont {Katsaros}}, \bibinfo {author} {\bibfnamefont {G.}~\bibnamefont {Isella}}, \bibinfo {author} {\bibfnamefont {D.}~\bibnamefont {Loss}},\ and\ \bibinfo {author} {\bibfnamefont {P.}~\bibnamefont {Scarlino}},\ }\bibfield  {title} {\bibinfo {title} {Strong hole-photon coupling in planar ge for probing charge degree and strongly correlated states},\ }\href {https://doi.org/10.1038/s41467-024-54520-7} {\bibfield  {journal} {\bibinfo  {journal} {Nature Communications}\ }\textbf {\bibinfo {volume} {15}},\ \bibinfo {pages} {10177} (\bibinfo {year} {2024})}\BibitemShut
  {NoStop}%
\bibitem [{\citenamefont {Jan{\'\i}k}\ \emph {et~al.}(2024)\citenamefont {Jan{\'\i}k}, \citenamefont {Roux}, \citenamefont {Espinosa}, \citenamefont {Sagi}, \citenamefont {Baghdadi}, \citenamefont {Adletzberger}, \citenamefont {Calcaterra}, \citenamefont {Botifoll}, \citenamefont {Manj{\'o}n}, \citenamefont {Arbiol} \emph {et~al.}}]{janik2024strong}%
  \BibitemOpen
  \bibfield  {author} {\bibinfo {author} {\bibfnamefont {M.}~\bibnamefont {Jan{\'\i}k}}, \bibinfo {author} {\bibfnamefont {K.}~\bibnamefont {Roux}}, \bibinfo {author} {\bibfnamefont {C.~B.}\ \bibnamefont {Espinosa}}, \bibinfo {author} {\bibfnamefont {O.}~\bibnamefont {Sagi}}, \bibinfo {author} {\bibfnamefont {A.}~\bibnamefont {Baghdadi}}, \bibinfo {author} {\bibfnamefont {T.}~\bibnamefont {Adletzberger}}, \bibinfo {author} {\bibfnamefont {S.}~\bibnamefont {Calcaterra}}, \bibinfo {author} {\bibfnamefont {M.}~\bibnamefont {Botifoll}}, \bibinfo {author} {\bibfnamefont {A.~G.}\ \bibnamefont {Manj{\'o}n}}, \bibinfo {author} {\bibfnamefont {J.}~\bibnamefont {Arbiol}}, \emph {et~al.},\ }\bibfield  {title} {\bibinfo {title} {Strong charge-photon coupling in planar germanium enabled by granular aluminium superinductors},\ }\bibfield  {journal} {\bibinfo  {journal} {arXiv preprint arXiv:2407.03079}\ }\href {https://doi.org/10.48550/arXiv.2407.03079} {10.48550/arXiv.2407.03079} (\bibinfo {year} {2024})\BibitemShut
  {NoStop}%
\bibitem [{\citenamefont {Michal}\ \emph {et~al.}(2023)\citenamefont {Michal}, \citenamefont {Abadillo-Uriel}, \citenamefont {Zihlmann}, \citenamefont {Maurand}, \citenamefont {Niquet},\ and\ \citenamefont {Filippone}}]{michal2023tunable}%
  \BibitemOpen
  \bibfield  {author} {\bibinfo {author} {\bibfnamefont {V.}~\bibnamefont {Michal}}, \bibinfo {author} {\bibfnamefont {J.}~\bibnamefont {Abadillo-Uriel}}, \bibinfo {author} {\bibfnamefont {S.}~\bibnamefont {Zihlmann}}, \bibinfo {author} {\bibfnamefont {R.}~\bibnamefont {Maurand}}, \bibinfo {author} {\bibfnamefont {Y.-M.}\ \bibnamefont {Niquet}},\ and\ \bibinfo {author} {\bibfnamefont {M.}~\bibnamefont {Filippone}},\ }\bibfield  {title} {\bibinfo {title} {Tunable hole spin-photon interaction based on g-matrix modulation},\ }\href {https://doi.org/10.1103/PhysRevB.107.L041303} {\bibfield  {journal} {\bibinfo  {journal} {Physical Review B}\ }\textbf {\bibinfo {volume} {107}},\ \bibinfo {pages} {L041303} (\bibinfo {year} {2023})}\BibitemShut {NoStop}%
\bibitem [{\citenamefont {Bosco}\ \emph {et~al.}(2022)\citenamefont {Bosco}, \citenamefont {Scarlino}, \citenamefont {Klinovaja},\ and\ \citenamefont {Loss}}]{bosco2022fully}%
  \BibitemOpen
  \bibfield  {author} {\bibinfo {author} {\bibfnamefont {S.}~\bibnamefont {Bosco}}, \bibinfo {author} {\bibfnamefont {P.}~\bibnamefont {Scarlino}}, \bibinfo {author} {\bibfnamefont {J.}~\bibnamefont {Klinovaja}},\ and\ \bibinfo {author} {\bibfnamefont {D.}~\bibnamefont {Loss}},\ }\bibfield  {title} {\bibinfo {title} {Fully tunable longitudinal spin-photon interactions in si and ge quantum dots},\ }\href {https://doi.org/10.1103/PhysRevLett.129.066801} {\bibfield  {journal} {\bibinfo  {journal} {Physical Review Letters}\ }\textbf {\bibinfo {volume} {129}},\ \bibinfo {pages} {066801} (\bibinfo {year} {2022})}\BibitemShut {NoStop}%
\bibitem [{\citenamefont {Tosato}\ \emph {et~al.}(2023)\citenamefont {Tosato}, \citenamefont {Levajac}, \citenamefont {Wang}, \citenamefont {Boor}, \citenamefont {Borsoi}, \citenamefont {Botifoll}, \citenamefont {Borja}, \citenamefont {Mart{\'\i}-S{\'a}nchez}, \citenamefont {Arbiol}, \citenamefont {Sammak}, \citenamefont {Veldhorst},\ and\ \citenamefont {Scappucci}}]{Tosato2023}%
  \BibitemOpen
  \bibfield  {author} {\bibinfo {author} {\bibfnamefont {A.}~\bibnamefont {Tosato}}, \bibinfo {author} {\bibfnamefont {V.}~\bibnamefont {Levajac}}, \bibinfo {author} {\bibfnamefont {J.-Y.}\ \bibnamefont {Wang}}, \bibinfo {author} {\bibfnamefont {C.~J.}\ \bibnamefont {Boor}}, \bibinfo {author} {\bibfnamefont {F.}~\bibnamefont {Borsoi}}, \bibinfo {author} {\bibfnamefont {M.}~\bibnamefont {Botifoll}}, \bibinfo {author} {\bibfnamefont {C.~N.}\ \bibnamefont {Borja}}, \bibinfo {author} {\bibfnamefont {S.}~\bibnamefont {Mart{\'\i}-S{\'a}nchez}}, \bibinfo {author} {\bibfnamefont {J.}~\bibnamefont {Arbiol}}, \bibinfo {author} {\bibfnamefont {A.}~\bibnamefont {Sammak}}, \bibinfo {author} {\bibfnamefont {M.}~\bibnamefont {Veldhorst}},\ and\ \bibinfo {author} {\bibfnamefont {G.}~\bibnamefont {Scappucci}},\ }\bibfield  {title} {\bibinfo {title} {Hard superconducting gap in germanium},\ }\href {https://www.nature.com/articles/s43246-023-00351-w} {\bibfield  {journal} {\bibinfo  {journal} {Communications Materials}\
  }\textbf {\bibinfo {volume} {4}},\ \bibinfo {pages} {23} (\bibinfo {year} {2023})}\BibitemShut {NoStop}%
\bibitem [{\citenamefont {Sagi}\ \emph {et~al.}(2024)\citenamefont {Sagi}, \citenamefont {Crippa}, \citenamefont {Valentini}, \citenamefont {Janik}, \citenamefont {Baghumyan}, \citenamefont {Fabris}, \citenamefont {Kapoor}, \citenamefont {Hassani}, \citenamefont {Fink}, \citenamefont {Calcaterra}, \citenamefont {Chrastina}, \citenamefont {Isella},\ and\ \citenamefont {Katsaros}}]{Sagi2024}%
  \BibitemOpen
  \bibfield  {author} {\bibinfo {author} {\bibfnamefont {O.}~\bibnamefont {Sagi}}, \bibinfo {author} {\bibfnamefont {A.}~\bibnamefont {Crippa}}, \bibinfo {author} {\bibfnamefont {M.}~\bibnamefont {Valentini}}, \bibinfo {author} {\bibfnamefont {M.}~\bibnamefont {Janik}}, \bibinfo {author} {\bibfnamefont {L.}~\bibnamefont {Baghumyan}}, \bibinfo {author} {\bibfnamefont {G.}~\bibnamefont {Fabris}}, \bibinfo {author} {\bibfnamefont {L.}~\bibnamefont {Kapoor}}, \bibinfo {author} {\bibfnamefont {F.}~\bibnamefont {Hassani}}, \bibinfo {author} {\bibfnamefont {J.}~\bibnamefont {Fink}}, \bibinfo {author} {\bibfnamefont {S.}~\bibnamefont {Calcaterra}}, \bibinfo {author} {\bibfnamefont {D.}~\bibnamefont {Chrastina}}, \bibinfo {author} {\bibfnamefont {G.}~\bibnamefont {Isella}},\ and\ \bibinfo {author} {\bibfnamefont {G.}~\bibnamefont {Katsaros}},\ }\bibfield  {title} {\bibinfo {title} {A gate tunable transmon qubit in planar ge},\ }\href {https://www.nature.com/articles/s41467-024-50763-6} {\bibfield  {journal} {\bibinfo
  {journal} {Nature Communications}\ }\textbf {\bibinfo {volume} {15}},\ \bibinfo {pages} {6400} (\bibinfo {year} {2024})}\BibitemShut {NoStop}%
\bibitem [{\citenamefont {Kiyooka}\ \emph {et~al.}(2024)\citenamefont {Kiyooka}, \citenamefont {Tangchingchai}, \citenamefont {Noirot}, \citenamefont {Leblanc}, \citenamefont {Brun}, \citenamefont {Zihlmann}, \citenamefont {Maurand}, \citenamefont {Schmitt}, \citenamefont {Dumur}, \citenamefont {Hartmann}, \citenamefont {Lefloch},\ and\ \citenamefont {De~Franceschi}}]{Kiyooka_2024}%
  \BibitemOpen
  \bibfield  {author} {\bibinfo {author} {\bibfnamefont {E.}~\bibnamefont {Kiyooka}}, \bibinfo {author} {\bibfnamefont {C.}~\bibnamefont {Tangchingchai}}, \bibinfo {author} {\bibfnamefont {L.}~\bibnamefont {Noirot}}, \bibinfo {author} {\bibfnamefont {A.}~\bibnamefont {Leblanc}}, \bibinfo {author} {\bibfnamefont {B.}~\bibnamefont {Brun}}, \bibinfo {author} {\bibfnamefont {S.}~\bibnamefont {Zihlmann}}, \bibinfo {author} {\bibfnamefont {R.}~\bibnamefont {Maurand}}, \bibinfo {author} {\bibfnamefont {V.}~\bibnamefont {Schmitt}}, \bibinfo {author} {\bibfnamefont {E.}~\bibnamefont {Dumur}}, \bibinfo {author} {\bibfnamefont {J.-M.}\ \bibnamefont {Hartmann}}, \bibinfo {author} {\bibfnamefont {F.}~\bibnamefont {Lefloch}},\ and\ \bibinfo {author} {\bibfnamefont {S.}~\bibnamefont {De~Franceschi}},\ }\bibfield  {title} {\bibinfo {title} {Gatemon qubit on a germanium quantum-well heterostructure},\ }\bibfield  {journal} {\bibinfo  {journal} {Nano Letters}\ }\href {https://doi.org/10.1021/acs.nanolett.4c05539}
  {10.1021/acs.nanolett.4c05539} (\bibinfo {year} {2024})\BibitemShut {NoStop}%
\bibitem [{\citenamefont {Valentini}\ \emph {et~al.}(2024)\citenamefont {Valentini}, \citenamefont {Sagi}, \citenamefont {Baghumyan}, \citenamefont {de~Gijsel}, \citenamefont {Jung}, \citenamefont {Calcaterra}, \citenamefont {Ballabio}, \citenamefont {Aguilera~Servin}, \citenamefont {Aggarwal}, \citenamefont {Janik}, \citenamefont {Adletzberger}, \citenamefont {Seoane~Souto}, \citenamefont {Leijnse}, \citenamefont {Danon}, \citenamefont {Schrade}, \citenamefont {Bakkers}, \citenamefont {Chrastina}, \citenamefont {Isella},\ and\ \citenamefont {Katsaros}}]{Valentini2024}%
  \BibitemOpen
  \bibfield  {author} {\bibinfo {author} {\bibfnamefont {M.}~\bibnamefont {Valentini}}, \bibinfo {author} {\bibfnamefont {O.}~\bibnamefont {Sagi}}, \bibinfo {author} {\bibfnamefont {L.}~\bibnamefont {Baghumyan}}, \bibinfo {author} {\bibfnamefont {T.}~\bibnamefont {de~Gijsel}}, \bibinfo {author} {\bibfnamefont {J.}~\bibnamefont {Jung}}, \bibinfo {author} {\bibfnamefont {S.}~\bibnamefont {Calcaterra}}, \bibinfo {author} {\bibfnamefont {A.}~\bibnamefont {Ballabio}}, \bibinfo {author} {\bibfnamefont {J.}~\bibnamefont {Aguilera~Servin}}, \bibinfo {author} {\bibfnamefont {K.}~\bibnamefont {Aggarwal}}, \bibinfo {author} {\bibfnamefont {M.}~\bibnamefont {Janik}}, \bibinfo {author} {\bibfnamefont {T.}~\bibnamefont {Adletzberger}}, \bibinfo {author} {\bibfnamefont {R.}~\bibnamefont {Seoane~Souto}}, \bibinfo {author} {\bibfnamefont {M.}~\bibnamefont {Leijnse}}, \bibinfo {author} {\bibfnamefont {J.}~\bibnamefont {Danon}}, \bibinfo {author} {\bibfnamefont {C.}~\bibnamefont {Schrade}}, \bibinfo {author} {\bibfnamefont
  {E.}~\bibnamefont {Bakkers}}, \bibinfo {author} {\bibfnamefont {D.}~\bibnamefont {Chrastina}}, \bibinfo {author} {\bibfnamefont {G.}~\bibnamefont {Isella}},\ and\ \bibinfo {author} {\bibfnamefont {G.}~\bibnamefont {Katsaros}},\ }\bibfield  {title} {\bibinfo {title} {Parity-conserving cooper-pair transport and ideal superconducting diode in planar germanium},\ }\href {https://doi.org/10.1038/s41467-023-44114-0} {\bibfield  {journal} {\bibinfo  {journal} {Nature Communications}\ }\textbf {\bibinfo {volume} {15}},\ \bibinfo {pages} {169} (\bibinfo {year} {2024})}\BibitemShut {NoStop}%
\bibitem [{\citenamefont {Leblanc}\ \emph {et~al.}(2024)\citenamefont {Leblanc}, \citenamefont {Tangchingchai}, \citenamefont {Momtaz}, \citenamefont {Kiyooka}, \citenamefont {Hartmann}, \citenamefont {Gustavo}, \citenamefont {Thomassin}, \citenamefont {Brun}, \citenamefont {Schmitt}, \citenamefont {Zihlmann} \emph {et~al.}}]{leblanc2024}%
  \BibitemOpen
  \bibfield  {author} {\bibinfo {author} {\bibfnamefont {A.}~\bibnamefont {Leblanc}}, \bibinfo {author} {\bibfnamefont {C.}~\bibnamefont {Tangchingchai}}, \bibinfo {author} {\bibfnamefont {Z.~S.}\ \bibnamefont {Momtaz}}, \bibinfo {author} {\bibfnamefont {E.}~\bibnamefont {Kiyooka}}, \bibinfo {author} {\bibfnamefont {J.-M.}\ \bibnamefont {Hartmann}}, \bibinfo {author} {\bibfnamefont {F.}~\bibnamefont {Gustavo}}, \bibinfo {author} {\bibfnamefont {J.-L.}\ \bibnamefont {Thomassin}}, \bibinfo {author} {\bibfnamefont {B.}~\bibnamefont {Brun}}, \bibinfo {author} {\bibfnamefont {V.}~\bibnamefont {Schmitt}}, \bibinfo {author} {\bibfnamefont {S.}~\bibnamefont {Zihlmann}}, \emph {et~al.},\ }\bibfield  {title} {\bibinfo {title} {Gate and flux tunable sin(2$\varphi$) josephson element in proximitized junctions},\ }\href {https://arxiv.org/abs/2405.14695} {\bibfield  {journal} {\bibinfo  {journal} {arXiv preprint arXiv:2405.14695}\ } (\bibinfo {year} {2024})}\BibitemShut {NoStop}%
\bibitem [{\citenamefont {Hinderling}\ \emph {et~al.}(2024)\citenamefont {Hinderling}, \citenamefont {ten Kate}, \citenamefont {Coraiola}, \citenamefont {Haxell}, \citenamefont {Stiefel}, \citenamefont {Mergenthaler}, \citenamefont {Paredes}, \citenamefont {Bedell}, \citenamefont {Sabonis},\ and\ \citenamefont {Nichele}}]{PRXQuantum.5.030357}%
  \BibitemOpen
  \bibfield  {author} {\bibinfo {author} {\bibfnamefont {M.}~\bibnamefont {Hinderling}}, \bibinfo {author} {\bibfnamefont {S.~C.}\ \bibnamefont {ten Kate}}, \bibinfo {author} {\bibfnamefont {M.}~\bibnamefont {Coraiola}}, \bibinfo {author} {\bibfnamefont {D.}~\bibnamefont {Haxell}}, \bibinfo {author} {\bibfnamefont {M.}~\bibnamefont {Stiefel}}, \bibinfo {author} {\bibfnamefont {M.}~\bibnamefont {Mergenthaler}}, \bibinfo {author} {\bibfnamefont {S.}~\bibnamefont {Paredes}}, \bibinfo {author} {\bibfnamefont {S.}~\bibnamefont {Bedell}}, \bibinfo {author} {\bibfnamefont {D.}~\bibnamefont {Sabonis}},\ and\ \bibinfo {author} {\bibfnamefont {F.}~\bibnamefont {Nichele}},\ }\bibfield  {title} {\bibinfo {title} {Direct microwave spectroscopy of andreev bound states in planar $\mathrm{Ge}$ josephson junctions},\ }\href {https://doi.org/10.1103/PRXQuantum.5.030357} {\bibfield  {journal} {\bibinfo  {journal} {PRX Quantum}\ }\textbf {\bibinfo {volume} {5}},\ \bibinfo {pages} {030357} (\bibinfo {year} {2024})}\BibitemShut
  {NoStop}%
\bibitem [{\citenamefont {Lakic}\ \emph {et~al.}(2024)\citenamefont {Lakic}, \citenamefont {Lawrie}, \citenamefont {van Driel}, \citenamefont {Stehouwer}, \citenamefont {Veldhorst}, \citenamefont {Scappucci}, \citenamefont {Kuemmeth},\ and\ \citenamefont {Chatterjee}}]{lakic2024}%
  \BibitemOpen
  \bibfield  {author} {\bibinfo {author} {\bibfnamefont {L.}~\bibnamefont {Lakic}}, \bibinfo {author} {\bibfnamefont {W.~I.}\ \bibnamefont {Lawrie}}, \bibinfo {author} {\bibfnamefont {D.}~\bibnamefont {van Driel}}, \bibinfo {author} {\bibfnamefont {L.~E.}\ \bibnamefont {Stehouwer}}, \bibinfo {author} {\bibfnamefont {M.}~\bibnamefont {Veldhorst}}, \bibinfo {author} {\bibfnamefont {G.}~\bibnamefont {Scappucci}}, \bibinfo {author} {\bibfnamefont {F.}~\bibnamefont {Kuemmeth}},\ and\ \bibinfo {author} {\bibfnamefont {A.}~\bibnamefont {Chatterjee}},\ }\bibfield  {title} {\bibinfo {title} {A proximitized quantum dot in germanium},\ }\href {https://arxiv.org/abs/2405.02013} {\bibfield  {journal} {\bibinfo  {journal} {arXiv preprint arXiv:2405.02013}\ } (\bibinfo {year} {2024})}\BibitemShut {NoStop}%
\bibitem [{\citenamefont {Laubscher}\ \emph {et~al.}(2024)\citenamefont {Laubscher}, \citenamefont {Sau},\ and\ \citenamefont {Das~Sarma}}]{laubscher2024germanium}%
  \BibitemOpen
  \bibfield  {author} {\bibinfo {author} {\bibfnamefont {K.}~\bibnamefont {Laubscher}}, \bibinfo {author} {\bibfnamefont {J.~D.}\ \bibnamefont {Sau}},\ and\ \bibinfo {author} {\bibfnamefont {S.}~\bibnamefont {Das~Sarma}},\ }\bibfield  {title} {\bibinfo {title} {Germanium-based hybrid semiconductor-superconductor topological quantum computing platforms: Disorder effects},\ }\href {10.1103/PhysRevB.110.155431} {\bibfield  {journal} {\bibinfo  {journal} {Physical Review B}\ }\textbf {\bibinfo {volume} {110}},\ \bibinfo {pages} {155431} (\bibinfo {year} {2024})}\BibitemShut {NoStop}%
\bibitem [{\citenamefont {Adelsberger}\ \emph {et~al.}(2023)\citenamefont {Adelsberger}, \citenamefont {Legg}, \citenamefont {Loss},\ and\ \citenamefont {Klinovaja}}]{adelsberger2023microscopic}%
  \BibitemOpen
  \bibfield  {author} {\bibinfo {author} {\bibfnamefont {C.}~\bibnamefont {Adelsberger}}, \bibinfo {author} {\bibfnamefont {H.~F.}\ \bibnamefont {Legg}}, \bibinfo {author} {\bibfnamefont {D.}~\bibnamefont {Loss}},\ and\ \bibinfo {author} {\bibfnamefont {J.}~\bibnamefont {Klinovaja}},\ }\bibfield  {title} {\bibinfo {title} {Microscopic analysis of proximity-induced superconductivity and metallization effects in superconductor-germanium hole nanowires},\ }\href {https://doi.org/10.1103/PhysRevB.108.155433} {\bibfield  {journal} {\bibinfo  {journal} {Physical Review B}\ }\textbf {\bibinfo {volume} {108}},\ \bibinfo {pages} {155433} (\bibinfo {year} {2023})}\BibitemShut {NoStop}%
\bibitem [{\citenamefont {Luethi}\ \emph {et~al.}(2023)\citenamefont {Luethi}, \citenamefont {Laubscher}, \citenamefont {Bosco}, \citenamefont {Loss},\ and\ \citenamefont {Klinovaja}}]{luethi2023planar}%
  \BibitemOpen
  \bibfield  {author} {\bibinfo {author} {\bibfnamefont {M.}~\bibnamefont {Luethi}}, \bibinfo {author} {\bibfnamefont {K.}~\bibnamefont {Laubscher}}, \bibinfo {author} {\bibfnamefont {S.}~\bibnamefont {Bosco}}, \bibinfo {author} {\bibfnamefont {D.}~\bibnamefont {Loss}},\ and\ \bibinfo {author} {\bibfnamefont {J.}~\bibnamefont {Klinovaja}},\ }\bibfield  {title} {\bibinfo {title} {Planar josephson junctions in germanium: Effect of cubic spin-orbit interaction},\ }\href {https://doi.org/10.1103/PhysRevB.107.035435} {\bibfield  {journal} {\bibinfo  {journal} {Phys. Rev. B}\ }\textbf {\bibinfo {volume} {107}},\ \bibinfo {pages} {035435} (\bibinfo {year} {2023})}\BibitemShut {NoStop}%
\bibitem [{\citenamefont {Moghaddam}\ \emph {et~al.}(2014)\citenamefont {Moghaddam}, \citenamefont {Kernreiter}, \citenamefont {Governale},\ and\ \citenamefont {Z{\"u}licke}}]{moghaddam2014exporting}%
  \BibitemOpen
  \bibfield  {author} {\bibinfo {author} {\bibfnamefont {A.}~\bibnamefont {Moghaddam}}, \bibinfo {author} {\bibfnamefont {T.}~\bibnamefont {Kernreiter}}, \bibinfo {author} {\bibfnamefont {M.}~\bibnamefont {Governale}},\ and\ \bibinfo {author} {\bibfnamefont {U.}~\bibnamefont {Z{\"u}licke}},\ }\bibfield  {title} {\bibinfo {title} {Exporting superconductivity across the gap: Proximity effect for semiconductor valence-band states due to contact with a simple-metal superconductor},\ }\href {https://doi.org/10.1103/PhysRevB.89.184507} {\bibfield  {journal} {\bibinfo  {journal} {Physical Review B}\ }\textbf {\bibinfo {volume} {89}},\ \bibinfo {pages} {184507} (\bibinfo {year} {2014})}\BibitemShut {NoStop}%
\bibitem [{\citenamefont {Phan}\ \emph {et~al.}(2022)\citenamefont {Phan}, \citenamefont {Senior}, \citenamefont {Ghazaryan}, \citenamefont {Hatefipour}, \citenamefont {Strickland}, \citenamefont {Shabani}, \citenamefont {Serbyn},\ and\ \citenamefont {Higginbotham}}]{PhysRevLett.128.107701}%
  \BibitemOpen
  \bibfield  {author} {\bibinfo {author} {\bibfnamefont {D.}~\bibnamefont {Phan}}, \bibinfo {author} {\bibfnamefont {J.}~\bibnamefont {Senior}}, \bibinfo {author} {\bibfnamefont {A.}~\bibnamefont {Ghazaryan}}, \bibinfo {author} {\bibfnamefont {M.}~\bibnamefont {Hatefipour}}, \bibinfo {author} {\bibfnamefont {W.~M.}\ \bibnamefont {Strickland}}, \bibinfo {author} {\bibfnamefont {J.}~\bibnamefont {Shabani}}, \bibinfo {author} {\bibfnamefont {M.}~\bibnamefont {Serbyn}},\ and\ \bibinfo {author} {\bibfnamefont {A.~P.}\ \bibnamefont {Higginbotham}},\ }\bibfield  {title} {\bibinfo {title} {Detecting induced $p\ifmmode\pm\else\textpm\fi{}ip$ pairing at the al-inas interface with a quantum microwave circuit},\ }\href {https://doi.org/10.1103/PhysRevLett.128.107701} {\bibfield  {journal} {\bibinfo  {journal} {Phys. Rev. Lett.}\ }\textbf {\bibinfo {volume} {128}},\ \bibinfo {pages} {107701} (\bibinfo {year} {2022})}\BibitemShut {NoStop}%
\bibitem [{\citenamefont {Zhu}\ \emph {et~al.}(2021)\citenamefont {Zhu}, \citenamefont {Papaj}, \citenamefont {Nie}, \citenamefont {Xu}, \citenamefont {Gu}, \citenamefont {Yang}, \citenamefont {Guan}, \citenamefont {Wang}, \citenamefont {Li}, \citenamefont {Liu}, \citenamefont {Luo}, \citenamefont {Xu}, \citenamefont {Zheng}, \citenamefont {Fu},\ and\ \citenamefont {Jia}}]{doi:10.1126/science.abf1077}%
  \BibitemOpen
  \bibfield  {author} {\bibinfo {author} {\bibfnamefont {Z.}~\bibnamefont {Zhu}}, \bibinfo {author} {\bibfnamefont {M.}~\bibnamefont {Papaj}}, \bibinfo {author} {\bibfnamefont {X.-A.}\ \bibnamefont {Nie}}, \bibinfo {author} {\bibfnamefont {H.-K.}\ \bibnamefont {Xu}}, \bibinfo {author} {\bibfnamefont {Y.-S.}\ \bibnamefont {Gu}}, \bibinfo {author} {\bibfnamefont {X.}~\bibnamefont {Yang}}, \bibinfo {author} {\bibfnamefont {D.}~\bibnamefont {Guan}}, \bibinfo {author} {\bibfnamefont {S.}~\bibnamefont {Wang}}, \bibinfo {author} {\bibfnamefont {Y.}~\bibnamefont {Li}}, \bibinfo {author} {\bibfnamefont {C.}~\bibnamefont {Liu}}, \bibinfo {author} {\bibfnamefont {J.}~\bibnamefont {Luo}}, \bibinfo {author} {\bibfnamefont {Z.-A.}\ \bibnamefont {Xu}}, \bibinfo {author} {\bibfnamefont {H.}~\bibnamefont {Zheng}}, \bibinfo {author} {\bibfnamefont {L.}~\bibnamefont {Fu}},\ and\ \bibinfo {author} {\bibfnamefont {J.-F.}\ \bibnamefont {Jia}},\ }\bibfield  {title} {\bibinfo {title} {Discovery of segmented fermi surface induced by
  cooper pair momentum},\ }\href {https://doi.org/10.1126/science.abf1077} {\bibfield  {journal} {\bibinfo  {journal} {Science}\ }\textbf {\bibinfo {volume} {374}},\ \bibinfo {pages} {1381} (\bibinfo {year} {2021})}\BibitemShut {NoStop}%
\bibitem [{\citenamefont {Yuan}\ and\ \citenamefont {Fu}(2018)}]{PhysRevB.97.115139}%
  \BibitemOpen
  \bibfield  {author} {\bibinfo {author} {\bibfnamefont {N.~F.~Q.}\ \bibnamefont {Yuan}}\ and\ \bibinfo {author} {\bibfnamefont {L.}~\bibnamefont {Fu}},\ }\bibfield  {title} {\bibinfo {title} {Zeeman-induced gapless superconductivity with a partial fermi surface},\ }\href {https://doi.org/10.1103/PhysRevB.97.115139} {\bibfield  {journal} {\bibinfo  {journal} {Phys. Rev. B}\ }\textbf {\bibinfo {volume} {97}},\ \bibinfo {pages} {115139} (\bibinfo {year} {2018})}\BibitemShut {NoStop}%
\bibitem [{\citenamefont {Babkin}\ \emph {et~al.}(2024)\citenamefont {Babkin}, \citenamefont {Higginbotham},\ and\ \citenamefont {Serbyn}}]{10.21468/SciPostPhys.16.5.115}%
  \BibitemOpen
  \bibfield  {author} {\bibinfo {author} {\bibfnamefont {S.~S.}\ \bibnamefont {Babkin}}, \bibinfo {author} {\bibfnamefont {A.~P.}\ \bibnamefont {Higginbotham}},\ and\ \bibinfo {author} {\bibfnamefont {M.}~\bibnamefont {Serbyn}},\ }\bibfield  {title} {\bibinfo {title} {{Proximity-induced gapless superconductivity in two-dimensional Rashba semiconductor in magnetic field}},\ }\href {https://doi.org/10.21468/SciPostPhys.16.5.115} {\bibfield  {journal} {\bibinfo  {journal} {SciPost Phys.}\ }\textbf {\bibinfo {volume} {16}},\ \bibinfo {pages} {115} (\bibinfo {year} {2024})}\BibitemShut {NoStop}%
\bibitem [{\citenamefont {Ares}\ \emph {et~al.}(2013)\citenamefont {Ares}, \citenamefont {Golovach}, \citenamefont {Katsaros}, \citenamefont {Stoffel}, \citenamefont {Fournel}, \citenamefont {Glazman}, \citenamefont {Schmidt},\ and\ \citenamefont {De~Franceschi}}]{ares2013nature}%
  \BibitemOpen
  \bibfield  {author} {\bibinfo {author} {\bibfnamefont {N.}~\bibnamefont {Ares}}, \bibinfo {author} {\bibfnamefont {V.~N.}\ \bibnamefont {Golovach}}, \bibinfo {author} {\bibfnamefont {G.}~\bibnamefont {Katsaros}}, \bibinfo {author} {\bibfnamefont {M.}~\bibnamefont {Stoffel}}, \bibinfo {author} {\bibfnamefont {F.}~\bibnamefont {Fournel}}, \bibinfo {author} {\bibfnamefont {L.~I.}\ \bibnamefont {Glazman}}, \bibinfo {author} {\bibfnamefont {O.~G.}\ \bibnamefont {Schmidt}},\ and\ \bibinfo {author} {\bibfnamefont {S.}~\bibnamefont {De~Franceschi}},\ }\bibfield  {title} {\bibinfo {title} {Nature of tunable hole g factors in quantum dots},\ }\href {https://doi.org/10.1103/PhysRevLett.110.046602} {\bibfield  {journal} {\bibinfo  {journal} {Physical review letters}\ }\textbf {\bibinfo {volume} {110}},\ \bibinfo {pages} {046602} (\bibinfo {year} {2013})}\BibitemShut {NoStop}%
\bibitem [{Note1()}]{Note1}%
  \BibitemOpen
  \bibinfo {note} {The terms 'heavy hole' and `light hole' refer to their respectively large and small effective masses for the motion in the growth direction. However, these effective masses are reversed in the in-plane motion~\cite {Winkler:684956}. The names HH (LH) usually refer to their $z$ component of angular momentum $m=\pm 3/2$ ($\pm 1/2$).}\BibitemShut {Stop}%
\bibitem [{\citenamefont {Scappucci}\ \emph {et~al.}(2021)\citenamefont {Scappucci}, \citenamefont {Kloeffel}, \citenamefont {Zwanenburg}, \citenamefont {Loss}, \citenamefont {Myronov}, \citenamefont {Zhang}, \citenamefont {De~Franceschi}, \citenamefont {Katsaros},\ and\ \citenamefont {Veldhorst}}]{Scappucci2021}%
  \BibitemOpen
  \bibfield  {author} {\bibinfo {author} {\bibfnamefont {G.}~\bibnamefont {Scappucci}}, \bibinfo {author} {\bibfnamefont {C.}~\bibnamefont {Kloeffel}}, \bibinfo {author} {\bibfnamefont {F.~A.}\ \bibnamefont {Zwanenburg}}, \bibinfo {author} {\bibfnamefont {D.}~\bibnamefont {Loss}}, \bibinfo {author} {\bibfnamefont {M.}~\bibnamefont {Myronov}}, \bibinfo {author} {\bibfnamefont {J.-J.}\ \bibnamefont {Zhang}}, \bibinfo {author} {\bibfnamefont {S.}~\bibnamefont {De~Franceschi}}, \bibinfo {author} {\bibfnamefont {G.}~\bibnamefont {Katsaros}},\ and\ \bibinfo {author} {\bibfnamefont {M.}~\bibnamefont {Veldhorst}},\ }\bibfield  {title} {\bibinfo {title} {The germanium quantum information route},\ }\href {https://doi.org/10.1038/s41578-020-00262-z} {\bibfield  {journal} {\bibinfo  {journal} {Nature Reviews Materials}\ }\textbf {\bibinfo {volume} {6}},\ \bibinfo {pages} {926} (\bibinfo {year} {2021})}\BibitemShut {NoStop}%
\bibitem [{Note2()}]{Note2}%
  \BibitemOpen
  \bibinfo {note} {Other materials, such as Si, have a much smaller split-off energy splitting and a 6KP theory is more adequate for their description.}\BibitemShut {Stop}%
\bibitem [{\citenamefont {Bir}\ and\ \citenamefont {Pikus}(1974)}]{bir1974symmetry}%
  \BibitemOpen
  \bibfield  {author} {\bibinfo {author} {\bibfnamefont {G.~L.}\ \bibnamefont {Bir}}\ and\ \bibinfo {author} {\bibfnamefont {G.~E.}\ \bibnamefont {Pikus}},\ }\bibfield  {title} {\bibinfo {title} {Symmetry and strain-induced effects in semiconductors},\ }\href@noop {} {\bibfield  {journal} {\bibinfo  {journal} {(No Title)}\ } (\bibinfo {year} {1974})}\BibitemShut {NoStop}%
\bibitem [{\citenamefont {wang}\ \emph {et~al.}(2021)\citenamefont {wang}, \citenamefont {Marcellina}, \citenamefont {Hamilton}, \citenamefont {Cullen}, \citenamefont {Rogge}, \citenamefont {Salfi},\ and\ \citenamefont {Culcer}}]{npj2021}%
  \BibitemOpen
  \bibfield  {author} {\bibinfo {author} {\bibfnamefont {Z.}~\bibnamefont {wang}}, \bibinfo {author} {\bibfnamefont {E.}~\bibnamefont {Marcellina}}, \bibinfo {author} {\bibfnamefont {A.~R.}\ \bibnamefont {Hamilton}}, \bibinfo {author} {\bibfnamefont {J.~H.}\ \bibnamefont {Cullen}}, \bibinfo {author} {\bibfnamefont {S.}~\bibnamefont {Rogge}}, \bibinfo {author} {\bibfnamefont {J.}~\bibnamefont {Salfi}},\ and\ \bibinfo {author} {\bibfnamefont {D.}~\bibnamefont {Culcer}},\ }\bibfield  {title} {\bibinfo {title} {Optimal operation points for ultrafast, highly coherent ge hole spin-orbit qubits},\ }\bibfield  {journal} {\bibinfo  {journal} {npj Quantum Information}\ }\textbf {\bibinfo {volume} {7}},\ \href {https://doi.org/10.1038/s41534-021-00386-2} {10.1038/s41534-021-00386-2} (\bibinfo {year} {2021})\BibitemShut {NoStop}%
\bibitem [{\citenamefont {Futterer}\ \emph {et~al.}(2011)\citenamefont {Futterer}, \citenamefont {Governale}, \citenamefont {Z{\"u}licke},\ and\ \citenamefont {K{\"o}nig}}]{futterer2011band}%
  \BibitemOpen
  \bibfield  {author} {\bibinfo {author} {\bibfnamefont {D.}~\bibnamefont {Futterer}}, \bibinfo {author} {\bibfnamefont {M.}~\bibnamefont {Governale}}, \bibinfo {author} {\bibfnamefont {U.}~\bibnamefont {Z{\"u}licke}},\ and\ \bibinfo {author} {\bibfnamefont {J.}~\bibnamefont {K{\"o}nig}},\ }\bibfield  {title} {\bibinfo {title} {Band-mixing-mediated andreev reflection of semiconductor holes},\ }\href {https://doi.org/10.1103/PhysRevB.84.104526} {\bibfield  {journal} {\bibinfo  {journal} {Physical Review B—Condensed Matter and Materials Physics}\ }\textbf {\bibinfo {volume} {84}},\ \bibinfo {pages} {104526} (\bibinfo {year} {2011})}\BibitemShut {NoStop}%
\bibitem [{\citenamefont {Ivchenko}\ \emph {et~al.}(1996)\citenamefont {Ivchenko}, \citenamefont {Kaminski},\ and\ \citenamefont {R{\"o}ssler}}]{ivchenko1996heavy}%
  \BibitemOpen
  \bibfield  {author} {\bibinfo {author} {\bibfnamefont {E.}~\bibnamefont {Ivchenko}}, \bibinfo {author} {\bibfnamefont {A.~Y.}\ \bibnamefont {Kaminski}},\ and\ \bibinfo {author} {\bibfnamefont {U.}~\bibnamefont {R{\"o}ssler}},\ }\bibfield  {title} {\bibinfo {title} {Heavy-light hole mixing at zinc-blende (001) interfaces under normal incidence},\ }\href {https://doi.org/10.1103/PhysRevB.54.5852} {\bibfield  {journal} {\bibinfo  {journal} {Physical Review B}\ }\textbf {\bibinfo {volume} {54}},\ \bibinfo {pages} {5852} (\bibinfo {year} {1996})}\BibitemShut {NoStop}%
\bibitem [{\citenamefont {Babkin}\ \emph {et~al.}(2025)\citenamefont {Babkin}, \citenamefont {Joecker}, \citenamefont {Flensberg}, \citenamefont {Serbyn},\ and\ \citenamefont {Danon}}]{babkin2024superconducting}%
  \BibitemOpen
  \bibfield  {author} {\bibinfo {author} {\bibfnamefont {S.~S.}\ \bibnamefont {Babkin}}, \bibinfo {author} {\bibfnamefont {B.}~\bibnamefont {Joecker}}, \bibinfo {author} {\bibfnamefont {K.}~\bibnamefont {Flensberg}}, \bibinfo {author} {\bibfnamefont {M.}~\bibnamefont {Serbyn}},\ and\ \bibinfo {author} {\bibfnamefont {J.}~\bibnamefont {Danon}},\ }\bibfield  {title} {\bibinfo {title} {Superconducting proximity effect in two-dimensional hole gases},\ }\href {https://doi.org/10.1103/k4jh-pnxy} {\bibfield  {journal} {\bibinfo  {journal} {Phys. Rev. B}\ }\textbf {\bibinfo {volume} {111}},\ \bibinfo {pages} {214518} (\bibinfo {year} {2025})}\BibitemShut {NoStop}%
\bibitem [{\citenamefont {Sau}\ \emph {et~al.}(2010)\citenamefont {Sau}, \citenamefont {Lutchyn}, \citenamefont {Tewari},\ and\ \citenamefont {Das~Sarma}}]{PhysRevLett.104.040502}%
  \BibitemOpen
  \bibfield  {author} {\bibinfo {author} {\bibfnamefont {J.~D.}\ \bibnamefont {Sau}}, \bibinfo {author} {\bibfnamefont {R.~M.}\ \bibnamefont {Lutchyn}}, \bibinfo {author} {\bibfnamefont {S.}~\bibnamefont {Tewari}},\ and\ \bibinfo {author} {\bibfnamefont {S.}~\bibnamefont {Das~Sarma}},\ }\bibfield  {title} {\bibinfo {title} {Generic new platform for topological quantum computation using semiconductor heterostructures},\ }\href {https://doi.org/10.1103/PhysRevLett.104.040502} {\bibfield  {journal} {\bibinfo  {journal} {Phys. Rev. Lett.}\ }\textbf {\bibinfo {volume} {104}},\ \bibinfo {pages} {040502} (\bibinfo {year} {2010})}\BibitemShut {NoStop}%
\bibitem [{\citenamefont {Alicea}(2010)}]{alicea2010majorana}%
  \BibitemOpen
  \bibfield  {author} {\bibinfo {author} {\bibfnamefont {J.}~\bibnamefont {Alicea}},\ }\bibfield  {title} {\bibinfo {title} {Majorana fermions in a tunable semiconductor device},\ }\href {https://doi.org/10.1103/PhysRevB.81.125318} {\bibfield  {journal} {\bibinfo  {journal} {Physical Review B—Condensed Matter and Materials Physics}\ }\textbf {\bibinfo {volume} {81}},\ \bibinfo {pages} {125318} (\bibinfo {year} {2010})}\BibitemShut {NoStop}%
\bibitem [{\citenamefont {Lutchyn}\ \emph {et~al.}(2010)\citenamefont {Lutchyn}, \citenamefont {Sau},\ and\ \citenamefont {Das~Sarma}}]{lutchyn2010majorana}%
  \BibitemOpen
  \bibfield  {author} {\bibinfo {author} {\bibfnamefont {R.~M.}\ \bibnamefont {Lutchyn}}, \bibinfo {author} {\bibfnamefont {J.~D.}\ \bibnamefont {Sau}},\ and\ \bibinfo {author} {\bibfnamefont {S.}~\bibnamefont {Das~Sarma}},\ }\bibfield  {title} {\bibinfo {title} {Majorana fermions and a topological phase transition in semiconductor-superconductor heterostructures},\ }\href {https://doi.org/10.1103/PhysRevLett.105.077001} {\bibfield  {journal} {\bibinfo  {journal} {Physical review letters}\ }\textbf {\bibinfo {volume} {105}},\ \bibinfo {pages} {077001} (\bibinfo {year} {2010})}\BibitemShut {NoStop}%
\bibitem [{\citenamefont {Oreg}\ \emph {et~al.}(2010)\citenamefont {Oreg}, \citenamefont {Refael},\ and\ \citenamefont {von Oppen}}]{PhysRevLett.105.177002}%
  \BibitemOpen
  \bibfield  {author} {\bibinfo {author} {\bibfnamefont {Y.}~\bibnamefont {Oreg}}, \bibinfo {author} {\bibfnamefont {G.}~\bibnamefont {Refael}},\ and\ \bibinfo {author} {\bibfnamefont {F.}~\bibnamefont {von Oppen}},\ }\bibfield  {title} {\bibinfo {title} {Helical liquids and majorana bound states in quantum wires},\ }\href {https://doi.org/10.1103/PhysRevLett.105.177002} {\bibfield  {journal} {\bibinfo  {journal} {Phys. Rev. Lett.}\ }\textbf {\bibinfo {volume} {105}},\ \bibinfo {pages} {177002} (\bibinfo {year} {2010})}\BibitemShut {NoStop}%
\bibitem [{Note3()}]{Note3}%
  \BibitemOpen
  \bibinfo {note} {Github repository for the code:~\protect \href {https://github.com/dmichelpino/superholes}{https://github.com/dmichelpino/superholes}}\BibitemShut {NoStop}%
\bibitem [{\citenamefont {Paul}(2016)}]{paul20168}%
  \BibitemOpen
  \bibfield  {author} {\bibinfo {author} {\bibfnamefont {D.}~\bibnamefont {Paul}},\ }\bibfield  {title} {\bibinfo {title} {8-band k{\textperiodcentered} p modelling of mid-infrared intersubband absorption in ge quantum wells},\ }\bibfield  {journal} {\bibinfo  {journal} {Journal of Applied Physics}\ }\textbf {\bibinfo {volume} {120}},\ \href {https://doi.org/10.1063/1.4959259} {10.1063/1.4959259} (\bibinfo {year} {2016})\BibitemShut {NoStop}%
\end{thebibliography}%

\end{document}